\documentclass{ws-ijbc}

\usepackage{graphicx}

\newcommand{\be}{\begin{equation}}
\newcommand{\ee}{\end{equation}}

\newcommand{\al}{\alpha}
\newcommand{\ra}{\rightarrow}

\newcommand{\gm}{\gamma}

\newcommand{\lbd}{\lambda}

\begin{document}

\catchline{}{}{}{}{} 

\markboth{V.I. Yukalov, E.P. Yukalova, D. Sornette}{Dynamic Transition in Symbiotic Evolution Induced by Growth Rate Variation}

\title{DYNAMIC TRANSITION IN SYMBIOTIC EVOLUTION INDUCED BY GROWTH RATE VARIATION}

\author{V.I. YUKALOV}

\address{Department of Management, Technology and Economics, \\
ETH Z\"urich, Swiss Federal Institute of Technology,
Z\"urich CH-8092, Switzerland\\ and \\
Bogolubov Laboratory of Theoretical Physics, \\
Joint Institute for Nuclear Research, Dubna 141980, Russia \\
yukalov@theor.jinr.ru}

\author{E.P. YUKALOVA}
\address{Department of Management, Technology and Economics, \\
ETH Z\"urich, Swiss Federal Institute of Technology,
Z\"urich CH-8092, Switzerland\\ and \\
Laboratory of Information Technologies, \\
Joint Institute for Nuclear Research, Dubna 141980, Russia \\
lyukalov@ethz.ch}

\author{D. SORNETTE}
\address{Department of Management, Technology and Economics, \\
ETH Z\"urich, Swiss Federal Institute of Technology,
Z\"urich CH-8092, Switzerland \\ and \\
Swiss Finance Institute, c/o University of Geneva, \\
40 blvd. Du Pont d'Arve, CH 1211 Geneva 4, Switzerland \\
dsornette@ethz.ch}

\maketitle

\begin{history}
\received{(to be inserted by publisher)}
\end{history}

\begin{abstract}
In a standard bifurcation of a dynamical system,  
the stationary points (or more generally attractors) change qualitatively
when varying a control parameter. Here we describe a novel unusual effect, 
when the change of a parameter, e.g. a growth rate, does not influence the stationary 
states, but nevertheless leads to a qualitative change of dynamics.  For instance, such a
dynamic transition can be between the convergence to a stationary state and a 
strong increase without stationary states, or between the convergence to one 
stationary state and that to a different state. This effect is illustrated for a dynamical 
system describing two symbiotic populations, one of which exhibits a growth rate 
larger than the other one. We show that, although the stationary states of the dynamical 
system do not depend on the growth rates, the latter influence the boundary of the 
basins of attraction. This change of the basins of attraction explains this unusual 
effect of the quantitative change of dynamics by growth rate variation.  
\end{abstract}

\keywords{Dynamics of symbiotic populations, growth rate, functional carrying capacity,
dynamic transitions, basin of attraction, bifurcation. }

\section{Introduction}

It is well known that varying the parameters controlling a dynamical system can 
change the existing fixed points. In the case of a bifurcation, this can 
qualitatively change the dynamical behavior of the system, leading to what is 
often called a {\it dynamic phase transition} or {\it bifurcation transition}
\cite{Schuster_1}. When the considered parameter characterizes a growth 
rate, its variance usually leads just to the acceleration or slowing down
of the convergence towards the stable fixed points, but does not induce
dynamic transitions. In the present paper, we show that this common wisdom 
is not always correct. It may happen that a varying growth rate, while
not influencing the fixed points, can nevertheless induce qualitative changes 
in the dynamics similar to a bifurcation transition, while no bifurcation of 
stationary states occurs. We demonstrate this unusual effect by considering 
an autonomous dynamical system describing co-evolving symbiotic populations. 

Qualitatively, the fact that the evolution of symbiotic species essentially 
depends on their proliferation rates has been discussed in many publications.
For example, it is known that, for optimal development, mutualistic symbiotic 
species ``must keep pace" between each other \cite{Bennett_2}. 
The growth rate of fungal endophites can either enhance or reduce plant 
reproduction \cite{Rodriguez_3}. Reef corals engage in symbiosis 
with single-celled Dinoflagelate Algae, from which they acquire photosynthetic 
products that support most of their energetic needs and help them build calcium 
carbonate skeletons that form the foundation of coral reefs. Unsufficient growth 
of the Algae results in the increased coral bleaching and mortality 
\cite{Cunning_4}. Intense proliferation of viral pathogens, such as the 
Deformed Wing Virus, undermines honey bee colonies and can lead to their 
collapse \cite{Prisco_5}. The symbiosis that is the most important for 
humans is the one between the human body and the multitudes of about 
$10^{14}$ microorganisms, consisting of bacteria, archaea, and fungi,  
participating in the synthesis of essential vitamins and amino acids, as 
well as in the degradation of otherwise indigestible plant material and of 
certain drugs and pollutants in the guts \cite{Ley_6}. It is 
now known that our gut microbiome coevolves with us and that their evolution
can have major consequences, both beneficial and harmful, for human health
\cite{Ley_7}. It is well established that mycorrhizal fungi symbiosis with 
plants is beneficial for plant growth and reproduction. However too fast 
proliferation of the fungi at the early stage of the plant seedling can have 
negative effects because of the carbon costs associated with sustaining
the fungi \cite{Varga_8}. 
         
Usually, in a symbiotic coexistence, the faster growth of species has just 
the effect of a faster convergence to the stationary states. Although in some 
cases, the change of a growth rate can result in a different state. We suggest 
a mathematical model demonstrating the existence of the unusual effect of 
a qualitative change of dynamical behavior induced by the variation of growth 
rates, while the stationary points are left untouched. Strictly speaking, 
this effect can occur in different nonlinear dynamical systems with feedbacks. 
We suggest a symbiotic interpretation for concretness and for explaining that 
the effect can really occur in nature. Section 2 presents the model. Section 3 
studies the stable stationary states. Section 4 reviews the cases where a change 
of a growth rate only modifies the rate of convergence to the stationary states. 
Section 5 covers the cases where the change of a growth rate leads to dynamic 
transitions. Section 6 describes the scale-separation approach that provides 
approximate solutions of the equations in the limit of large differences between 
the growth rates of the two species. Section 7 summarizes the article and concludes 
by suggesting a biological fungi-plant system in which the reported effect could 
be at work.

\section{Symbiosis with Functional Carrying Capacity}

Symbiotic species interact with each other through influencing their carrying 
capacities \cite{Boucher_9,Douglas_10,Sapp_11,Ahmadjian_12}. 
A mathematical model characterizing these interactions has been suggested in
\cite{YYS_13,YYS_14,YYS_30,YYS_15,YYS_16}, where a detailed justification 
and discussions on numerous possible applications for biological and social 
symbiotic systems can be found. In these previous articles, symbiotic
species were assumed to enjoy the same growth rate. Here, we analyze the
influence of the birth rates on the behavior of the populations. It turns out
that changing birth rates not merely modifies the velocity of the growth processes, 
but can also lead to the unexpected effect of a drastic change in the dynamics of  
populations.     
   
Let us consider symbiotic species, enumerated by the index $i$ and whose 
populations are denoted by $N_i$. Each population satisfies the logistic-type
equation
\be
\label{1}
 \frac{dN_i}{dt} = \gm_i \left ( N_i - \; \frac{N_i^2}{K_i} \right ) \;  ,
\ee
where $\gamma_i$ is a birth rate and $K_i$ is the carrying capacity, generally
being a functional of the populations \cite{YYS_13,YYS_14,YYS_30}.
By employing a scaling parameter $C_i$, it is always possible to introduce 
dimensionless quantities for each of the populations and for the related 
carrying capacity, respectively,
\be
\label{2}
x_i \equiv \frac{N_i}{C_i} \; , \qquad y_i \equiv \frac{K_i}{C_i} \;    .
\ee
Then equation (\ref{1}) reads as
\be
\label{3}
 \frac{dx_i}{dt} = \gm_i \left ( x_i - \; \frac{x_i^2}{y_i} \right ) \;  .
\ee

The explicit expression for the carrying capacity can be derived in the 
following way. Keeping in mind that the carrying capacity $y_i$ is a function 
of the dimensionless populations $x_i$, it is possible to express it as a 
Taylor expansion
$$
y_i = y_i(x_1,x_2,\ldots ) \; , \qquad
y_i \simeq 1 + \sum_j c_j x_j + \sum_{kl} c_{kl} x_k x_l  \;   .
$$
Note that the first term of the expansion can be made equal to $1$ by the 
appropriate choice of the scaling parameter $C_i$. When the values of $x_i$ are 
small, it is admissible to limit oneself to a finite number of terms in 
the above expansion. However, the assumption of the smallness of $x_i$ 
is too restrictive. The generalization to arbitrary values of the 
variables $x_i$ can be accomplished by resorting to the self-similar 
approximation theory \cite{Yukalov_17,Yukalov_18}, providing an effective 
summation of the infinite series. Using exponential self-similar 
summation \cite{Yukalov_19} we obtain
\be
\label{4}
y_i = \exp \left ( \sum_j a_j x_j \right ) \;   .
\ee

The growth rate $\gamma_i$ can be presented as the difference 
$\gamma_i = \gamma_{birth} - \gamma_{death}$ of a birth rate and a death rate.
In what follows, we assume that the birth rate surpasses the death rate, 
so that the growth rate is positive, $\gamma_i > 0$. 

We consider the symbiosis of two species and define the relative growth rate
\be
\label{5}
 \al \equiv \frac{\gm_1}{\gm_2} \;  .
\ee
To simplify the notation, we denote
\be
\label{6}
  x \equiv x_1 \; , \qquad z \equiv x_2  
\ee
and measure time in units of $1/\gamma_2$. Thus we come to the 
two-dymensional dynamical system describing the symbiosis of the 
dimensionless populations $x$ and $z$, with the equations
\be
\label{7}
 \frac{dx}{dt} = \al \left ( x - \; \frac{x^2}{y_1} \right )  
\ee
and
\be
\label{8}
\frac{dz}{dt} =  z - \; \frac{z^2}{y_2}  \; .
\ee

The mutual carrying capacities, in the case of symbiosis, depend on the 
populations of the other species. The species self-action is excluded,
since it is related to other effects influencing the carrying capacity by
self-improvement or self- destruction, which are not connected to symbiosis
\cite{YYS_29,YYS_14}. We set the notation $a_{12} = b$ and $a_{21} = g$. 
Then the carrying capacities take the form
\be
\label{9}
  y_1 = e^{bz} \; , \qquad y_2 = e^{gx} \; .
\ee

Since our aim is to analyze the dynamics under different growth rates,
we can assume, without loss of generality, that $\gamma_1$ is larger 
than $\gamma_2$, so that
\be
\label{10}
\al > 1 \qquad (\gm_1 > \gm_2) \;  .
\ee
In particular, $\alpha$ can be much larger than one, which would classify
the variable $x$ as fast and $z$ as slow. 
 
It is useful to emphasize that the system of equations (\ref{7}) and (\ref{8})
describes all types of symbiosis, depending on the symbiotic parameters 
$b$ and $g$. Thus, mutualism corresponds to the case
$$
b > 0 , ~~   g > 0 \qquad (mutualism) \;  . 
$$
Parasitic symbiosis is characterized by one of the inequalities
\begin{eqnarray}
\nonumber
\left.
\begin{array}{ll}
b > 0, ~ & ~ g < 0 \\
b < 0, ~ & ~ g > 0 \\
b < 0, ~ & ~ g < 0
\end{array} \right \} \qquad (parasitism) \; .
\end{eqnarray}
And commensalism happens under one of the conditions
\begin{eqnarray}
\nonumber
\left.
\begin{array}{ll}
~~~~~ b > 0, ~ & ~ g = 0 \\
~~~~~ b =0, ~ & ~ g > 0 
\end{array} \right \} \qquad (commensalism) \; .
\end{eqnarray}
This classification derives from the fact that the signs of the parameters $b$
and $g$ define whether the mutual influence on the carrying capacities
is beneficial (positive sign) or destructive (negative sign). While a zero
parameter signifies the absence of influence.

\section{Evolutionary Stable Stationary States}

The dynamical system under consideration is given by the equations
\be
\label{11}
\frac{dx}{dt} = \al \left ( x - x^2 e^{-bz} \right ) \;   , \qquad
\frac{dz}{dt} =  z - z^2 e^{-gx}  \;   ,
\ee
with the parameters spanning the following intervals
\be  
\label{12}
- \infty < b < \infty \; , \qquad - \infty < g < \infty \; , \qquad
\al > 1 \;  .
\ee
We are looking for non-negative solutions $x = x(t) \geq 0$ and 
$z = z(t) \geq 0$, with initial conditions
$$
x_0 = x(0) \; , \qquad z_0 = z(0) \;   .
$$

There are three trivial fixed points: the unstable node $\{0,0\}$, with the
characteristic exponents $\lambda_1 = 1$ and $\lambda_2 = \alpha$; a saddle
$\{1,0\}$, with the characteristic exponents $\lambda_1 = 1$ and 
$\lambda_2 = - \alpha$; and the saddle $\{0,1\}$, with the characteristic 
exponents $\lambda_1 = - 1$ and $\lambda_2 = \alpha$. 

The nontrivial stationary states are defined by the equations
\be
\label{13}
x^* = e^{bz^*} \; , \qquad z^* = e^{gx^*} \;   ,
\ee
which can also be represented as
$$
 x^* = \exp \left ( be^{gx^*} \right ) \; , \qquad
 z^* = \exp \left ( ge^{bz^*} \right ) \;  .
$$

It is important to stress that the stationary states, defined by equations (\ref{13}), 
do not depend on the growth rate $\alpha$.  

The characteristic exponents are the solutions to the equation
$$
 \lbd^2 + c_1 \lbd + c_0 = 0 \;  ,
$$
where
$$
 c_0 = \lbd_1 \lbd_2 = \al\left ( 1 - bg x^* z^* \right ) \; , 
\qquad
c_1 = - (\lbd_1 + \lbd_2) = 1 + \al \;  .
$$
Thus
\be
\label{14}
\lbd_{1,2} = -\; \frac{1}{2} \; (1 +\al) \pm \; 
\sqrt{ ( 1-\al)^2 + 4\al bg x^*z^* } \;    .
\ee

The plane of the parameters $b$ and $g$ is separated into five regions with
different behavior of the solutions. 

In the region of {\it strong mutualism}
\be
\label{15}
 A = \{ b>0 \; , ~ g > g_c(b) \} \;  ,
\ee
there are no fixed points. 

In the region of {\it moderate mutualism}
\be
\label{16}
 B = \{ b>0 \; , ~ 0< g < g_c(b) \} \;  ,
\ee
there are two fixed points, a stable node $\{x_1^*,z_1^*\}$, with a limited
basin of attraction, and a saddle $\{x_2^*,z_2^*\}$, such that
$$
1 < x_1^* < x_2^* \; , \qquad  1 < z_1^* < z_2^* \; .
$$
 
The region of {\it one-side parasitism}
\begin{eqnarray}
\label{17}
C = \left \{ \begin{array}{ll}
b < 0 , ~ & ~ g > 0 \\
b > 0 , ~ & ~ g < 0 
\end{array} \right. \; ,
\end{eqnarray}
where one of the species is parasitic, while the other is not, contains
a stable focus $\{x^*,z^*\}$, with the basin of attraction being the whole
plane of initial conditions $x_0$ and $z_0$.

The region of {\it two-side parasitism}
\be
\label{18}
D = \{ b < 0 \; , ~ g < 0 \} = D_1 \bigcup D_2 \;   ,
\ee
where both species are parasitic, is divided into two subregions. In the
subregion
\begin{eqnarray}
\label{19}
D_1 = \left \{ \begin{array}{rr}
b < -e , ~ & ~ g < g_1(b) < -e \\
b < -e , ~ & ~ g_2(b) < g  < 0 \\
- e < b < 0 ,  ~ & ~ g < 0
\end{array} \right.    \; ,
\end{eqnarray}
there exists only one stable node, with the basin of attraction being the 
whole plane of initial conditions $x_0$ and $z_0$. While the subregion
\be
\label{20}
D_2 = \{ b < -e \; , ~~~ g_1(b) < g < g_2(b) < - e \} = D \setminus D_1 \;   
\ee
contains a stable node $\{x_1^*,z_1^*\}$, with a limited basin of attraction,
a saddle $\{x_2^*,z_2^*\}$, and another stable node $\{x_3^*,z_3^*\}$, with
a limited basin of attraction. The fixed points are related by the 
inequalities
$$
1 > x_1^* > x_2^* > x_3^* \; , \qquad  
z_1^* < z_2^* < z_3^* < 1 \;  .
$$
These regions are shown in Fig. 1.  

\begin{figure}[ht]
\vspace{9pt}
\centerline{
\hbox{ \includegraphics[width=5.5cm]{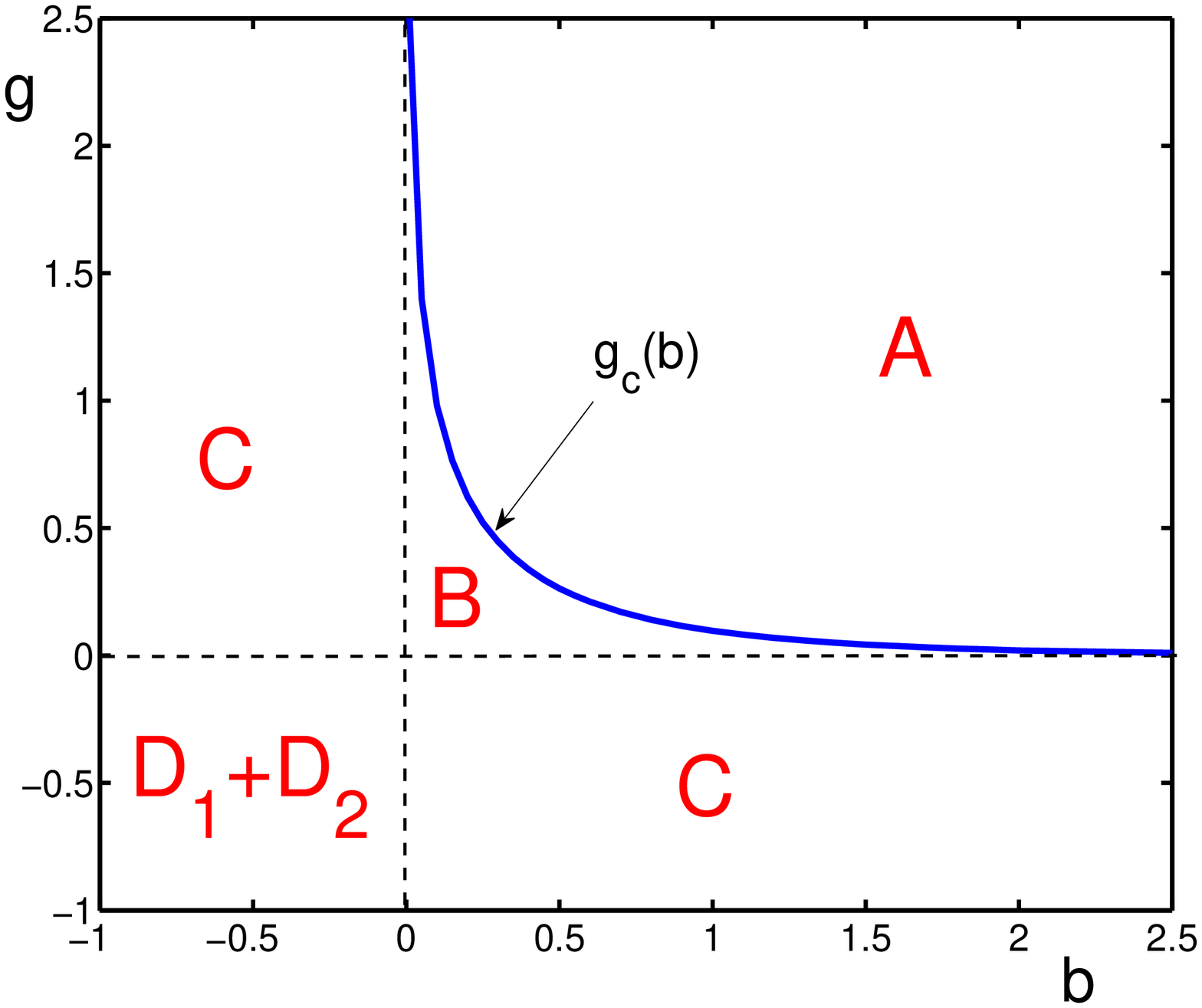}
\hspace{2.5cm}
\includegraphics[width=5.5cm]{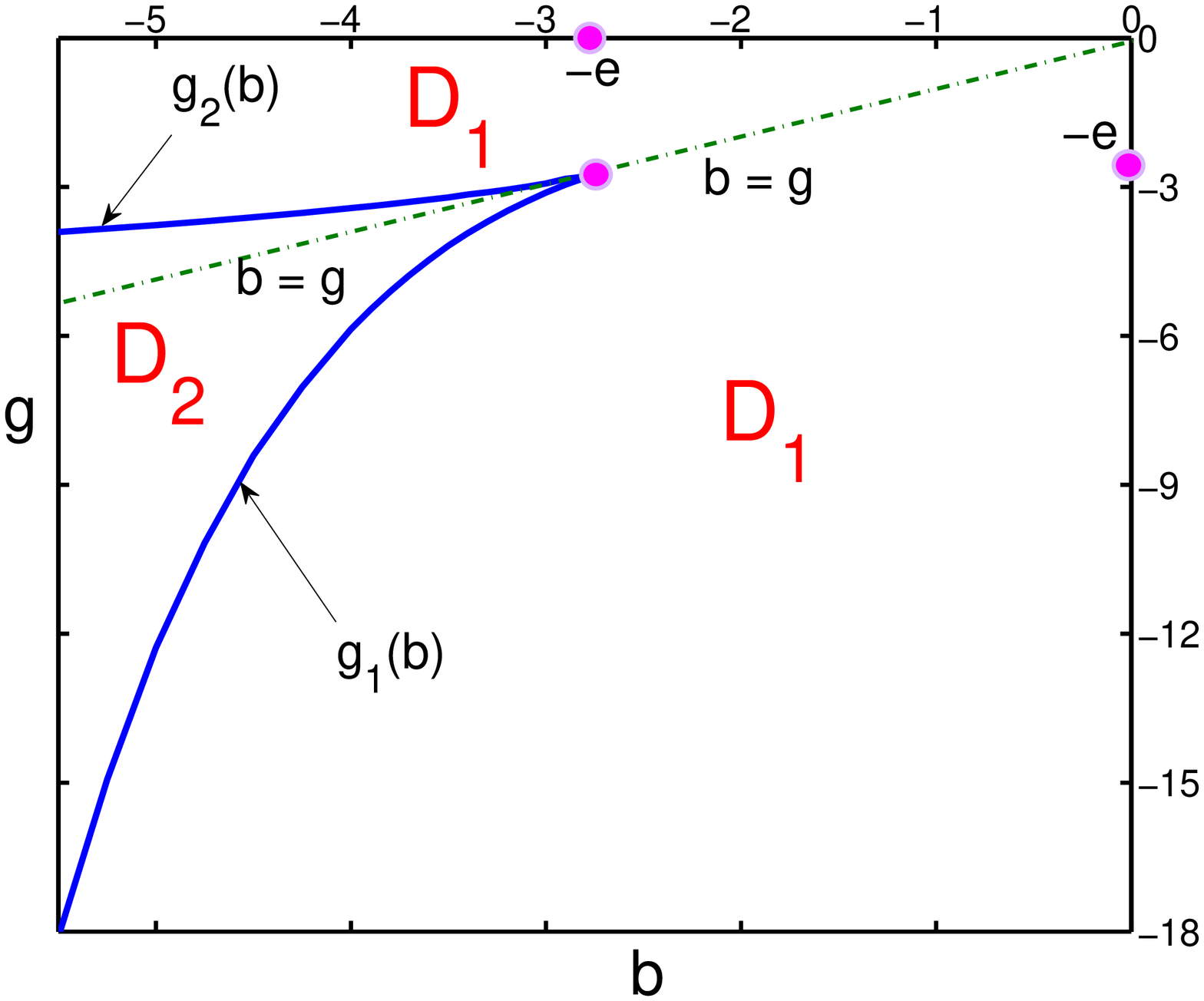} } }
\caption{Regions on the plane $b-g$, as discussed in the text. In
region $A$, there are no fixed points. In region $B$, there exist two
fixed points, one being a stable node, while the other is a saddle. Region $C$
contains one fixed point being a stable focus. Region $D$ is subdivided into
two subregions shown in more details in the right panel. In region $D_1$,
there is one stable fixed point, being a stable node. In region $D_2$,
there are three fixed points, two of them being stable nodes, while the third is
a saddle.
}
\label{fig:Fig.1}
\end{figure}

\section{Growth-Rate Acceleration of Population Dynamics}

Since the stationary states, defined by equations (\ref{13}), do not depend 
on the growth rate $\alpha$, it is reasonable to expect that the increase 
of the latter should result only in the acceleration of the temporal dynamics
of the symbiotic populations, without qualitative changes in the overall 
picture. In many cases, it is really so, as is explained below.

\subsection{Strong mutualistic growth of populations}

In region $A$, where there are no fixed points, mutualistic populations 
grow faster when increasing the growth rate $\alpha$, displaying the same 
qualitative behavior, as is illustrated in Fig. 2.

\begin{figure}[ht]
\vspace{9pt}
\centerline{
\hbox{ \includegraphics[width=5.5cm]{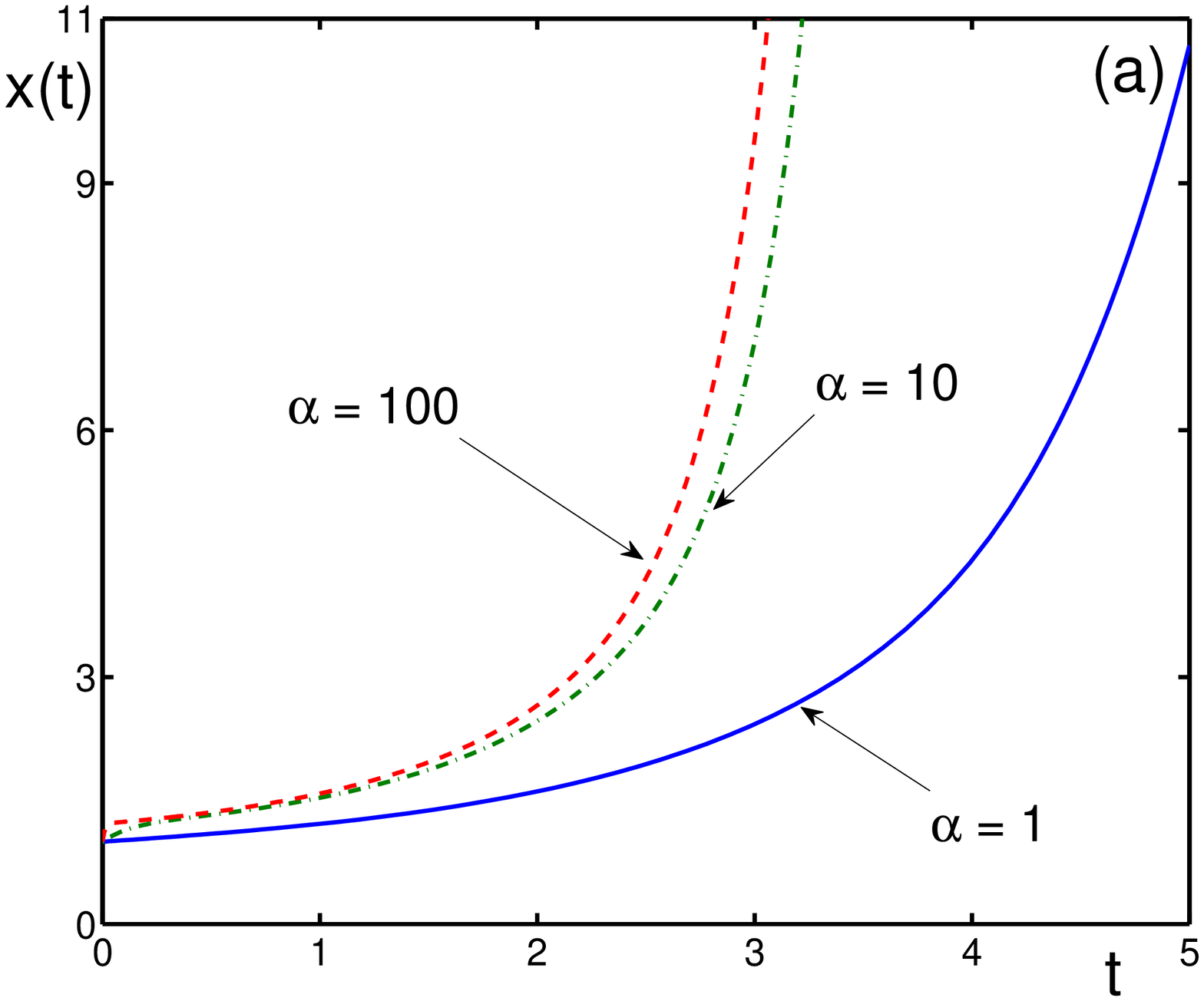}
\hspace{2.5cm}
\includegraphics[width=5.5cm]{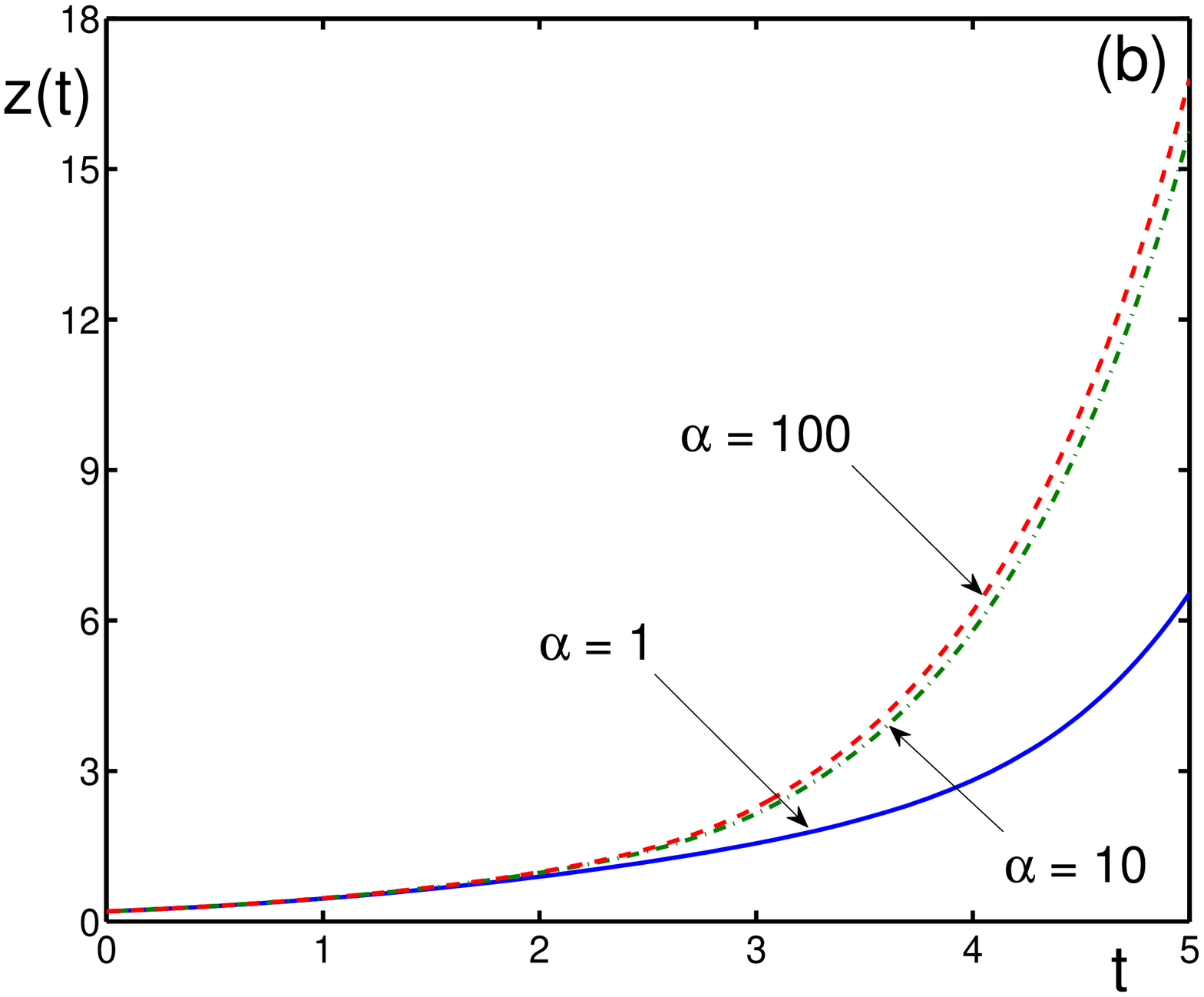} } }
\caption{Dynamics of populations $x(t)$ and $z(t)$ in the parametric
region $A$, for different growth rates. Here $b = 1$ and
$g = 0.5 > g_c \approx 0.0973$. The initial conditions are
$\{x_0 = 1, z_0 = 0.2\}$. (a) Population $x(t)$ for $\al = 1$ (solid line),
$\al = 10$ (dashed-dotted line), and $\al = 100$ (dashed line);
(b) population $z(t)$ for $\al = 1$ (solid line), $\al = 10$ (dashed-dotted line),
and $\al = 100$ (dashed line).
}
\label{fig:Fig.2}
\end{figure}

\subsection{Convergence to single stationary states}

In region $C$, there is just a single fixed point, being a stable focus.
The convergence to the stationary state can be of slightly different type, as
is shown in Figs. 3 and 4, but it is always faster when the parameter 
$\alpha$ is larger. The phase portrait is presented in Fig. 5.        

The region $D_1$ contains a single stationary state, a stable node. Again,
the convergence to the stationary state is faster when the growth
rate $\alpha$ is larger, as is shown in Fig. 6.

\begin{figure}[ht]
\centerline{
\hbox{ \includegraphics[width=5.5cm]{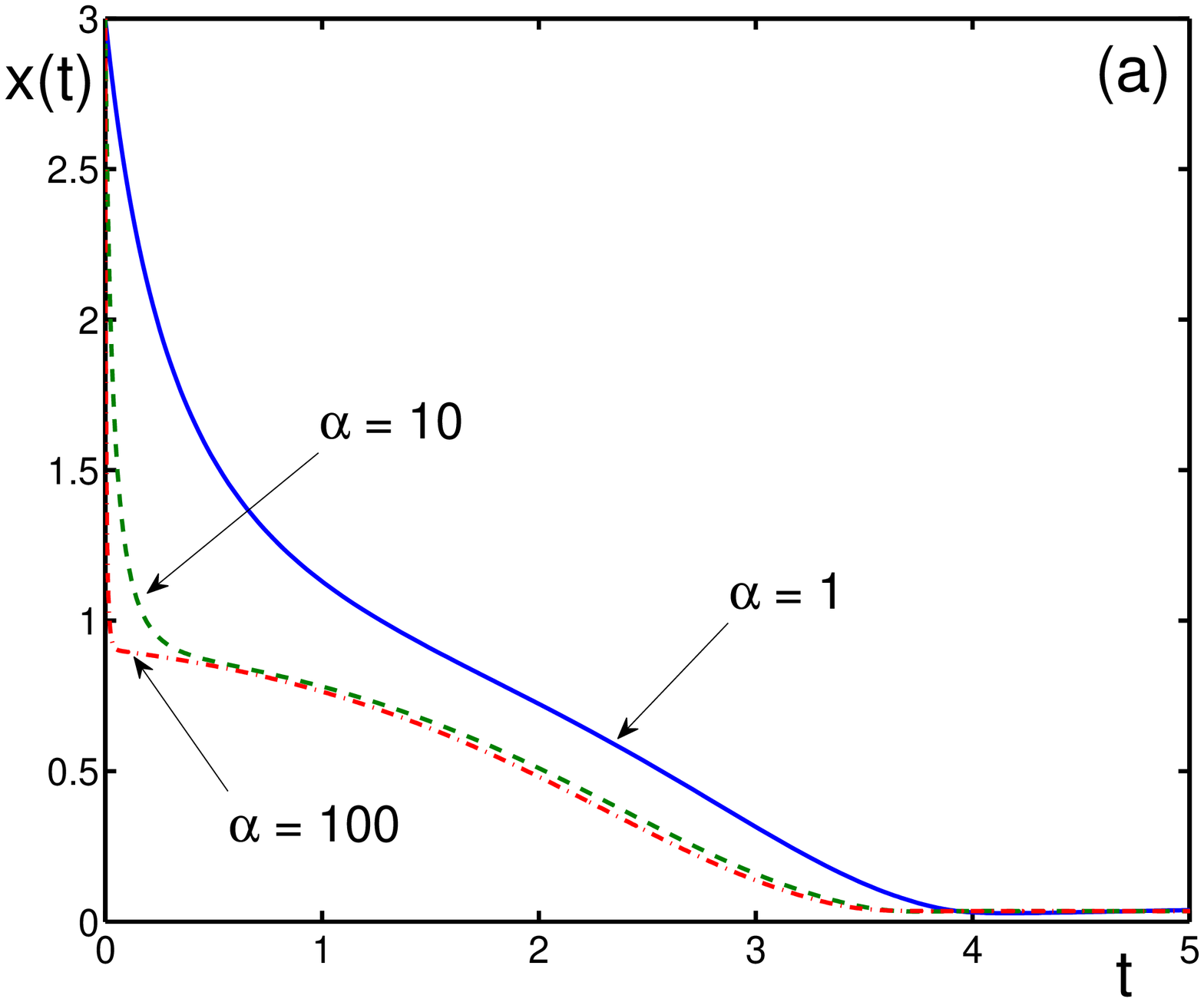} \hspace{2.5cm}
\includegraphics[width=5.5cm]{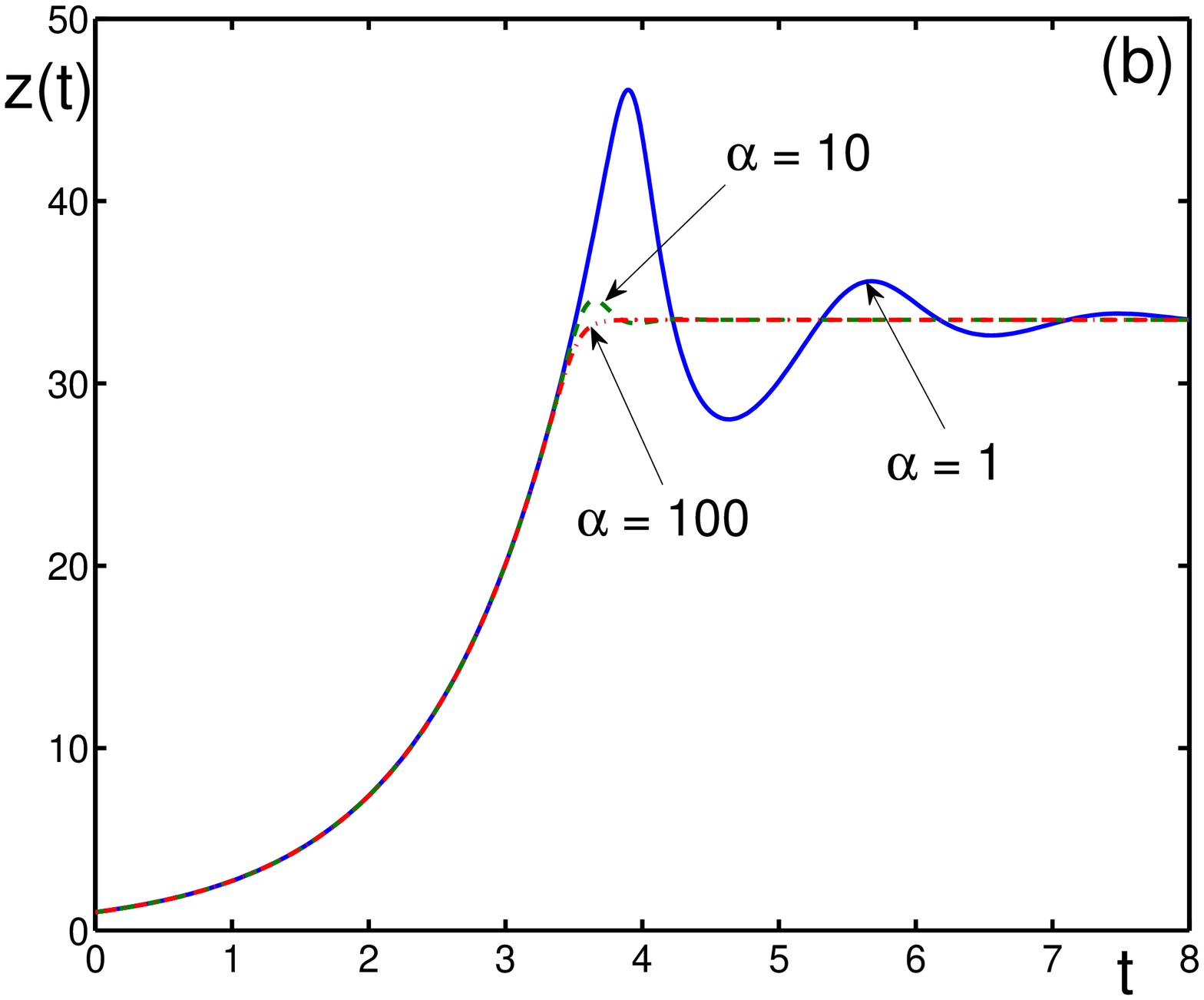} }  }
\vskip 9pt
\centerline{
\hbox{ \includegraphics[width=5.5cm]{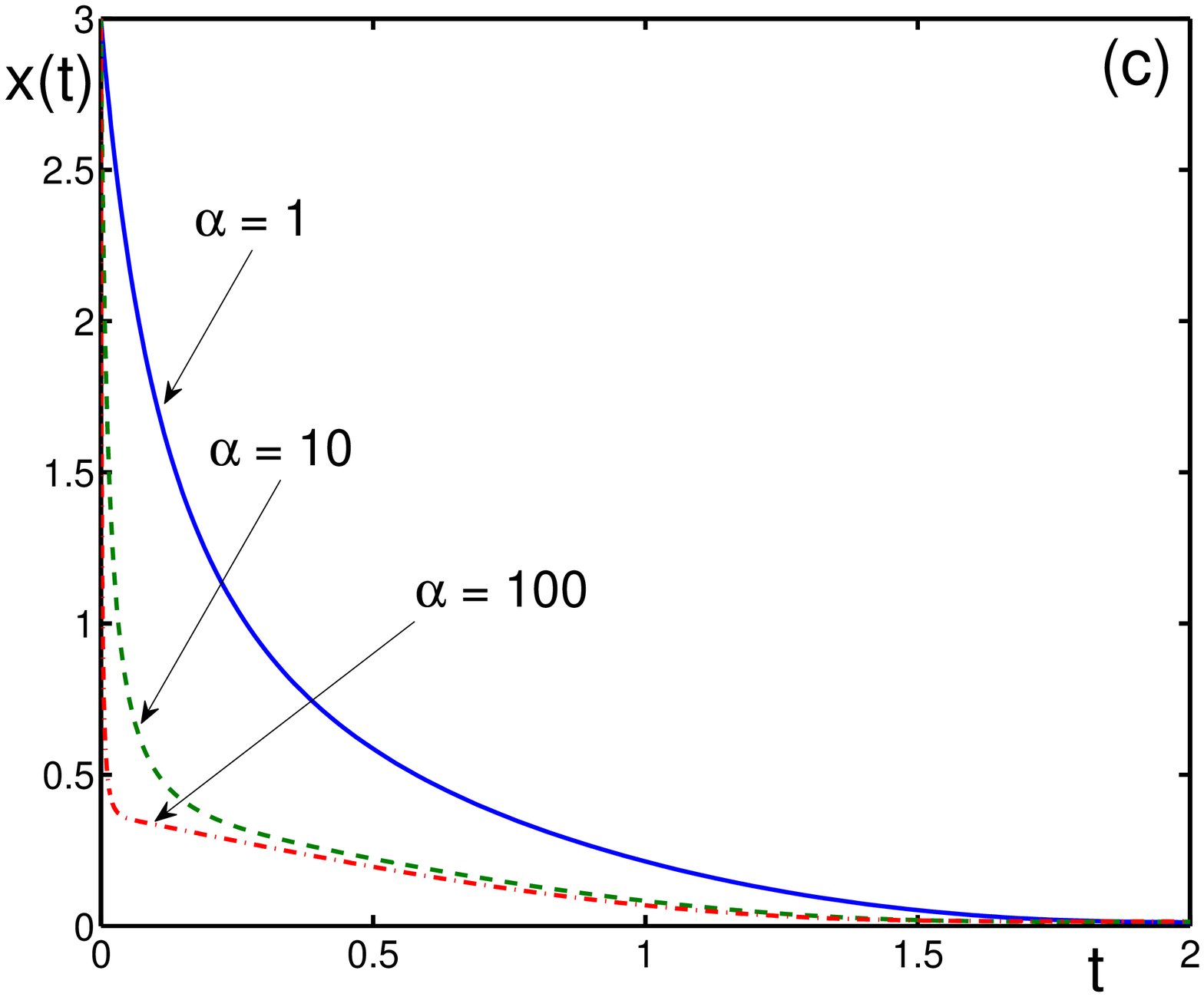} \hspace{2.5cm}
\includegraphics[width=5.5cm]{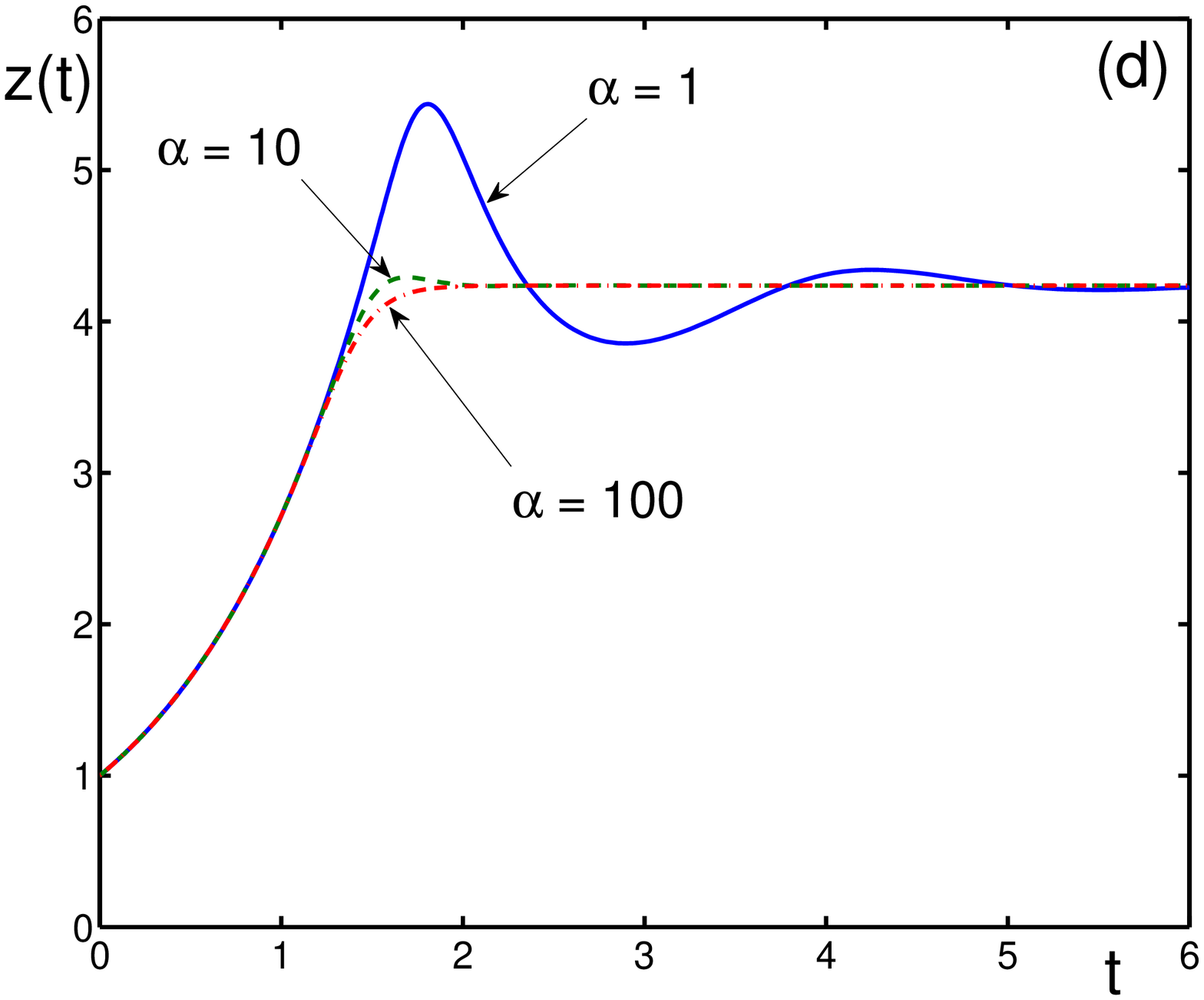} } }
\caption{Dynamics of populations $x(t)$ and $z(t)$ in the parametric
region $C$, for different $b < 0$ and $g = 100 > 0$. The initial conditions
are $x_0 = 3$ and $z_0 = 1$. (a) Population $x(t)$, with $b = -0.1$, for
$\al = 1$ (solid line), $\al = 10$ (dashed line), and $\al = 100$
(dashed-dotted line). The stable fixed point is $x^* = 0.035113$.
(b) Population $z(t)$, with $b = -0.1$, for $\al = 1$ (solid line), $\al = 10$
(dashed line), and $\al = 100$ (dashed-dotted line). The stable fixed point
is $z^* = 33.4918$. (c) Population $x(t)$, with $b = -1$, for $\al = 1$
(solid line), $\al = 10$ (dashed line), and $\al = 100$ (dashed-dotted line).
The stable fixed point is $x^* = 0.01444$. (d) Population $z(t)$, with $b = -1$,
for $\al = 1$ (solid line), $\al = 10$ (dashed line), and $\al = 100$
(dashed-dotted line). The stable fixed point is $z^* = 4.23773$.
}
\label{fig:Fig.3}
\end{figure}

\begin{figure}[ht]
\centerline{
\hbox{ \includegraphics[width=5.5cm]{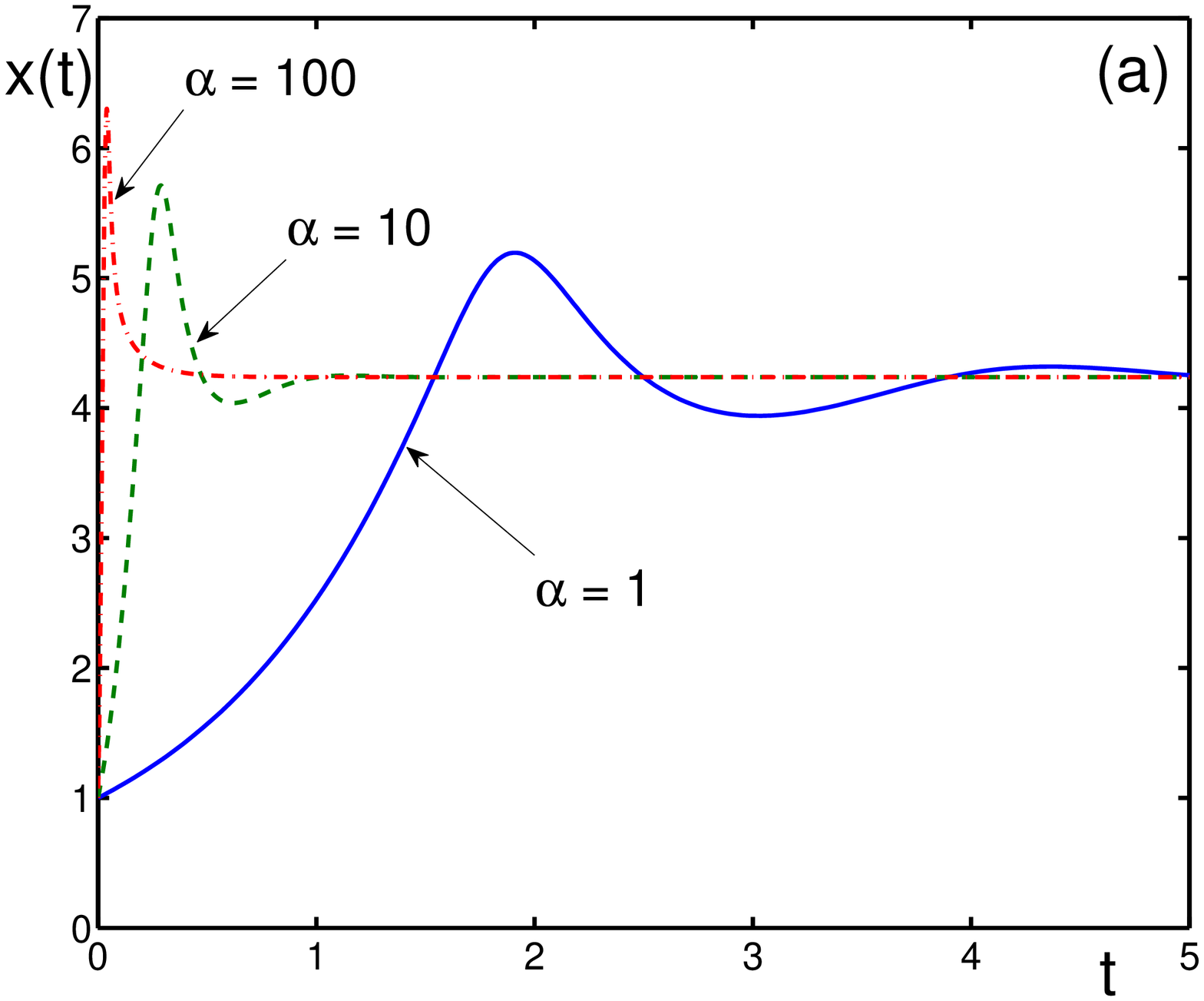} \hspace{2.5cm}
\includegraphics[width=5.5cm]{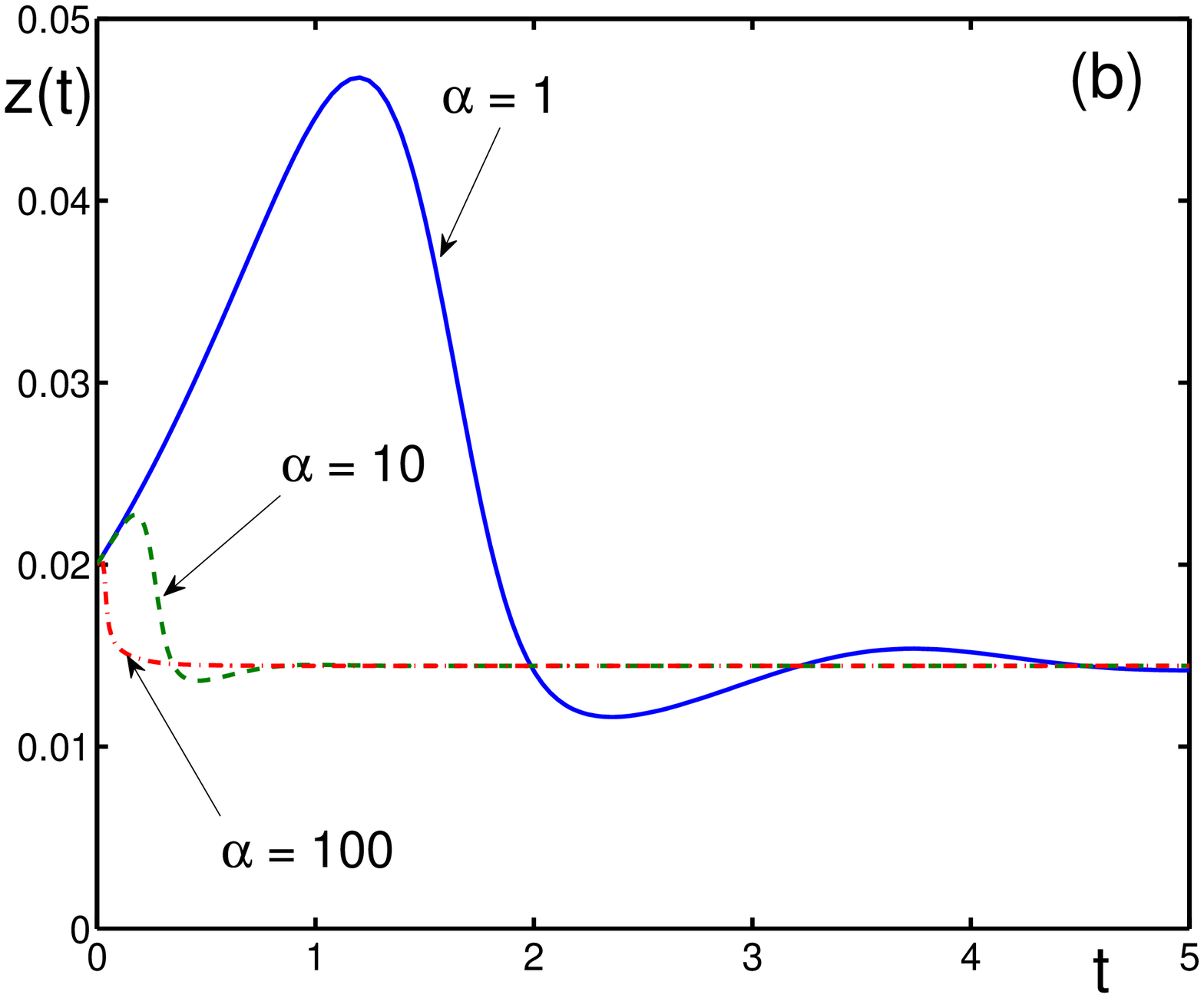} }  }
\vskip 9pt
\centerline{
\hbox{ \includegraphics[width=5.5cm]{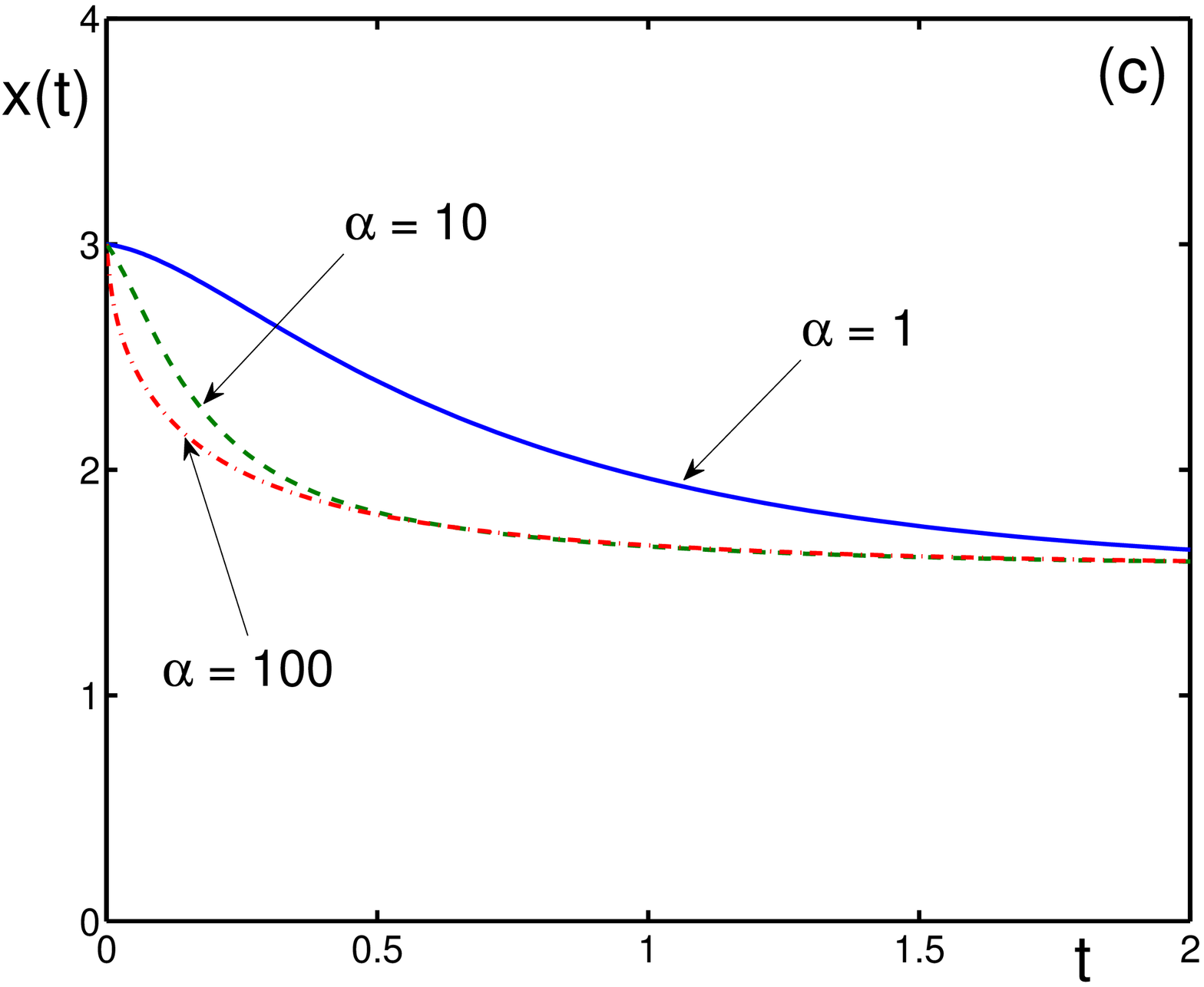} \hspace{2.5cm}
\includegraphics[width=5.5cm]{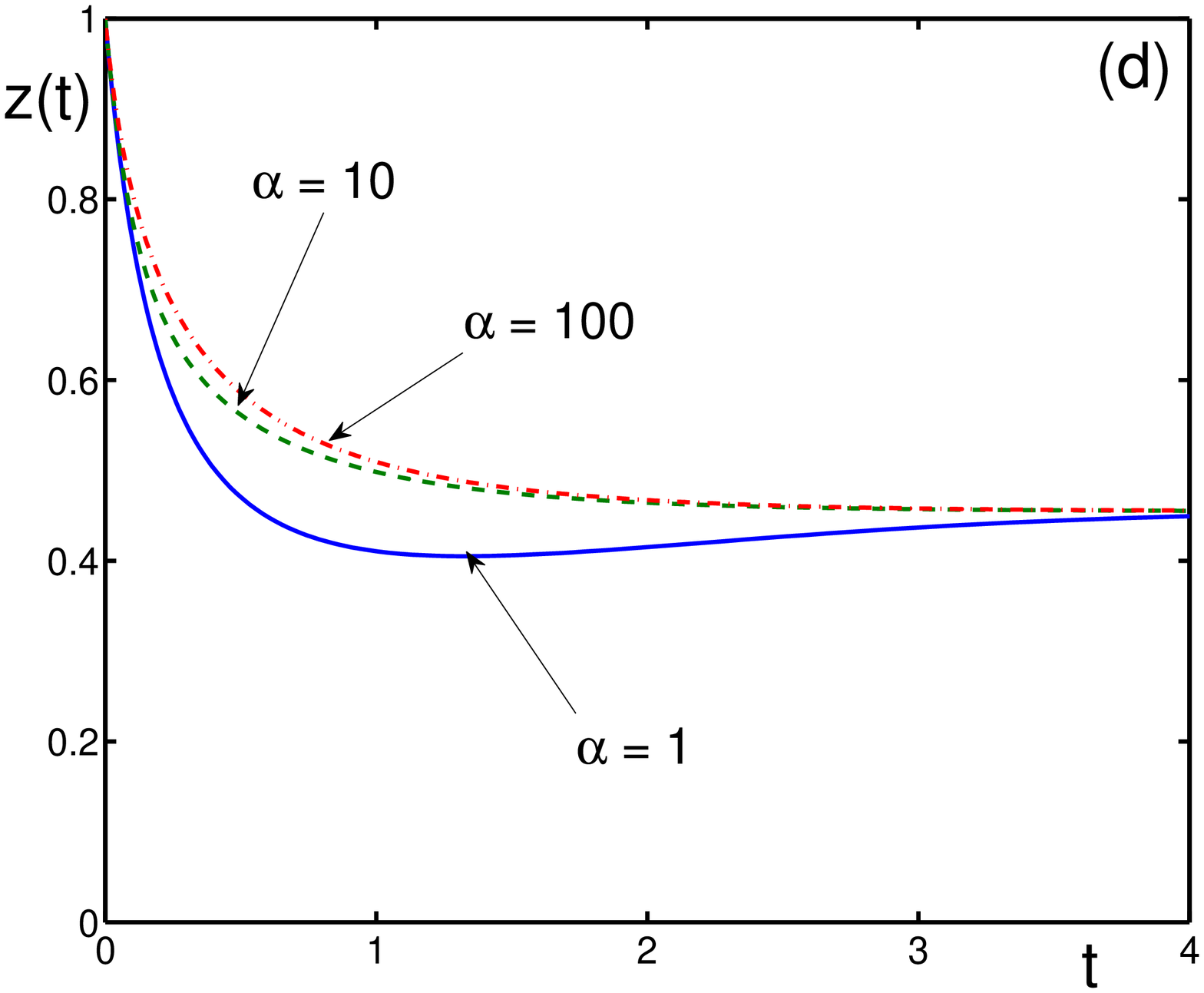} } }
\caption{Dynamics of populations $x(t)$ and $z(t)$ in the parametric
region $C$ for different initial conditions and different growth rates,
when $b > 0$, while $g < 0$. (a) Population $x(t)$, with the symbiotic parameters
$b = 100$ and $g = -1$. The initial conditions are $x_0 = 1$ and $z_0 = 0.02$.
The growth rate is $\al = 1$ (solid line), $\al = 10$ (dashed line), and
$\al = 100$ (dashed-dotted line). The stable fixed point is $x^* = 4.2377$.
(b) Population $z(t)$ for the same symbiotic parameters and initial conditions
as in (a) for the growth rates $\al = 1$ (solid line), $\al = 10$ (dashed line),
and $\al = 100$ (dashed-dotted line). The stationary state is $z^* = 0.01444$.
(c) Population $x(t)$, with the symbiotic parameters $b = 1$ and $g = - 0.5$.
The initial conditions are $x_0 = 3$ and $z_0 = 1$. Growth rate is $\al = 1$
(solid line), $\al = 10$ (dashed line), and $\al = 100$ (dashed-dotted line).
The stationary state is $x^* = 1.5758$. (d) Population $z(t)$ for the same
symbiotic parameters and initial conditions, as in (c), for $\al = 1$
(solid line), $\al = 10$ (dashed line), and $\al = 100$ (dashed-dotted line).
The stationary state is $z^* = 0.45479$.
}
\label{fig:Fig.4}
\end{figure}

\begin{figure}[ht]
\centerline{
\hbox{ \includegraphics[width=5.5cm]{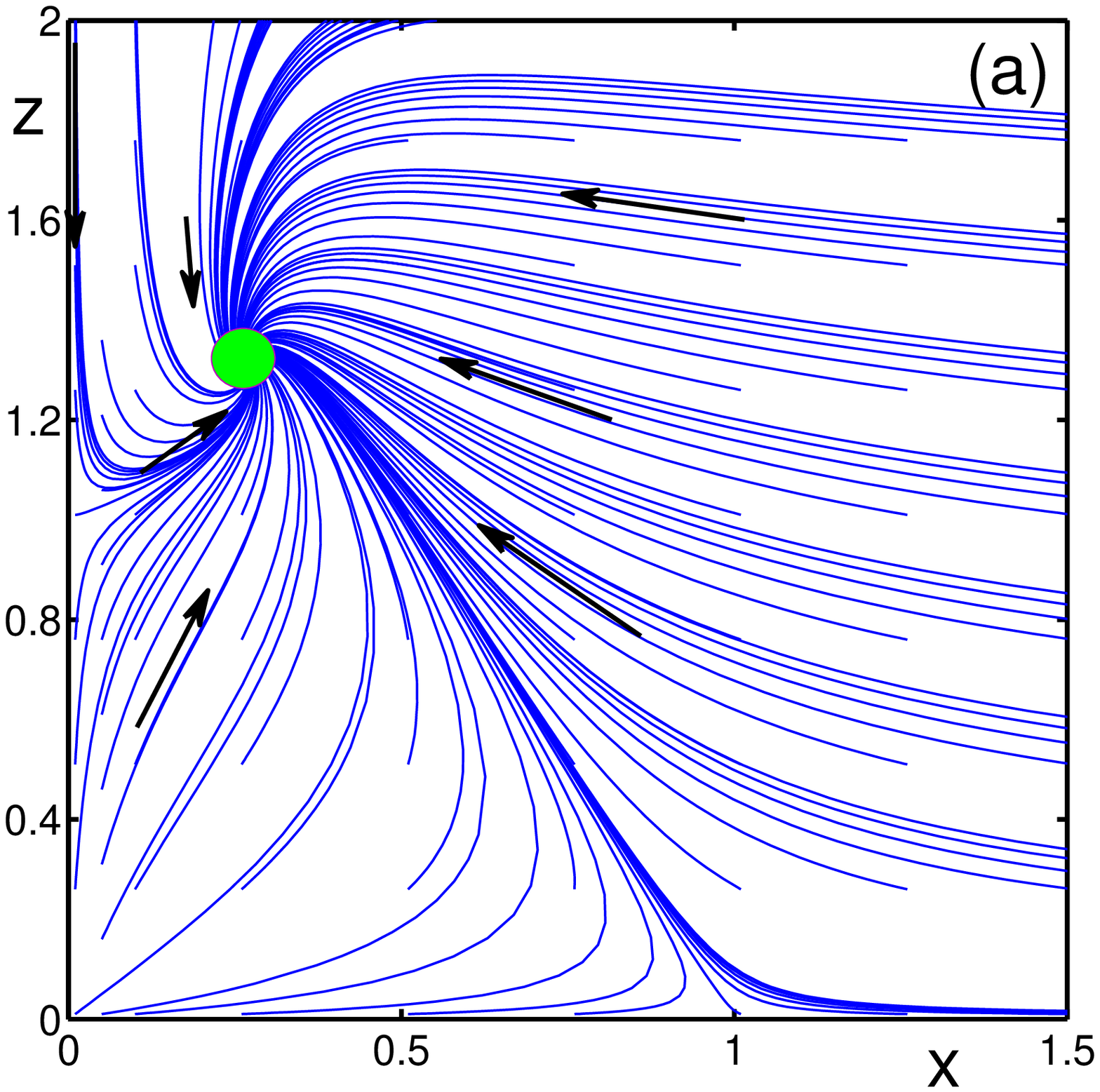} \hspace{2.5cm}
\includegraphics[width=5.5cm]{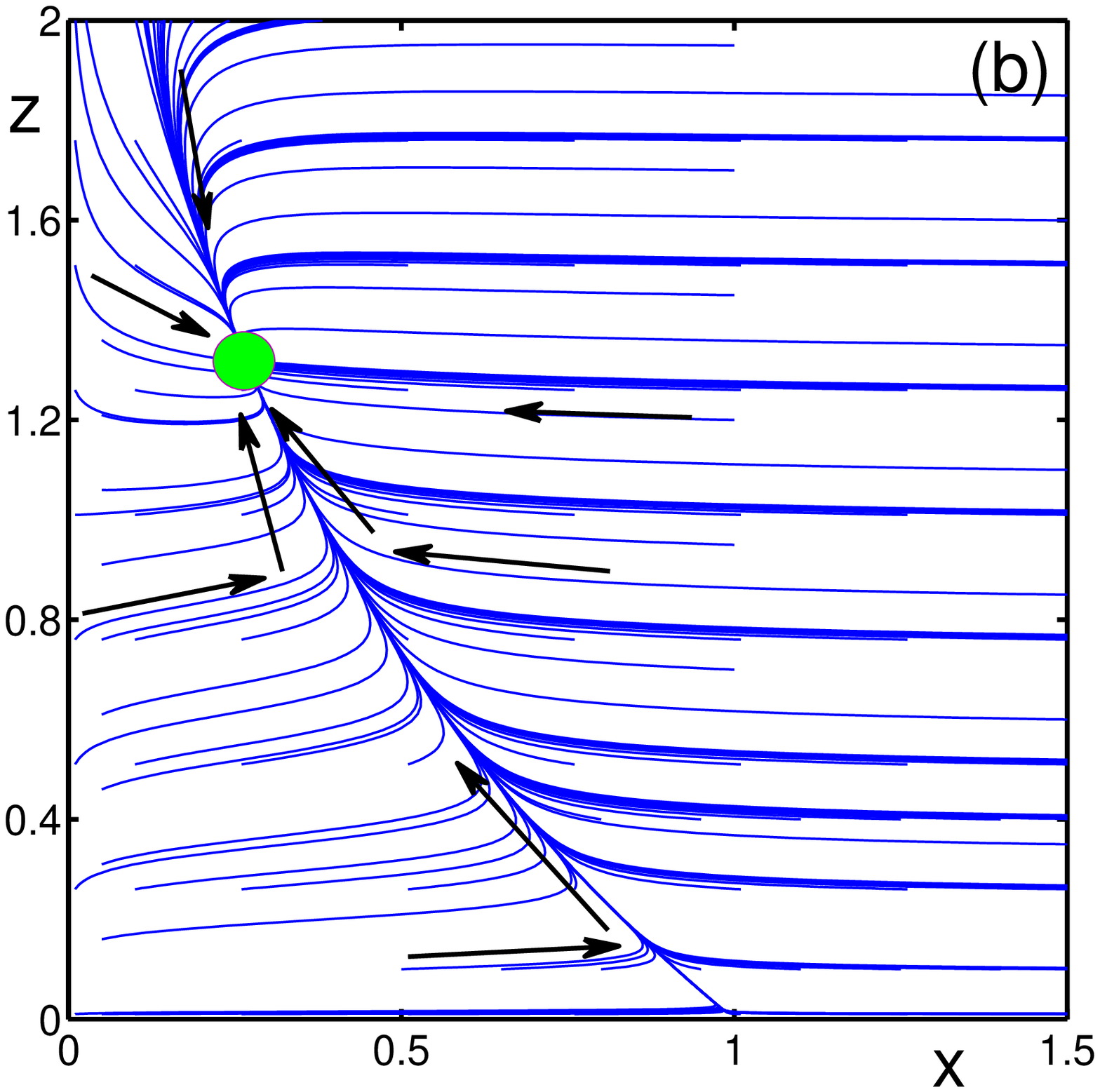} }  }
\caption{Phase portrait on the plane $x-z$, for the parametric region $C$,
for $b = -1$ and $g = 1$, and different growth rates $\al$. There exists a
single fixed point, a stable focus, shown by the filled green disc. The fixed point
is $\{x_1^* = 0.26987, z_1^* = 1.3098\}$. (a) Phase portrait for $\al = 1$;
(b) phase portrait for $\al = 10$.
}
\label{fig:Fig.5}
\end{figure}

\begin{figure}[ht]
\centerline{
\hbox{ \includegraphics[width=5.5cm]{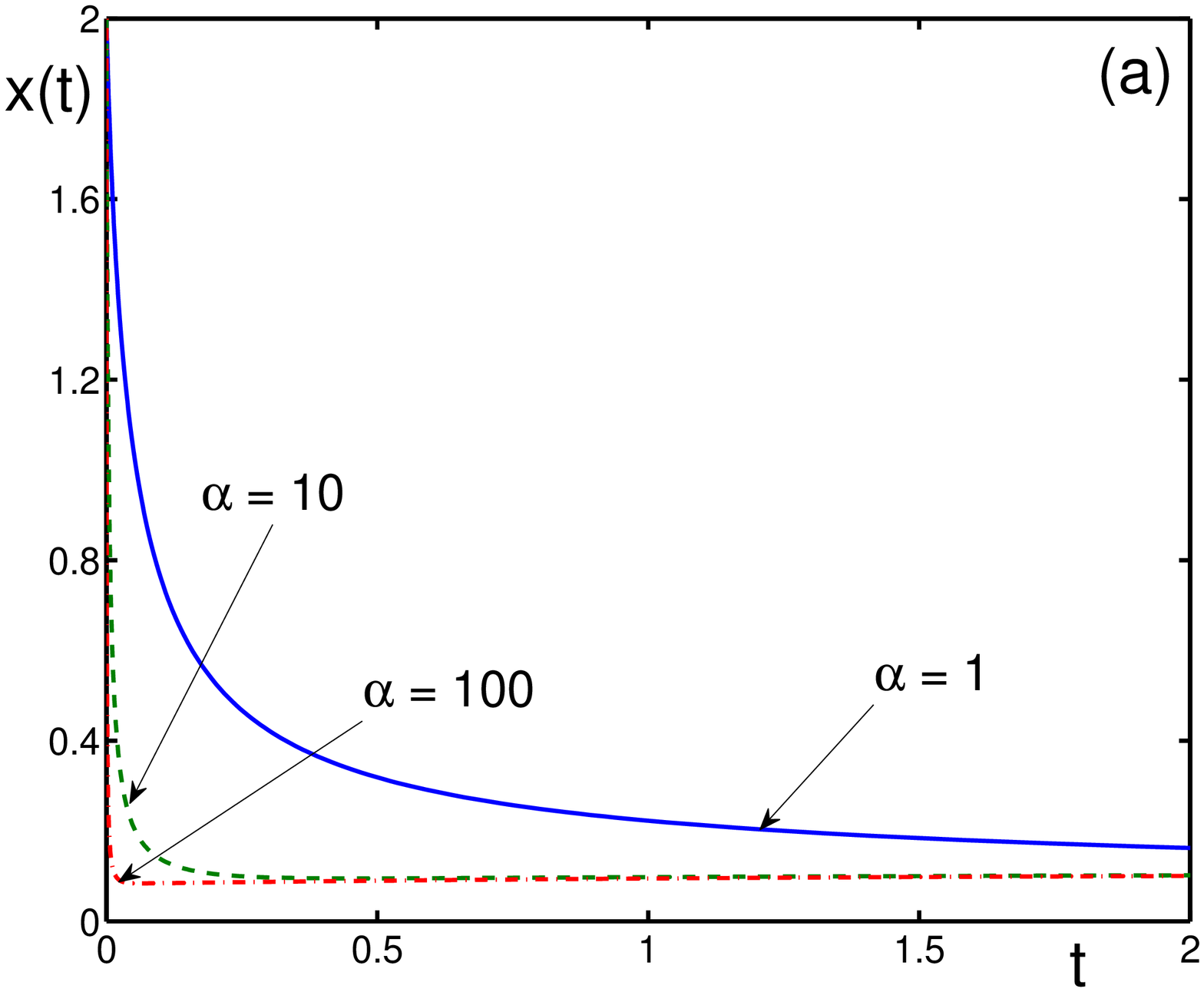} \hspace{2.5cm}
\includegraphics[width=5.5cm]{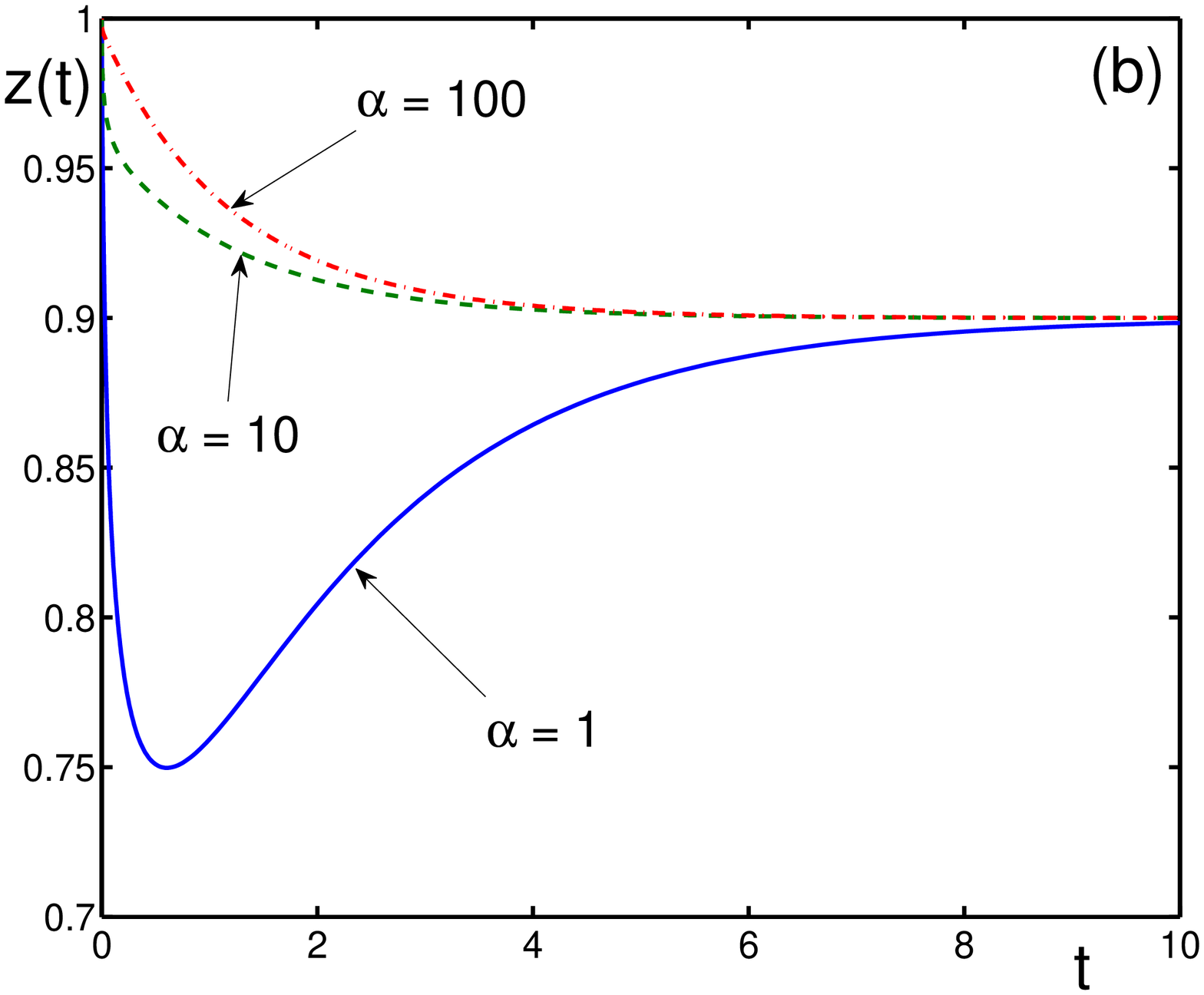} }  }
\vskip 9pt
\centerline{
\hbox{ \includegraphics[width=5.5cm]{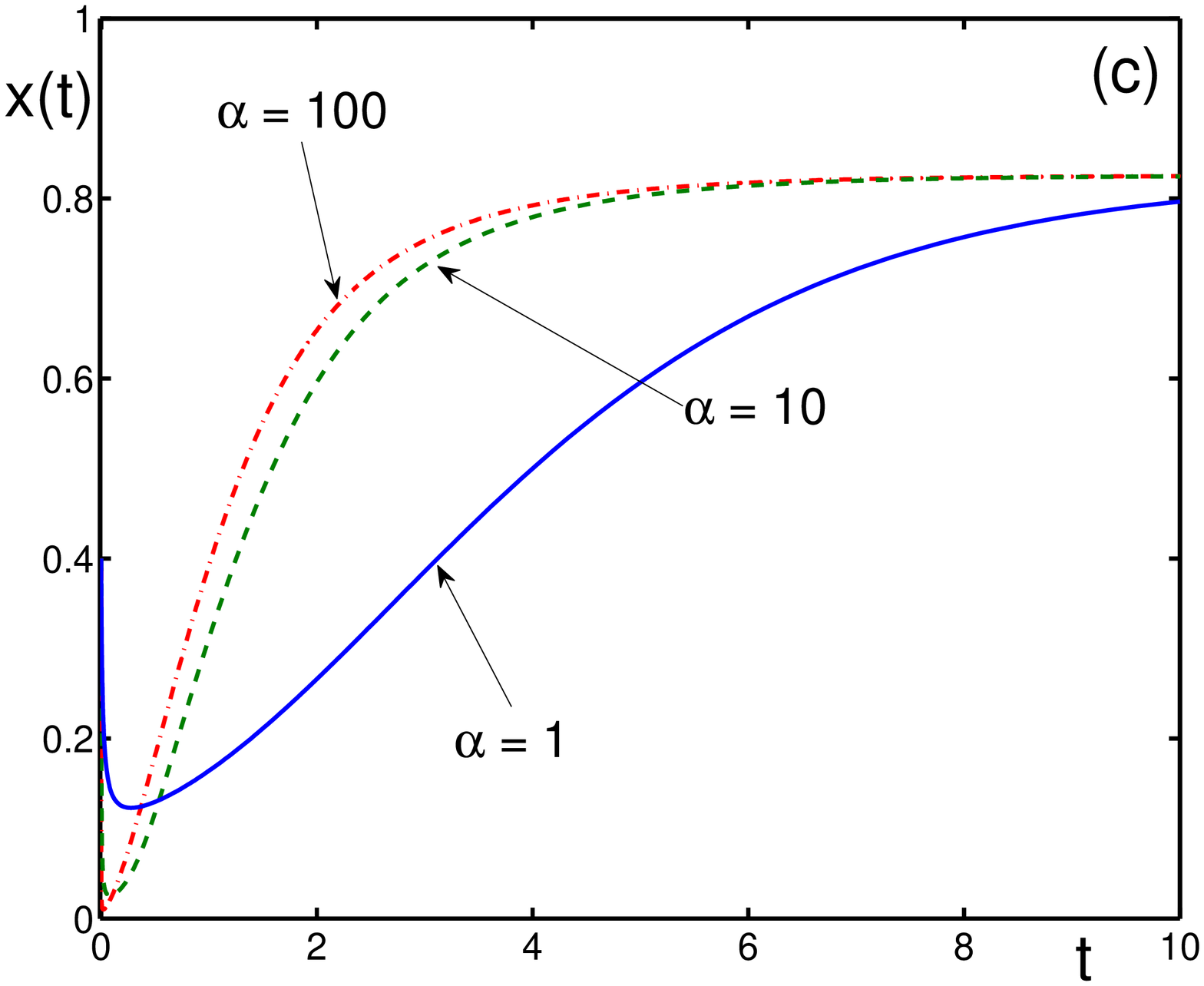} \hspace{2.5cm}
\includegraphics[width=5.5cm]{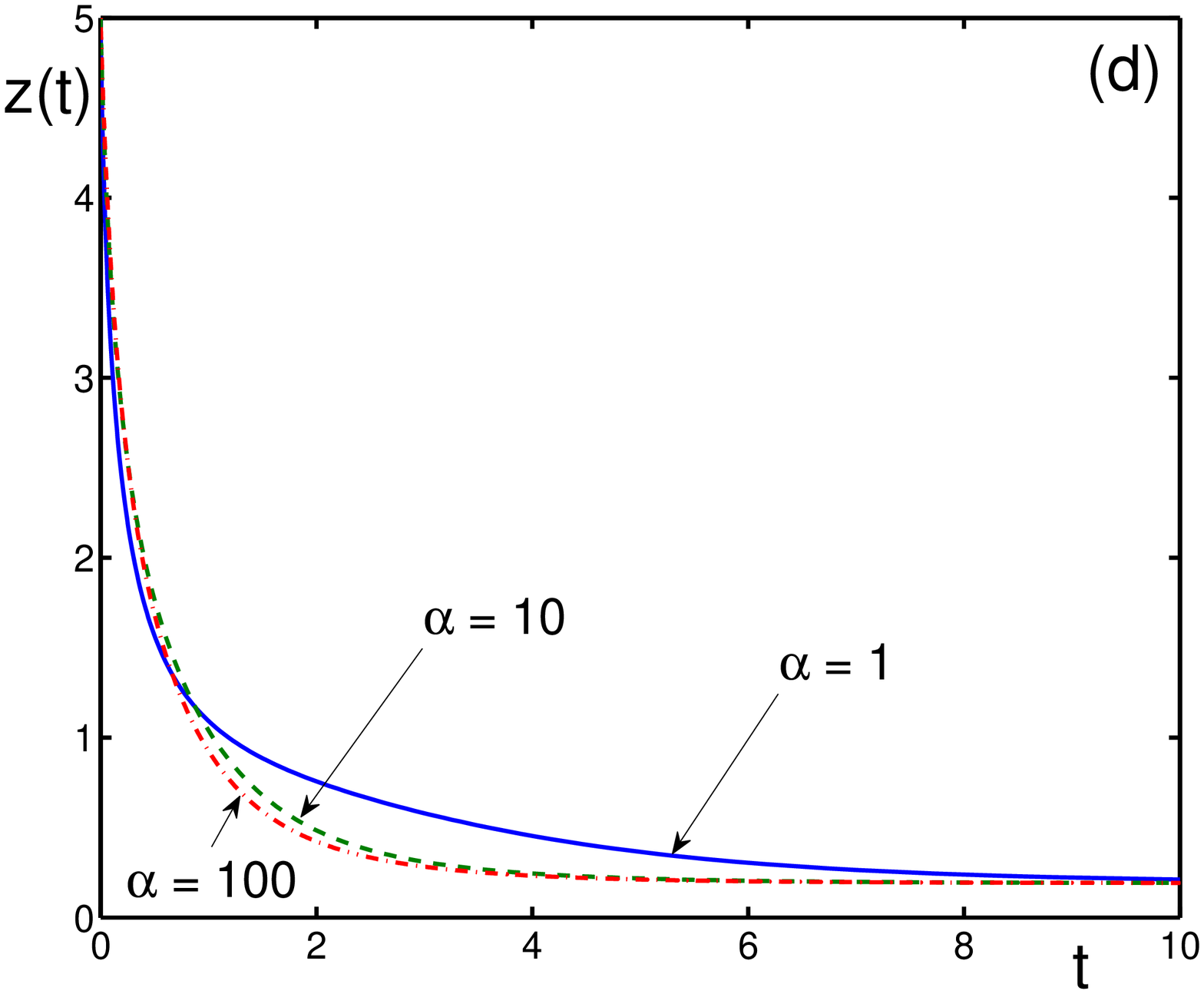} }  }
\caption{Dynamics of populations $x(t)$ and $z(t)$ in the region $D_1$,
for different symbiotic parameters and initial conditions. (a) Population $x(t)$,
with $b = -2.5$ and $g = - 1$, and the initial conditions $\{x_0 = 2, z_0 = 1\}$,
for $\al = 1$ (solid line), $\al = 10$ (dashed line), and $\al = 100$
(dashed-dotted line). The stable fixed point is $x^* = 0.10541$.
(b) Population $z(t)$, for the same symbiotic parameters and initial conditions,
as in (a), for $\al = 1$ (solid line), $\al = 10$ (dashed line), and $\al = 100$
(dashed-dotted line). The fixed point is $z^* = 0.89995$. (c) Population $x(t)$,
with $b = - 1$ and $g = - 2$, initial conditions $\{x_0 = 0.4, z_0 = 5\}$, for
$\al = 1$ (solid line), $\al = 10$ (dashed line), and $\al = 100$
(dashed-dotted line). The fixed point is $x^* = 0.82539$. (d) Population $z(t)$
for the same symbiotic parameters and initial conditions, as in (c), for
$\al = 1$ (solid line), $\al = 10$ (dashed line), and $\al = 100$
(dashed-dotted line). The fixed point is $z^* = 0.19190$.
}
\label{fig:Fig.6}
\end{figure}

\subsection{Convergent behavior in marginal cases}

The marginal cases correspond to zero values of symbiotic parameters. Thus,
if $b = 0$, while $g$ is arbitrary, the sole stationary state is the stable node 
$$
x^* = 1 \; , \qquad z^* = e^g \qquad (b = 0 , ~ -\infty < g < \infty ) \; ,
$$
with the characteristic exponents $\lambda_1 = - 1$ and $\lambda_2 = - \alpha$. 
The population $x$ is described by the explicit formula
$$
x = \frac{x_0}{x_0 + (1-x_0)\exp(-\al t)} \;   .
$$
The convergence to the stationary state is faster for larger $\alpha$.

When $b$ is arbitrary, while $g = 0$, then again there exits just a single
fixed point, a stable node
$$
 x^* = e^b \; , \qquad z^* = 1 \;   ,
$$
with the characteristic exponents $\lambda_1 = - \alpha$ and $\lambda_2 = - 1$. 
The population $z$ does not depend on $\alpha$, being given by the expression
$$
z = \frac{z_0}{z_0 + (1-z_0)\exp(- t)} \;   ,
$$
while the population $x$ converges to the stationary state faster for larger
$\alpha$. 

The examples of the present section illustrate the expected situation, 
where the growth rate $\alpha$ directly influences the time scales of the dynamics
of the symbiotic populations, but does not qualitatively distort the overall picture.

\section{Growth-Rate Induced Dynamic Transitions}

In the present section, we show that there may happen unexpected situations,
when the variation of the growth rate, although not influencing the 
stationary states, can lead to dramatic changes in the population dynamics.

\subsection{Dynamic transition under mutualism}
 
In the parametric region $B$, where $b > 0$ and $0 < g < g_c(b)$, there 
exist two fixed points, $\{x^*_1,z^*_1\}$ and $\{x^*_2,z^*_2\}$, with 
$1 < x^*_1 < x^*_2$ and $1 < z^*_1 < z^*_2$. The fixed point 
$\{x^*_1,z^*_1\}$ is a stable node and $\{x^*_2,z^*_2\}$ is a saddle. 
The stable node possesses a basin of attraction, whose boundary passes
through the saddle. The behavior of the populations $x(t)$ and $z(t)$ 
depends on whether the initial conditions are taken inside the basin 
of attraction or not. 

On the line $\{b,g_c(b)\}$, the stable node $\{x^*_1,z^*_1\}$, and the 
saddle $\{x^*_2,z^*_2\}$ merge together and disappear for $g > g_c(b)$.
When $g = g_c(b)$, then $x_1^* = x_2^*$ and $z_1^* = z_2^*$. 

It turns out that the growth rate $\alpha$, although not influencing
the stationary states as such, does influence the boundary of the 
attraction basin. Therefore, it may happen that the same initial 
conditions, depending on the value of $\alpha$, can occur inside the
attraction basin or outside it. This delicate situation is illustrated 
in Figs. 7 to 9.    

Figure 7 demonstrates the convergence of the populations $x(t)$ and $z(t)$ 
for $g$ taken close to the line $\{b,g_c(b)\}$, with initial conditions 
that are inside the attraction basin of the stable fixed point for all 
$\al = 1, 10, 100$. But in Fig. 8, the initial conditions $\{x_0,z_0\}$ 
are such that they are {\it outside} of the basin of attraction for $\al = 1$, 
but {\it inside} it, when $\al = 10$ and $\al = 100$. Contrary to Fig. 8,
in Fig. 9, we show the situation when the initial conditions $\{x_0,z_0\}$ 
are {\it inside} the basin of attraction of the stable fixed point for 
$\al = 1$, but {\it outside} of it for $\al = 10$ and $\al = 100$. Phase 
portraits for region $B$, under different $\al$, are presented in Fig. 10. 
The boundary of the attraction basin essentially depends on the symbiotic 
parameters $b$ and $g$, as well as on the growth rate $\al$.

\begin{figure}[ht]
\centerline{
\hbox{ \includegraphics[width=5.5cm]{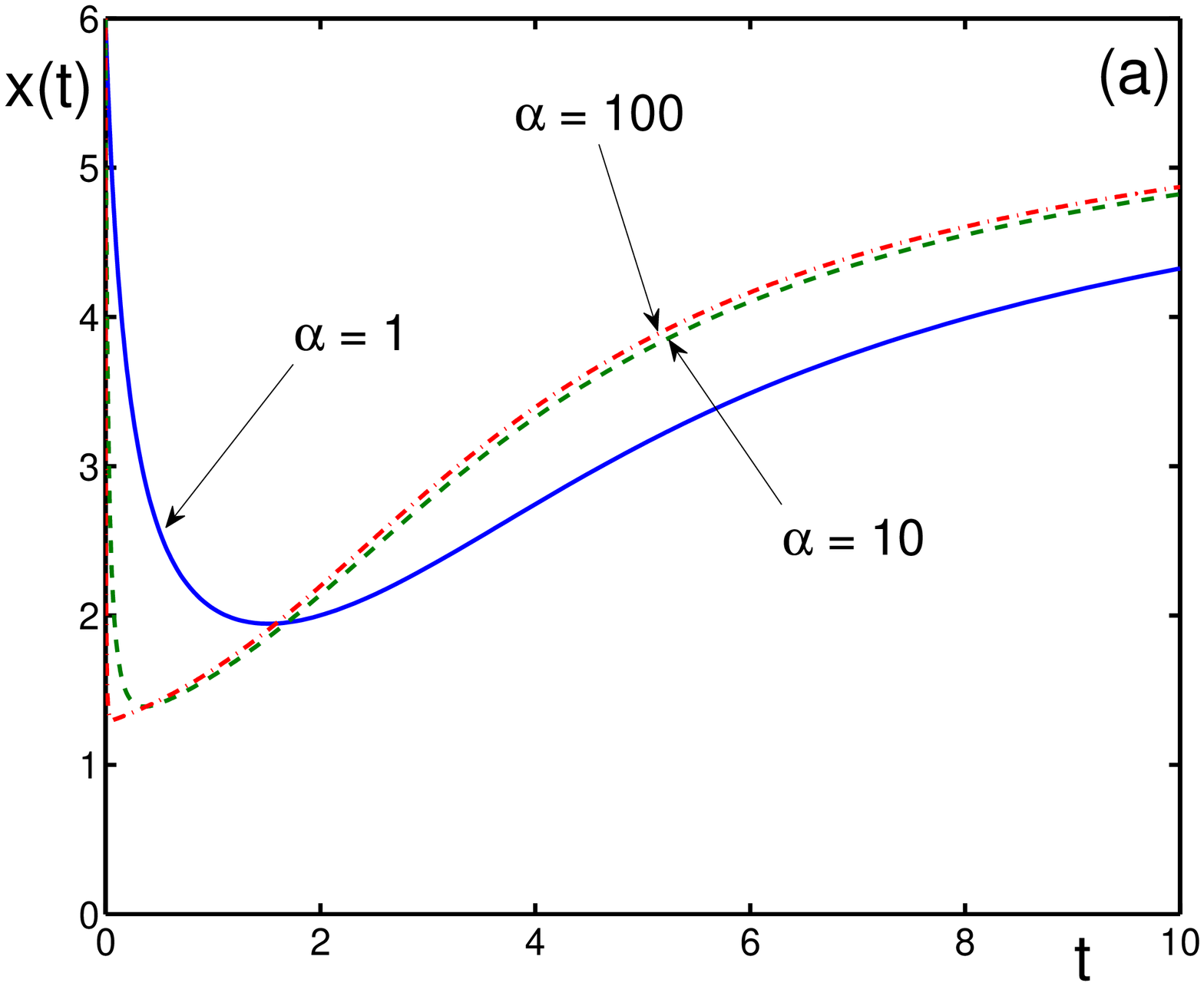} \hspace{2.5cm}
\includegraphics[width=5.5cm]{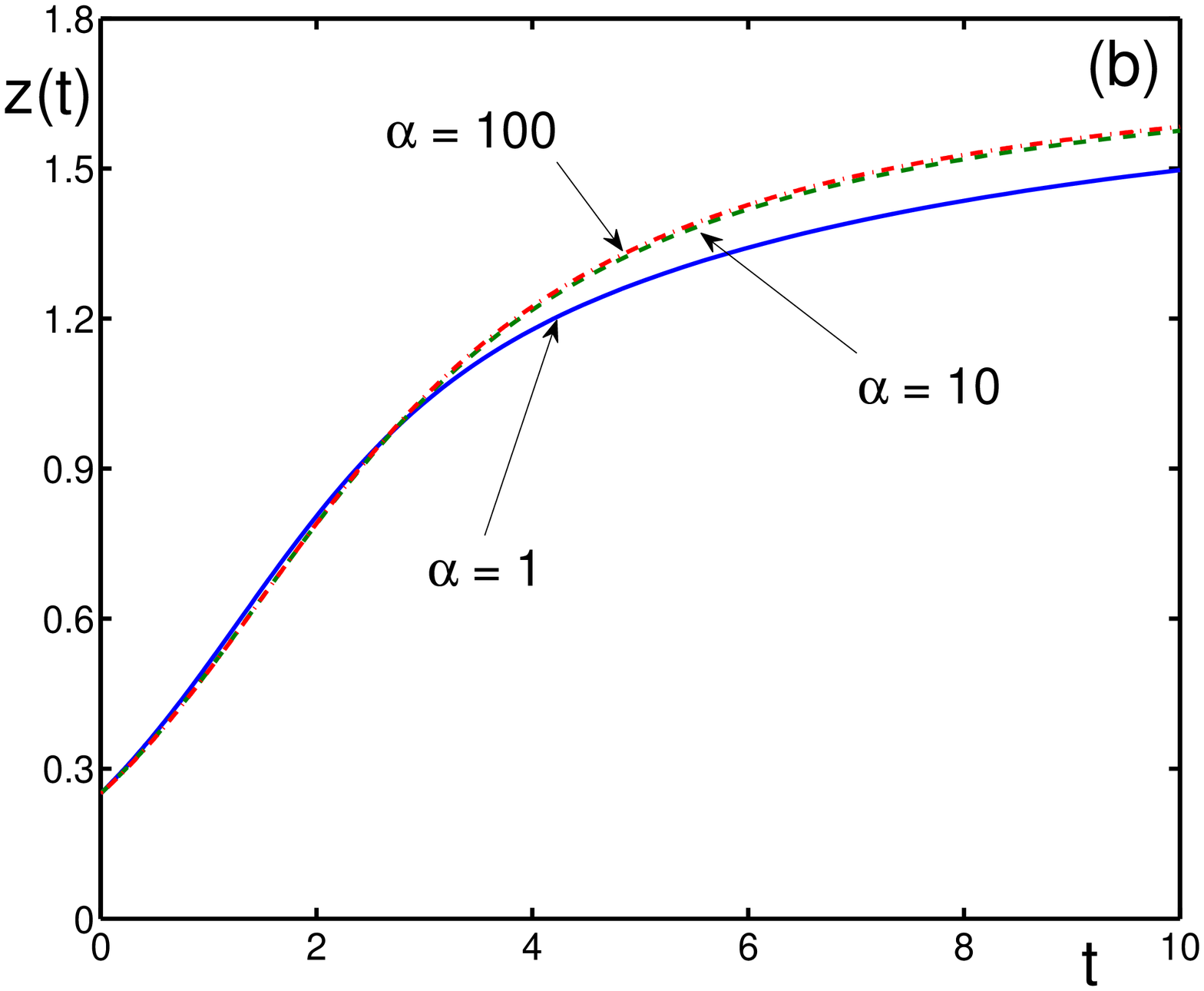} }  }
\caption{Dynamics of populations $x(t)$ and $z(t)$ in the parametric
region $B$, with $b = 1$ and $g = 0.0972 < g_c \approx 0.0973$, and the initial
conditions $\{x_0 = 6, z_0 = 0.25 \}$, such that the initial point is inside
the attraction basin for all $\al = 1, 10, 100$. The stable fixed point is
$\{x^*_1 = 5.6717, z^*_1 = 1.7355\}$, and the saddle is
$\{x^*_2 = 5.9995, z^*_2 = 1.7917\}$. (a) Population $x(t)$ for $\al = 1$
(solid line), $\al = 10$ (dashed line), and $\al = 100$ (dashed-dotted line);
(b) population $z(t)$ for $\al = 1$ (solid line), $\al = 10$ (dashed line),
and $\al = 100$ (dashed-dotted line).
}
\label{fig:Fig.7}
\end{figure}

\begin{figure}[ht]
\centerline{
\hbox{ \includegraphics[width=5.5cm]{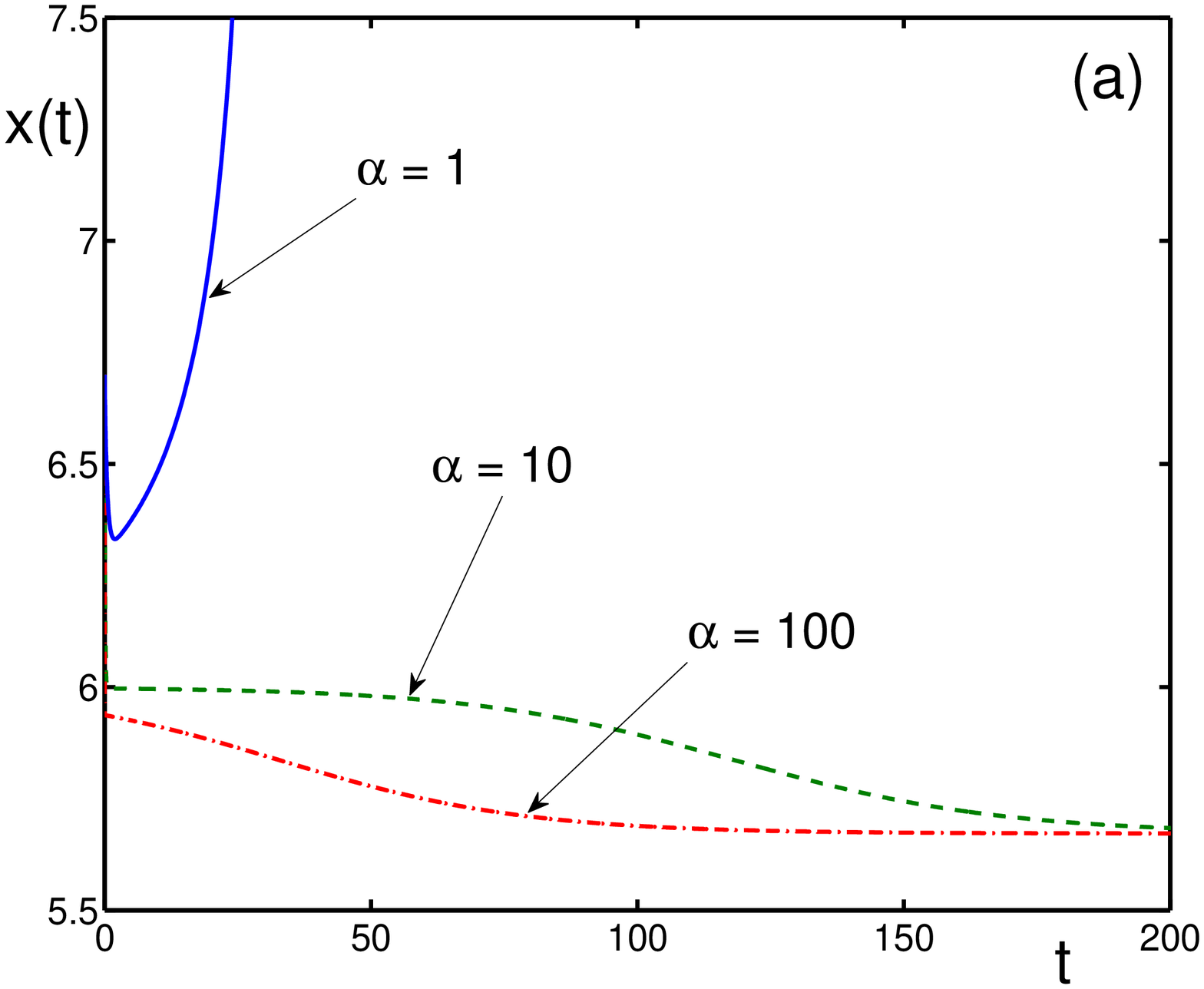} \hspace{2.5cm}
\includegraphics[width=5.5cm]{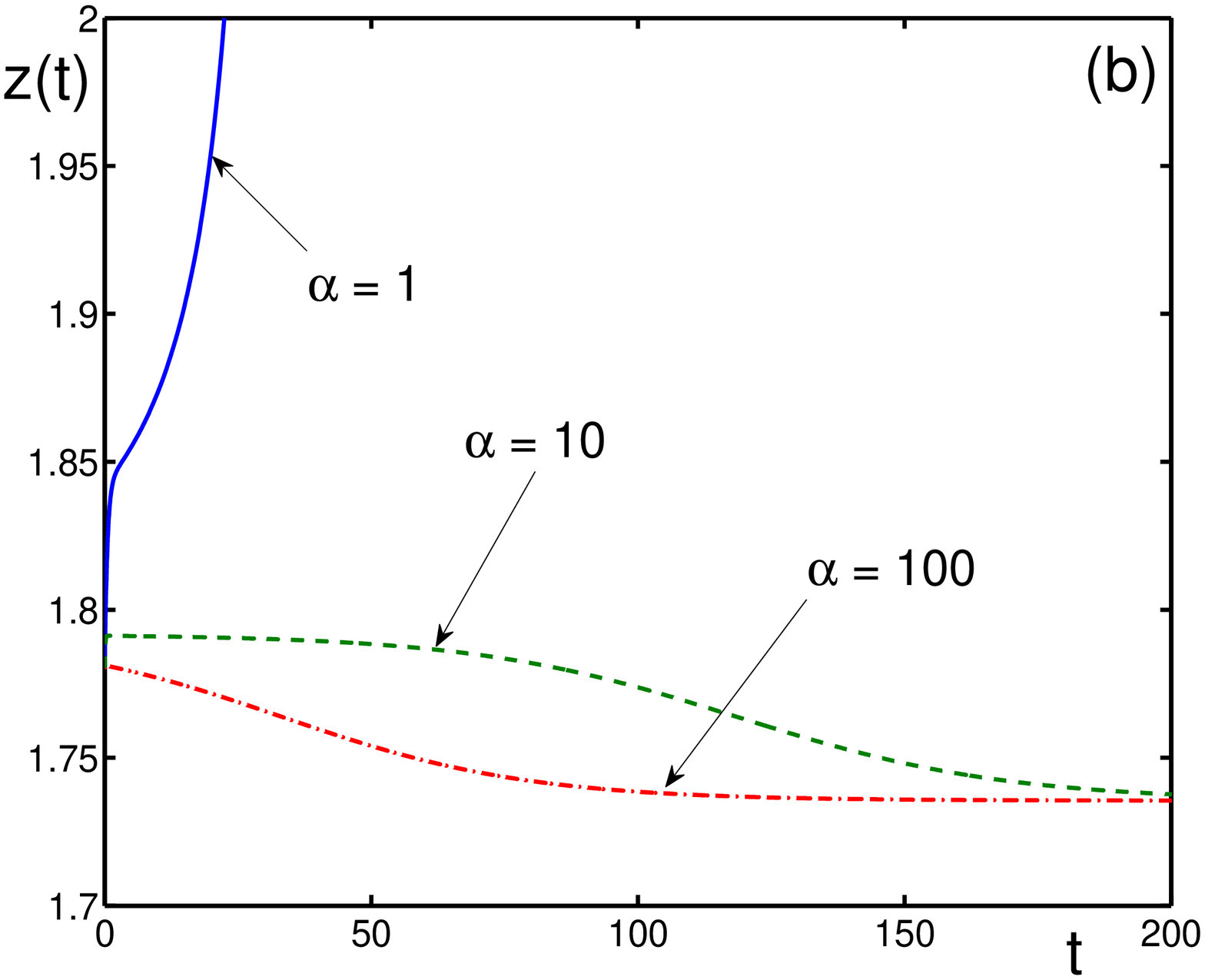} }  }
\caption{Dynamics of populations $x(t)$ and $z(t)$ for the parametric
region $B$, with $b = 1$ and $g = 0.0972 < g_c \approx 0.0973$, as in figure 7. The
stable fixed point is $\{x^*_1 = 5.6717, z^*_1 = 1.7355\}$. The saddle is
$\{x^*_2 = 5.9995, z^*_2 = 1.7917\}$, also as in figure 7. The initial conditions
$\{x_0 = 6.7, z_0 = 1.78\}$ are such that they are outside of the attraction
basin for $\al = 1$, but inside for $\al = 10$ and $\al = 100$.
(a) Population $x(t)$ for $\al = 1$ (solid line), $\al = 10$ (dashed line),
and $\al = 100$ (dashed-dotted line); (b) population $z(t)$ for $\al = 1$
(solid line), $\al = 10$ (dashed line), and $\al = 100$ (dashed-dotted line).
}
\label{fig:Fig.8}
\end{figure}

\begin{figure}[ht]
\centerline{
\hbox{ \includegraphics[width=5.5cm]{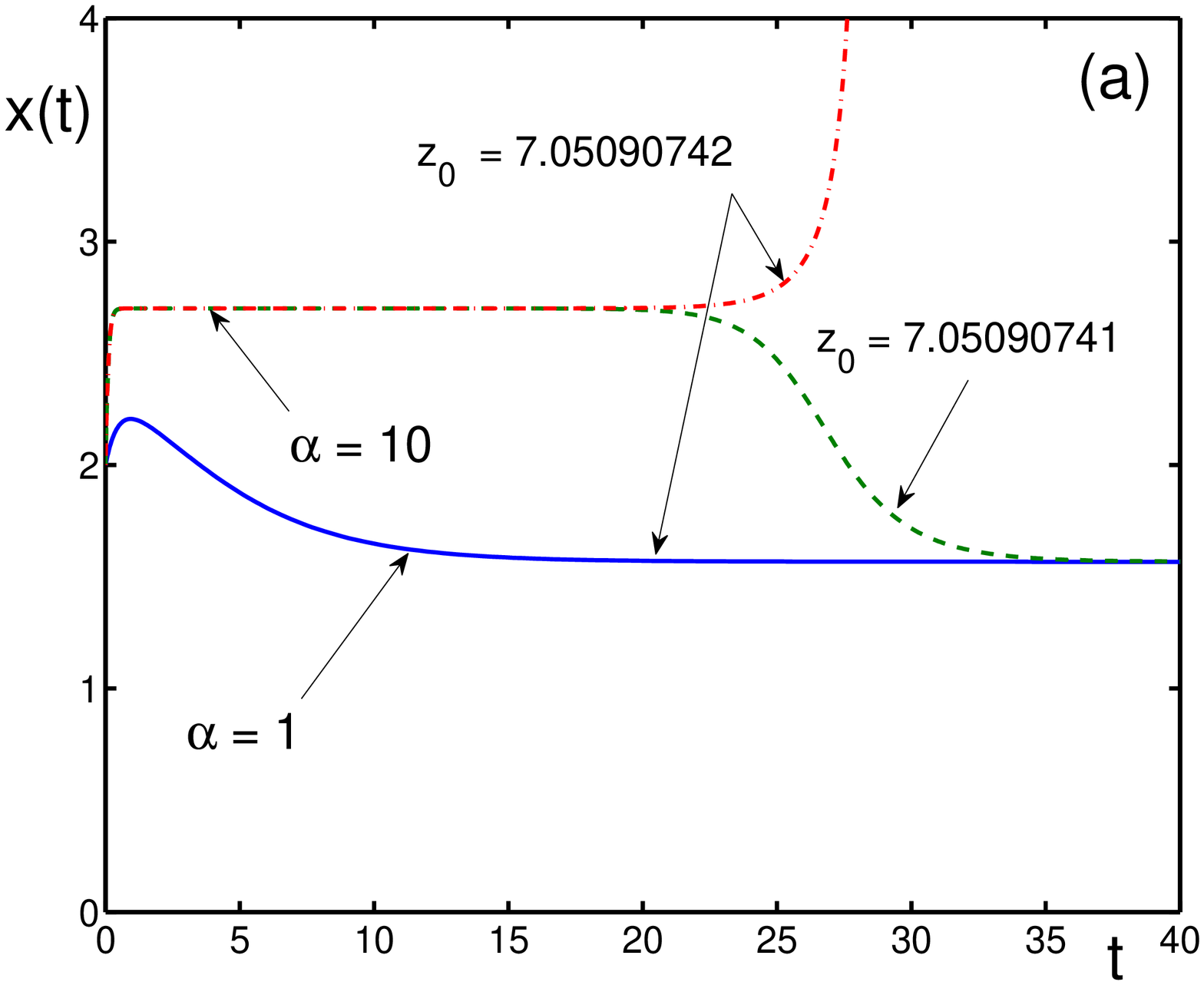} \hspace{2.5cm}
\includegraphics[width=5.5cm]{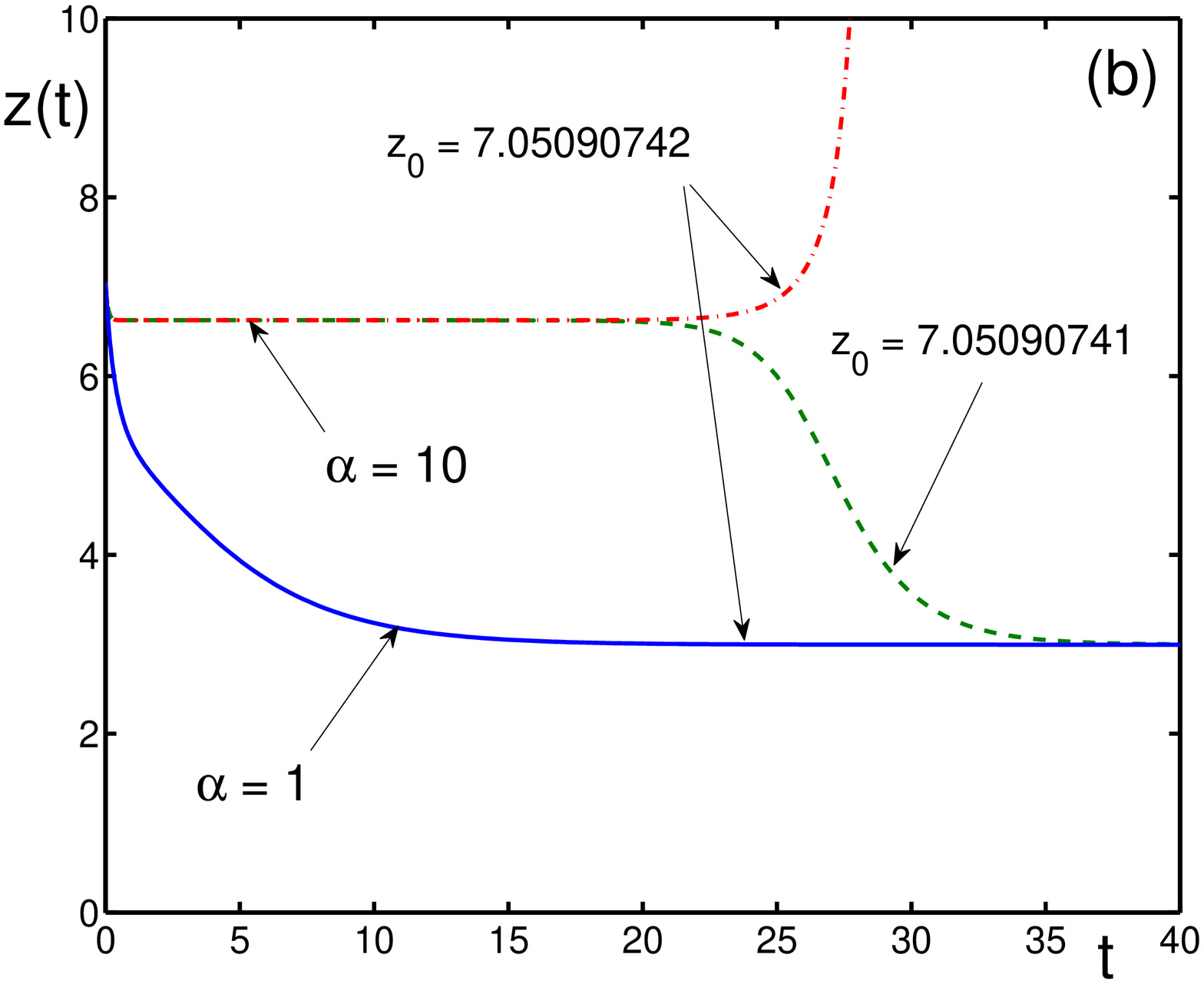} }  }
\caption{Dynamics of populations $x(t)$ and $z(t)$ in the parametric
region $B$, with $b = 0.15$ and $g = 0.7 < g_c \approx 0.76584$. The initial
condition $x_0 = 2$ is fixed and $z_0$ is varied close to the boundaries
of the attraction basins. The stable fixed point is
$\{x^*_1 = 1.56721, z^*_1 = 2.9953\}$, and the saddle is
$\{x^*_2 = 2.7011, z^*_2 = 6.6243\}$. (a) Population $x(t)$ for
$\al = 1$ and $z_0 = 7.05090742$ (solid line), for $\al = 10$ and
$z_0 = 7.05090742$ (dashed-dotted line), and for $\al = 10$ but
$z_0 = 7.05090741$ (dashed line); (b) population $z(t)$ for $\al = 1$
and $z_0 = 7.05090742$ (solid line), for $\al = 10$ and $z_0 = 7.05090742$
(dashed-dotted line), and for $\al = 10$ but $z_0 = 7.05090741$ (dashed line).
}
\label{fig:Fig.9}
\end{figure}

\begin{figure}[ht]
\centerline{
\hbox{ \includegraphics[width=5cm]{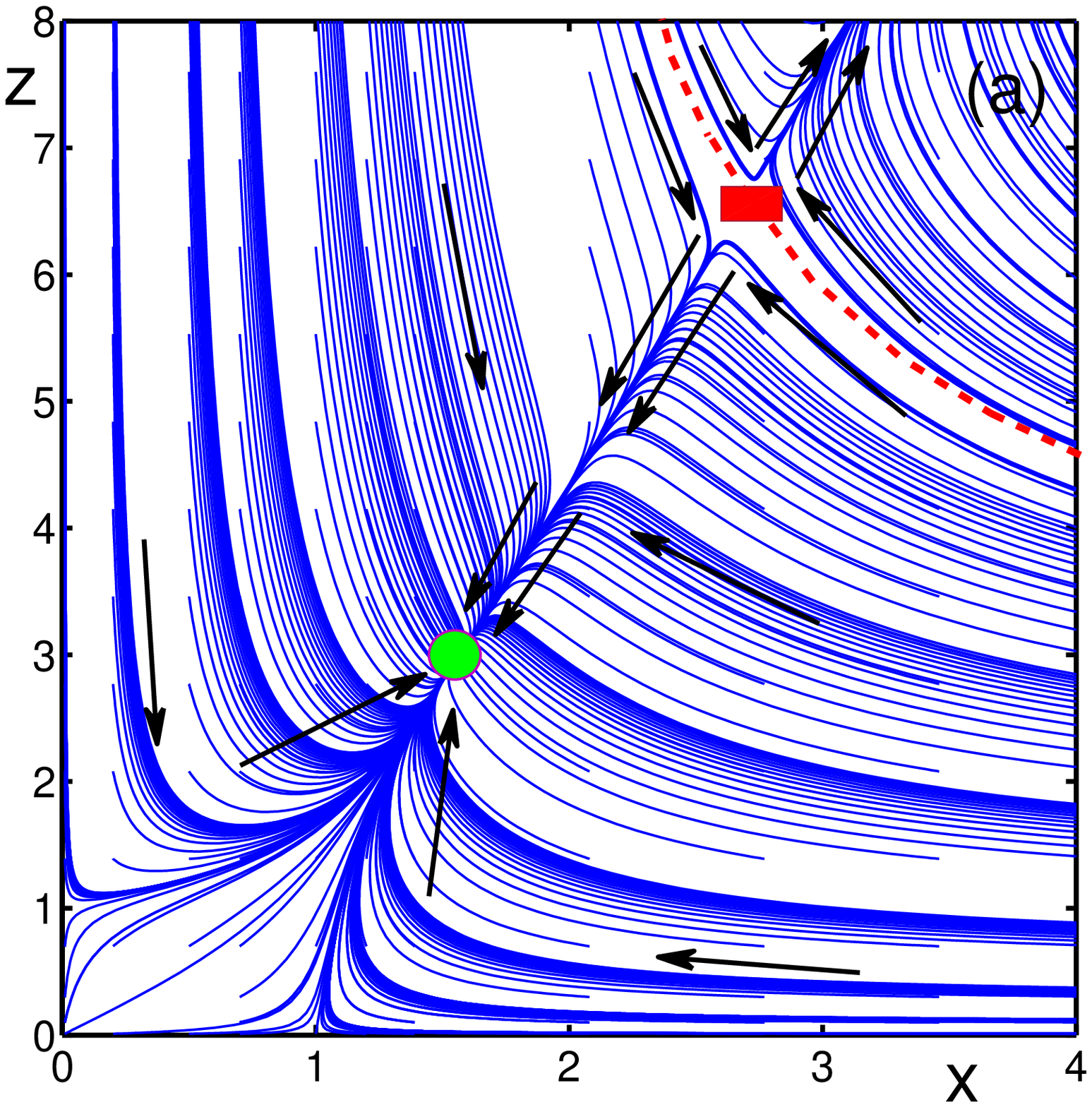} \hspace{1cm}
\includegraphics[width=5cm]{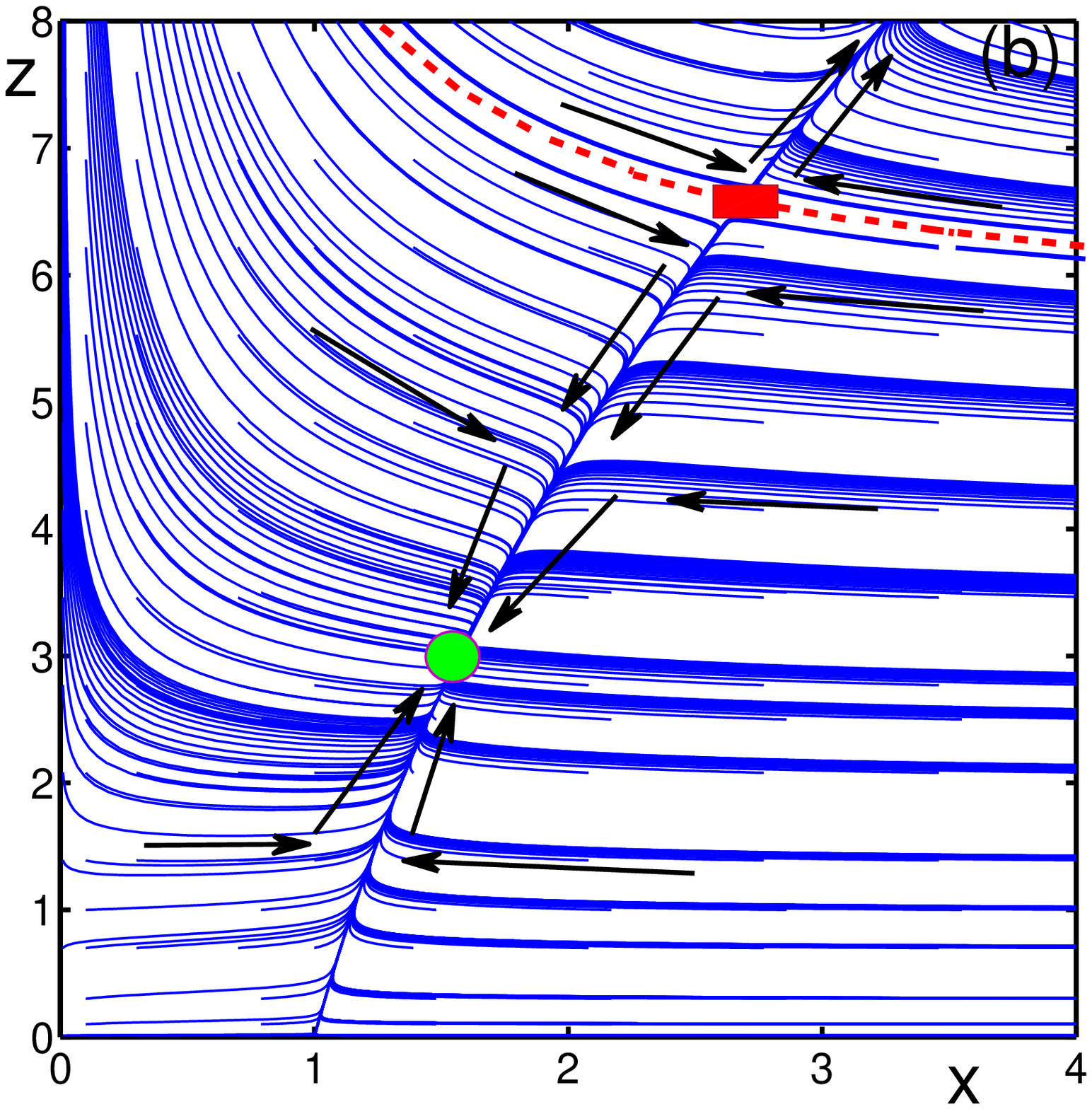} \hspace{1cm}
\includegraphics[width=5cm]{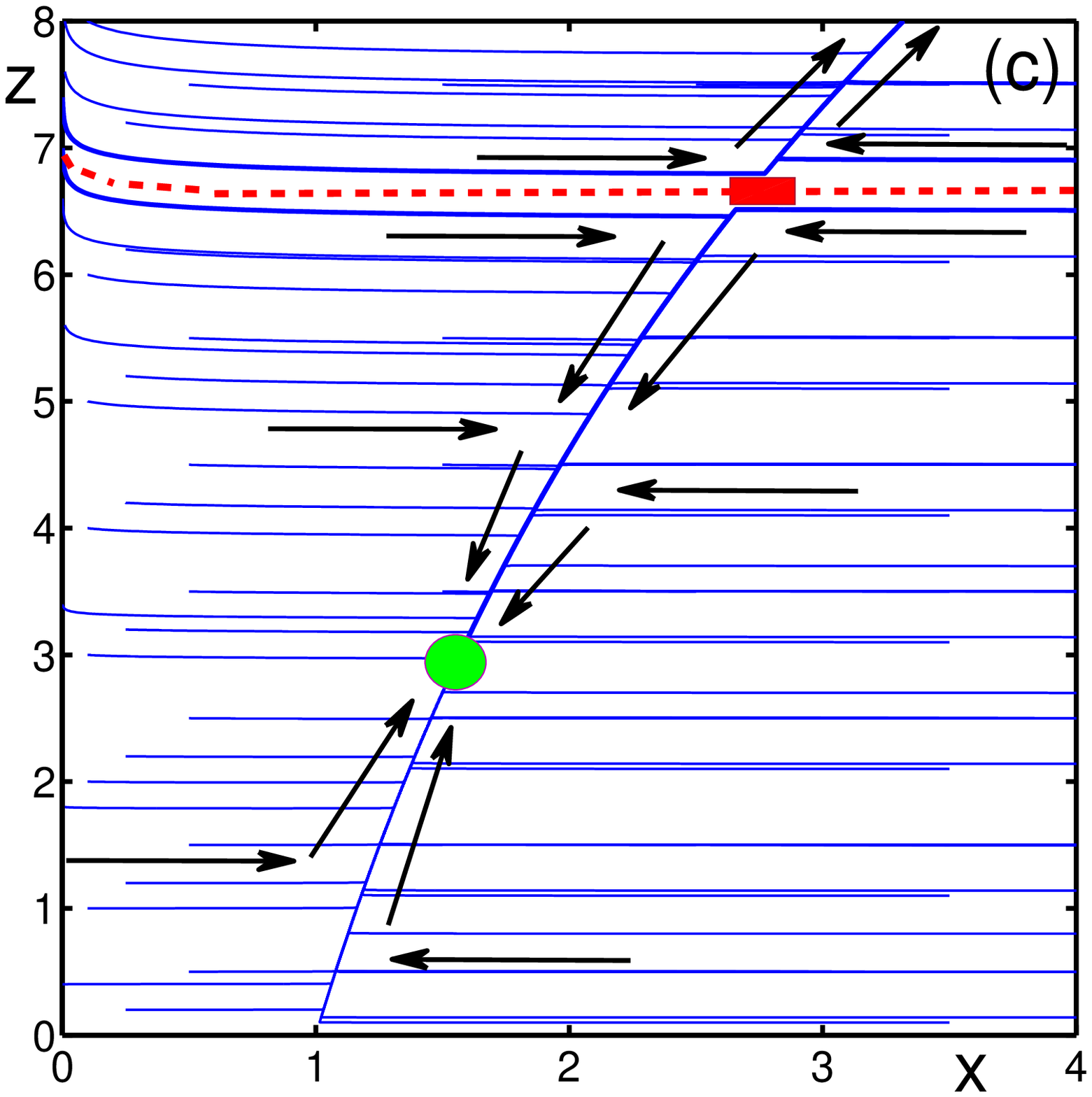} }  }
\caption{Phase portraits in the plane $x-z$ for the parametric region $B$,
for the symbiotic parameters $b = 0.15$ and $g = 0.7$ and different growth
rates $\al$. The stable node $\{x_1^* = 1.56721, z_1^* = 2.9953\}$ is
denoted by the green filled disc, and the saddle $\{x_2^* = 2.7011, z_2^* = 6.6243\}$,
by a red filled rectangle. The boundary of the basin of attraction is shown by the
dashed line. (a) Phase portrait for $\al = 1$; (b) phase portrait for
$\al = 10$; (c) phase portrait for $\al = 500$.
}
\label{fig:Fig.10}
\end{figure}

\subsection{Dynamic transition under parasitism}

In the parametric region $D_2$, there exist three fixed points, such that
$1 > x_1^* > x_2^* > x_3^*$ and $z_1^* < z_2^* < z_3^* < 1$. The points
$\{x_1^*, z_1^*\}$ and $\{x_3^*, z_3^*\}$ are stable nodes, while
$\{x_2^*,z_2^*\}$ is a saddle. In the region $D_2$, the behavior of 
populations depends on initial conditions $\{x_0, z_0\}$ and on the growth
rate $\al$. With increasing time, the populations $x(t)$ and $z(t)$ can 
tend either to $\{x^*_1, z^*_1\}$ or to $\{x^*_3,z^*_3\}$, depending on 
the chosen initial conditions. The location of the boundary between the 
attraction basins, corresponding to different fixed points, strongly 
depends on the growth rate $\al$. The temporal behavior of the 
symbiotic populations is illustrated in Figs. 11, 12 and 13. Figures 14 
and 15 present the related phase portraits for different growth rates.

\begin{figure}[ht]
\centerline{
\hbox{ \includegraphics[width=5.5cm]{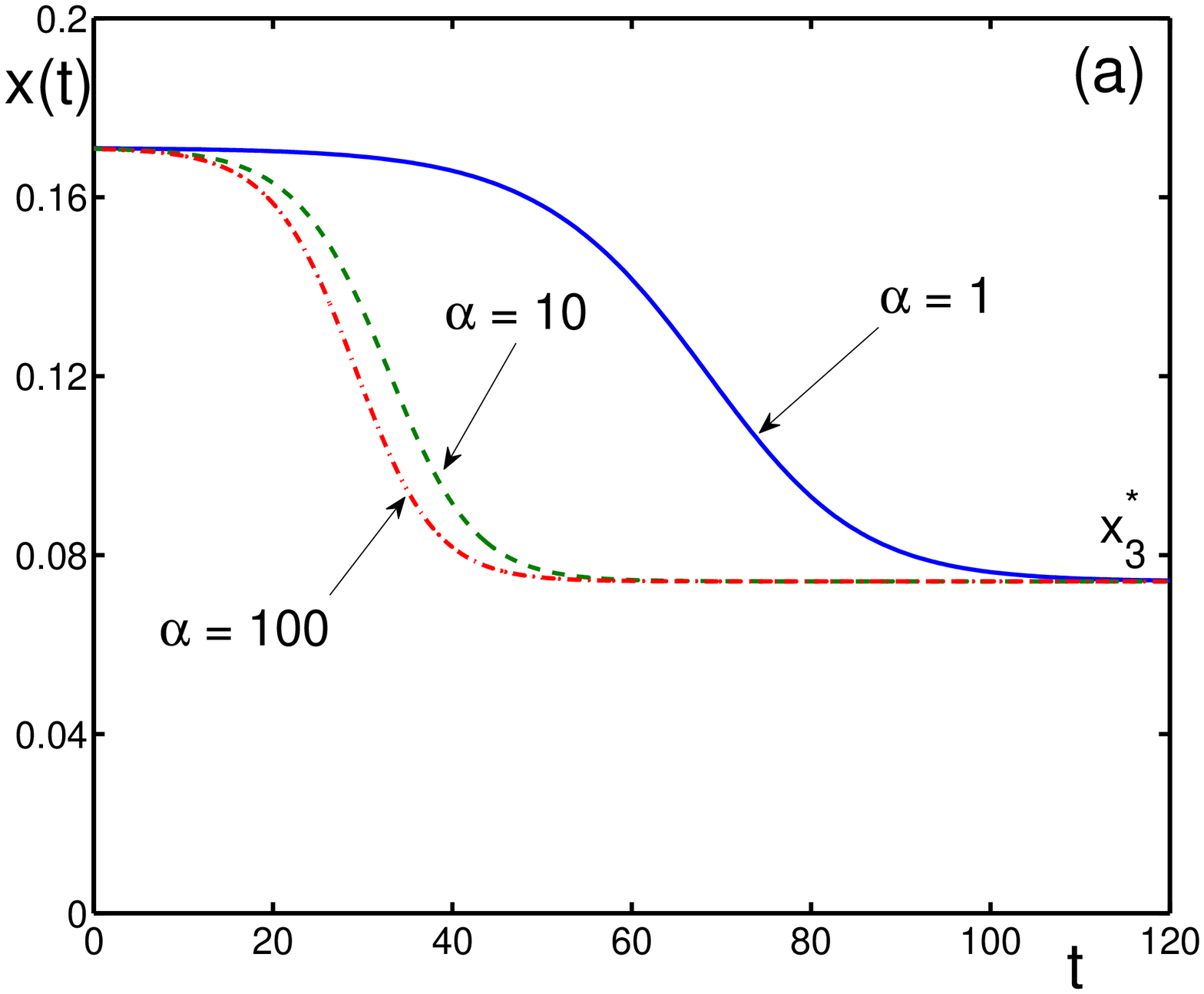} \hspace{2.5cm}
\includegraphics[width=5.5cm]{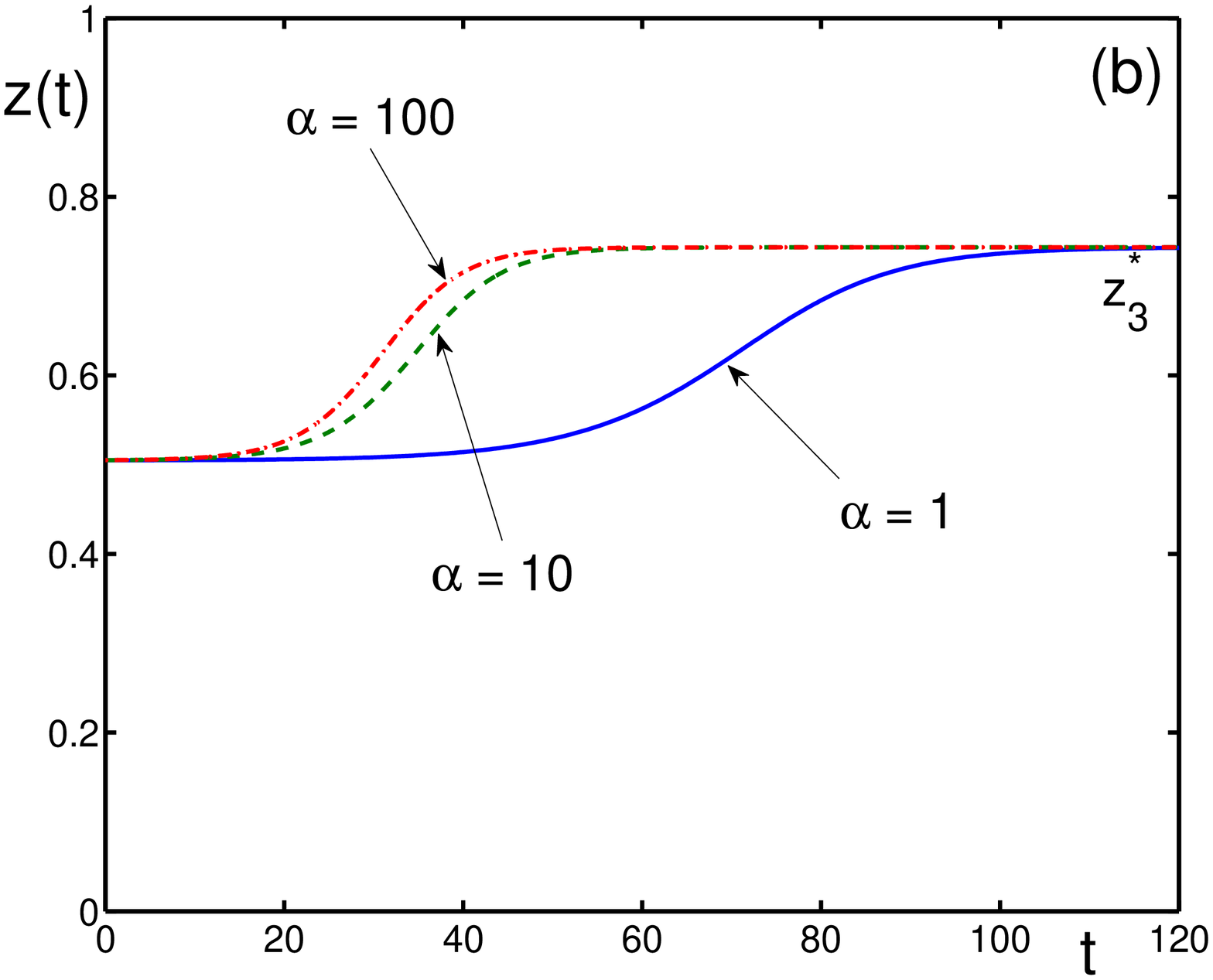} }  }
\vskip 9pt
\centerline{
\hbox{ \includegraphics[width=5.5cm]{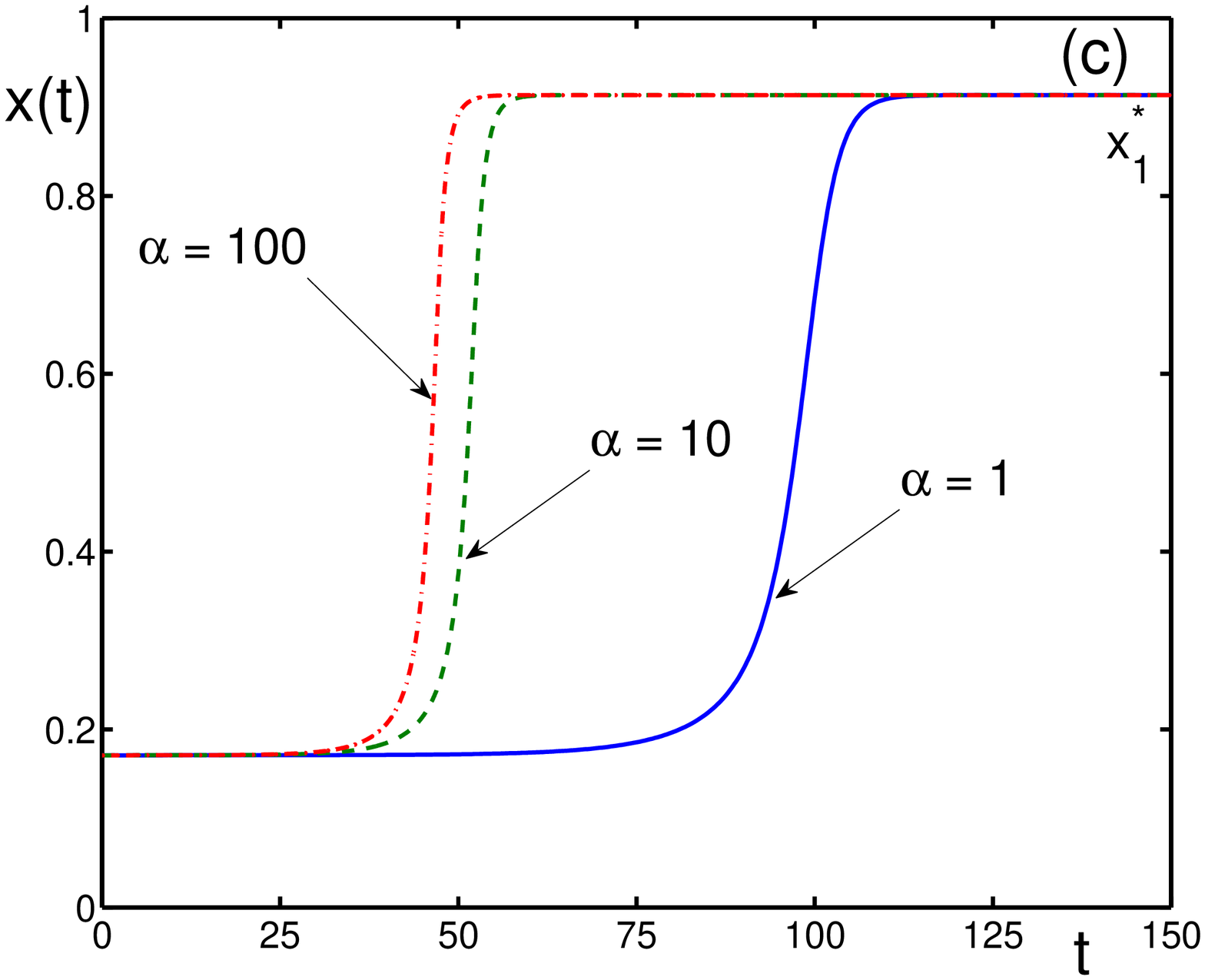} \hspace{2.5cm}
\includegraphics[width=5.5cm]{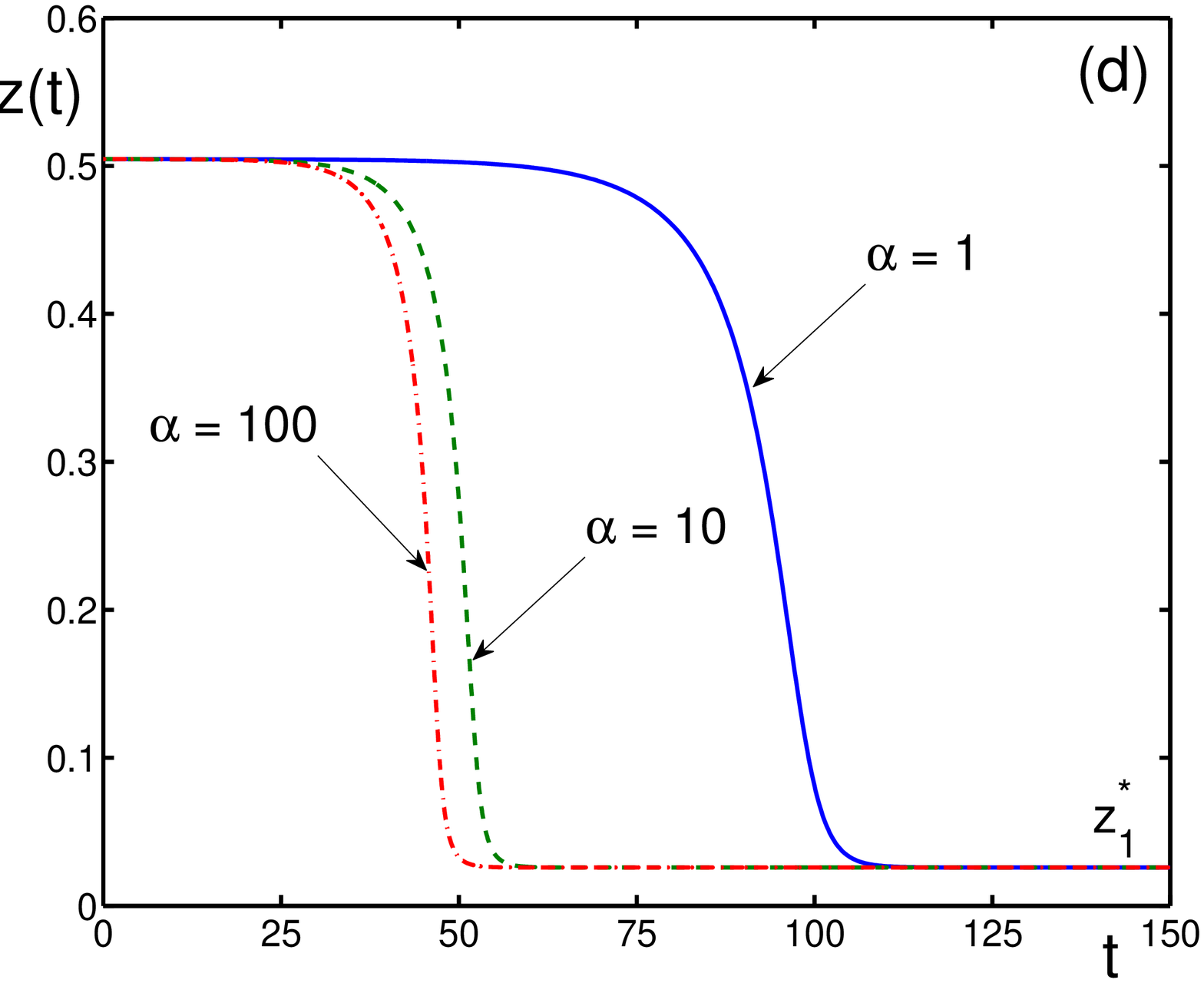} } }
\caption{Dynamics of populations $x(t)$ and $z(t)$, in the parametric
region $D_2$, for the symbiotic parameters $b = -3.5$ and $g = - 4$, with
the initial condition $x_0 = 0.171$ and different $z_0$ and $\al$. There
are two stable nodes, $\{x_1^* = 0.91332, z_1^* = 0.02591\}$ and
$\{x_3^* = 0.074141, z_3^*=0.74337\}$, and the saddle
$\{x_2^* = 0.17099, z_2^* = 0.50461\}$. (a) Population $x(t)$, with
$z_0 = 0.505 > z_2^*$, for $\al = 1$ (solid line), $\al = 10$ (dashed line),
and $\al = 100$ (dashed-dotted line); (b) population $z(t)$, with the
same $z_0$, as in (a), for $\al = 1$ (solid line), $\al = 10$ (dashed line),
and $\al = 100$ (dashed-dotted line); (c) population $x(t)$, with
$z_0 = 0.5046 < z_2^*$, for $\al = 1$ (solid line), $\al = 10$ (dashed line),
and $\al = 100$ (dashed-dotted line); (d) population $z(t)$, with the same
$z_0$, as in (c), for $\al = 1$ (solid line), $\al = 10$ (dashed line), and
$\al = 100$ (dashed-dotted line).
}
\label{fig:Fig.11}
\end{figure}

\begin{figure}[ht]
\centerline{
\hbox{ \includegraphics[width=5.5cm]{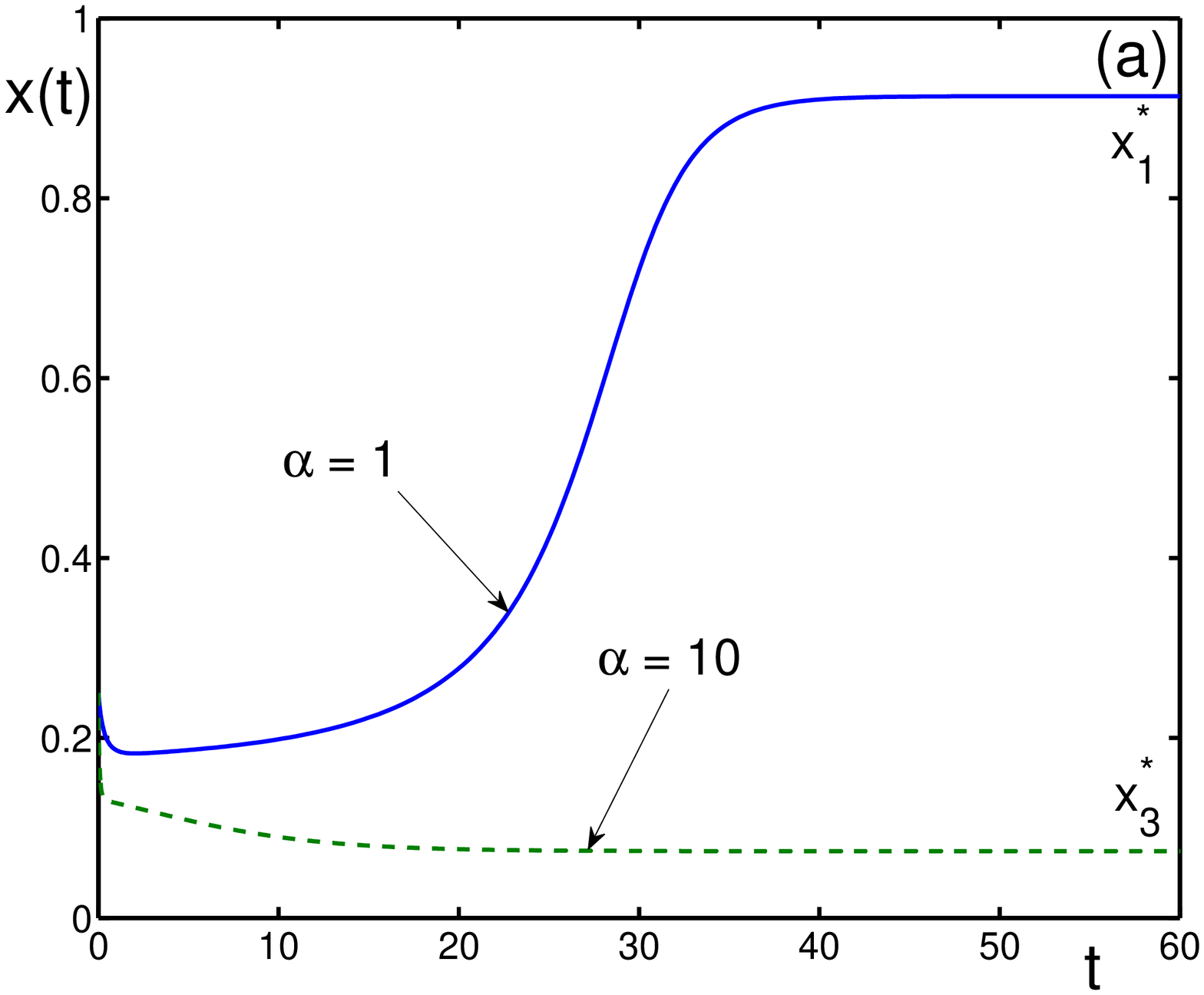} \hspace{2.5cm}
\includegraphics[width=5.5cm]{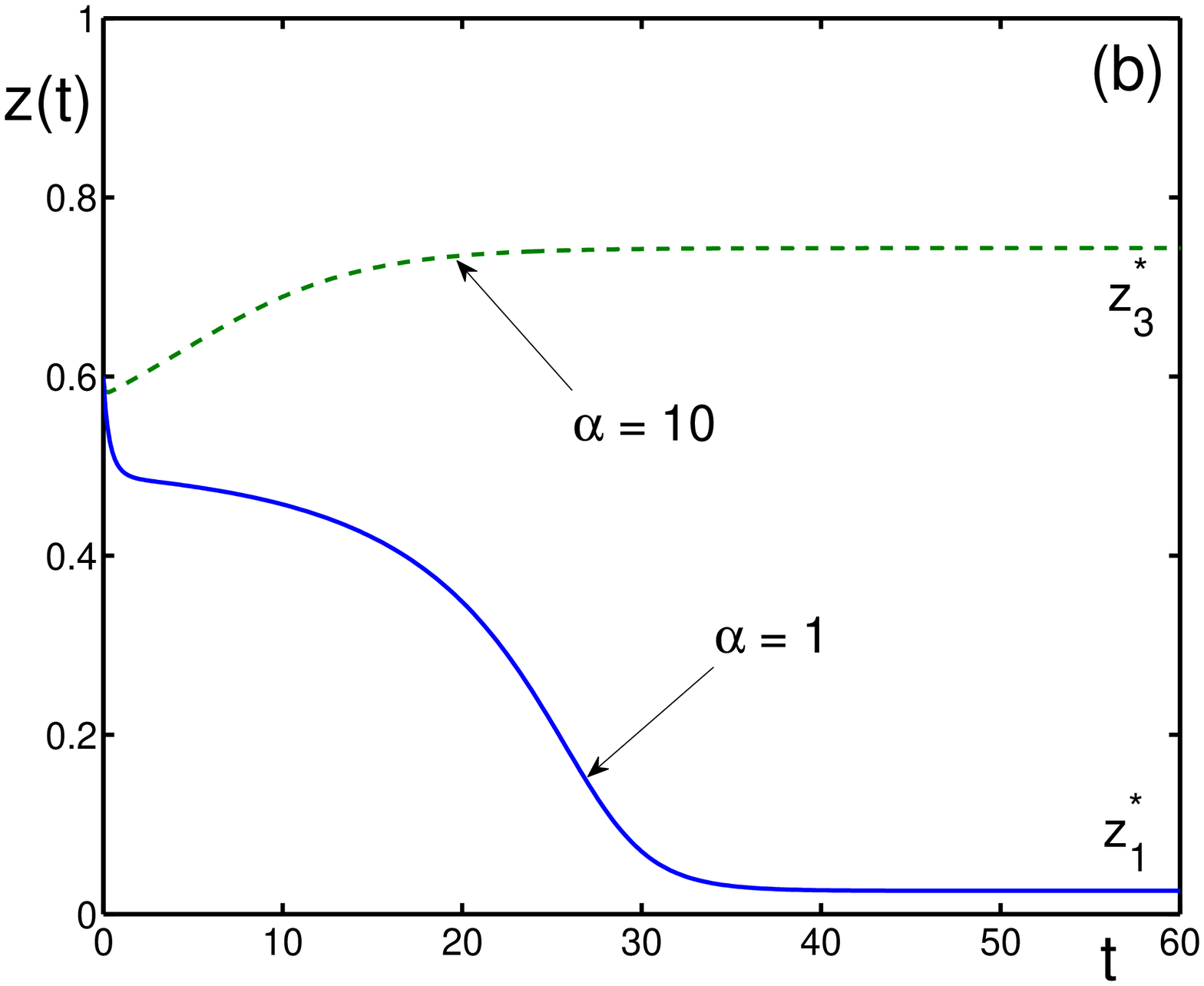} }  }
\vskip 9pt
\centerline{
\hbox{ \includegraphics[width=5.5cm]{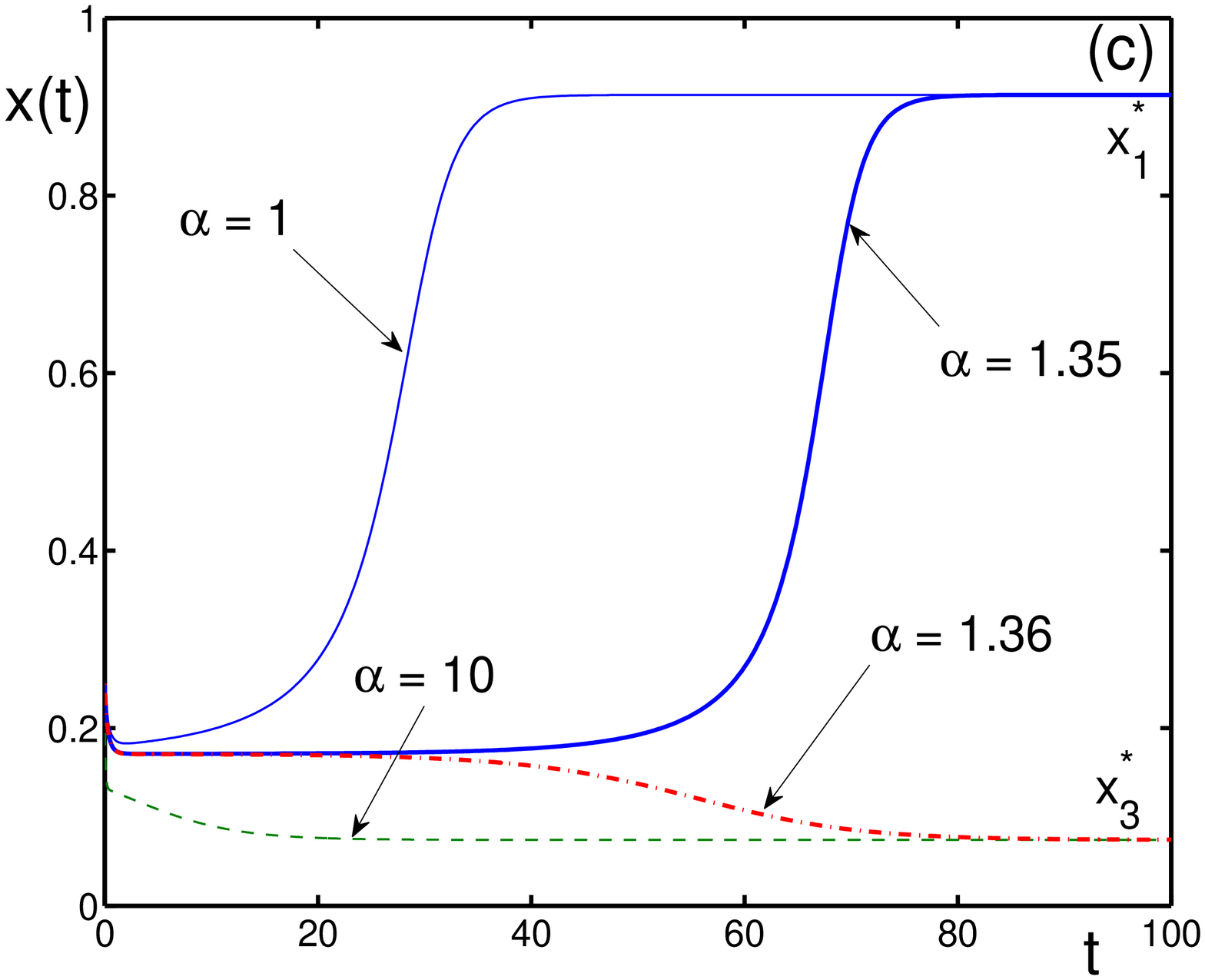} \hspace{2.5cm}
\includegraphics[width=5.5cm]{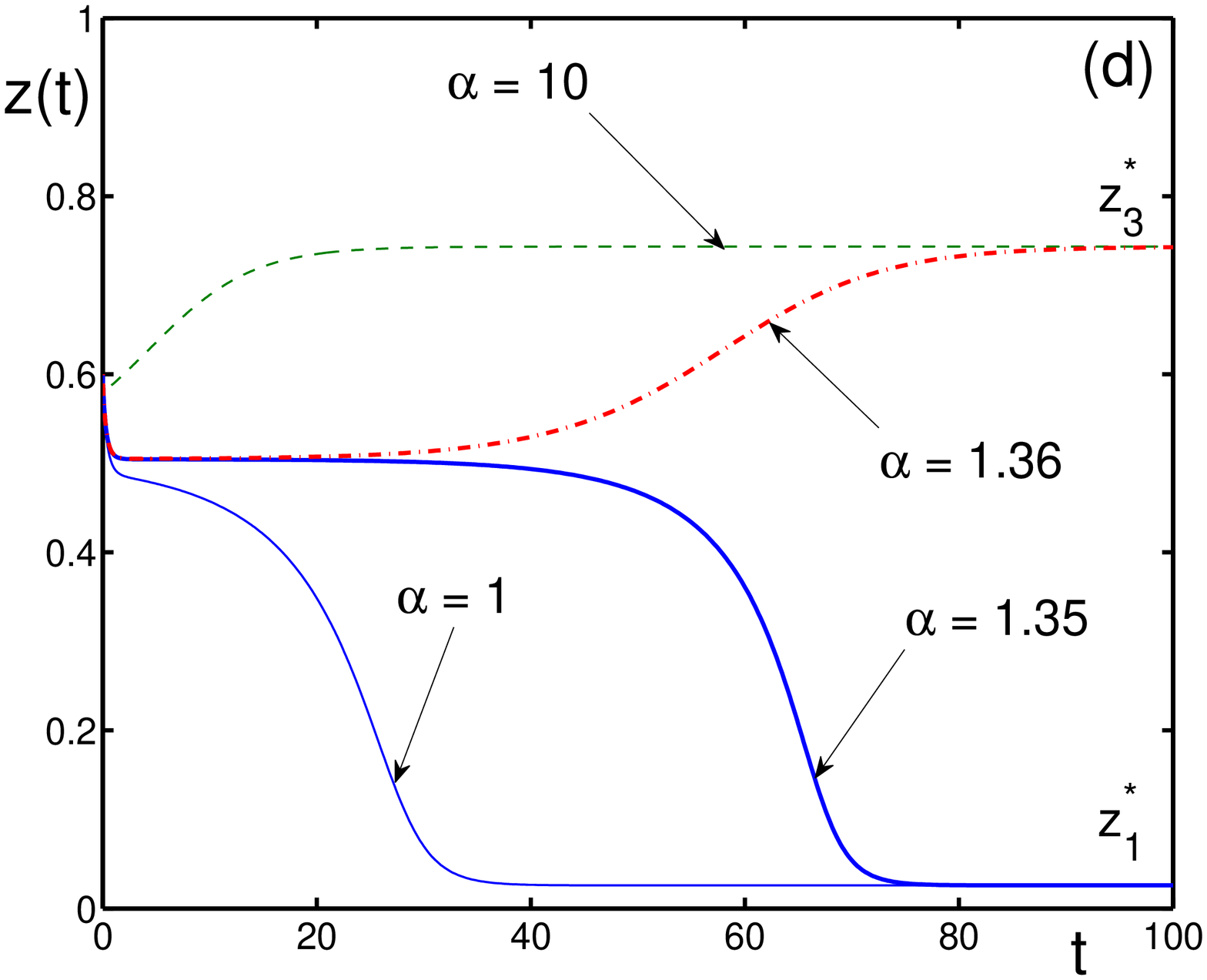} } }
\caption{Dynamics of populations $x(t)$ and $z(t)$, in the parametric
region $D_2$, with $b = - 3.5$ and $g = - 4$ as in figure 11, the initial conditions
$x_0 = 0.25$ and $z_0 = 0.6$ for different growth rates $\al$. The stable
nodes are $\{x_1^* = 0.91332, z_1^* = 0.02591\}$ and
$\{x_3^* = 0.074141, z_3^* = 0.74337\}$ and the saddle is
$\{x_2^*=0.17099, z_2^*=0.50461\}$, as in figure 11. (a) Population $x(t)$ 
for $\al=1$ (solid line) and $\al = 10$ (dashed line); (b) population $z(t)$ 
for $\al = 1$ (solid line) and $\al = 10$ (dashed line); (c) population $x(t)$
for $\al = 1$ (thin solid line), $\al = 1.35$ (solid line), $\al = 1.36$
(dashed-dotted line), and $\al = 10$ (thin dashed line); (d) population $z(t)$
for $\al = 1$ (thin solid line), $\al = 1.35$ (solid line), $\al = 1.36$
(dashed-dotted line), and $\al = 10$ (thin dashed line). For the given
parameters, there exists the critical value $\al_{crit}$ of the growth rate,
such that $1.35 < \al_{crit} < 1.36$. Under the same initial conditions,
if $1 \leq \al < \al_{crit}$, then the populations tend to the node
$\{x_1^* = 0.91332, z_1^* = 0.02591\}$, while when $\al > \al_{crit}$,
the populations converge to the other node
$\{x_3^* = 0.074141, z_3^* = 0.74337\}$.
}
\label{fig:Fig.12}
\end{figure}

\begin{figure}[ht]
\centerline{
\hbox{ \includegraphics[width=5.5cm]{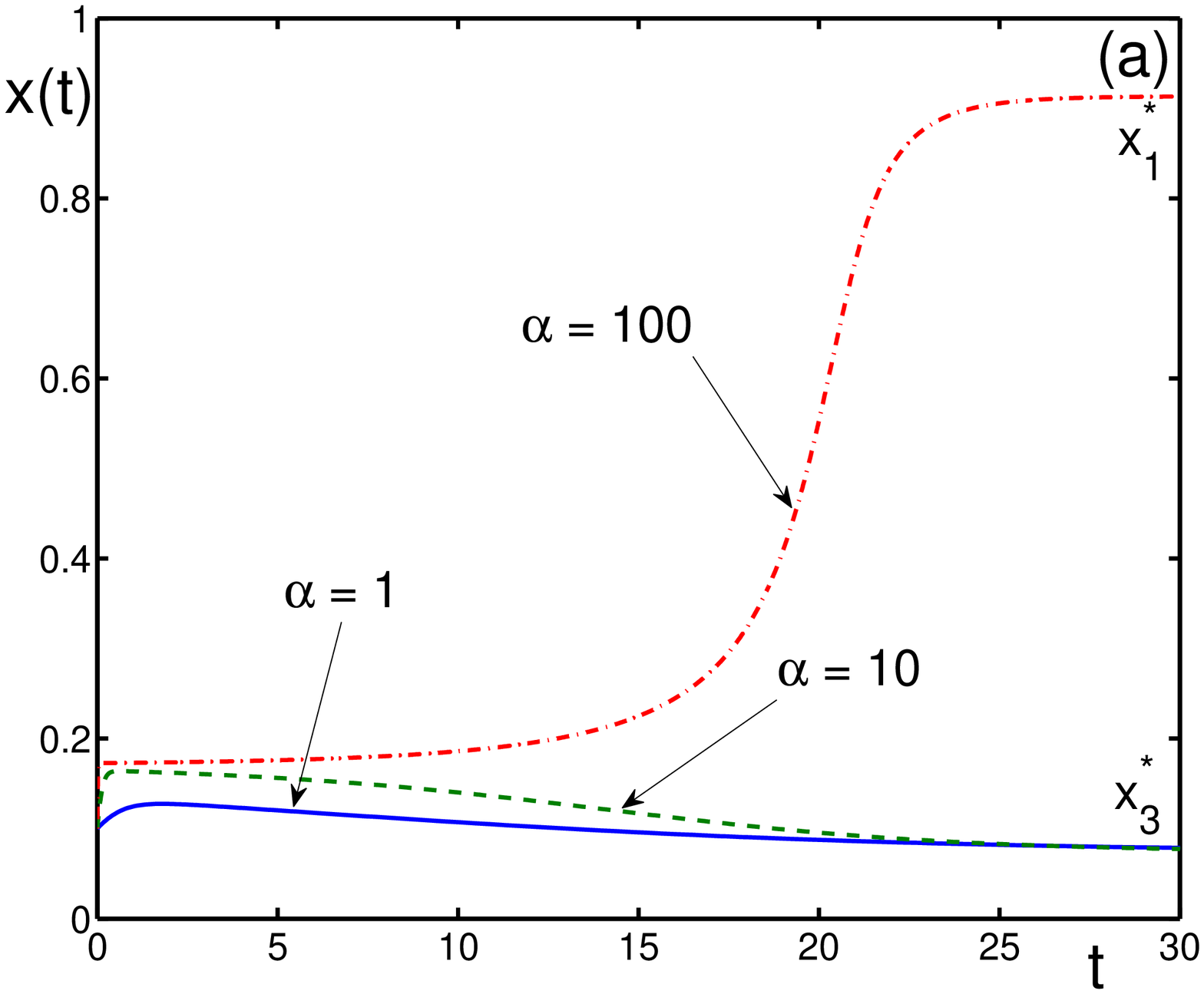} \hspace{2.5cm}
\includegraphics[width=5.5cm]{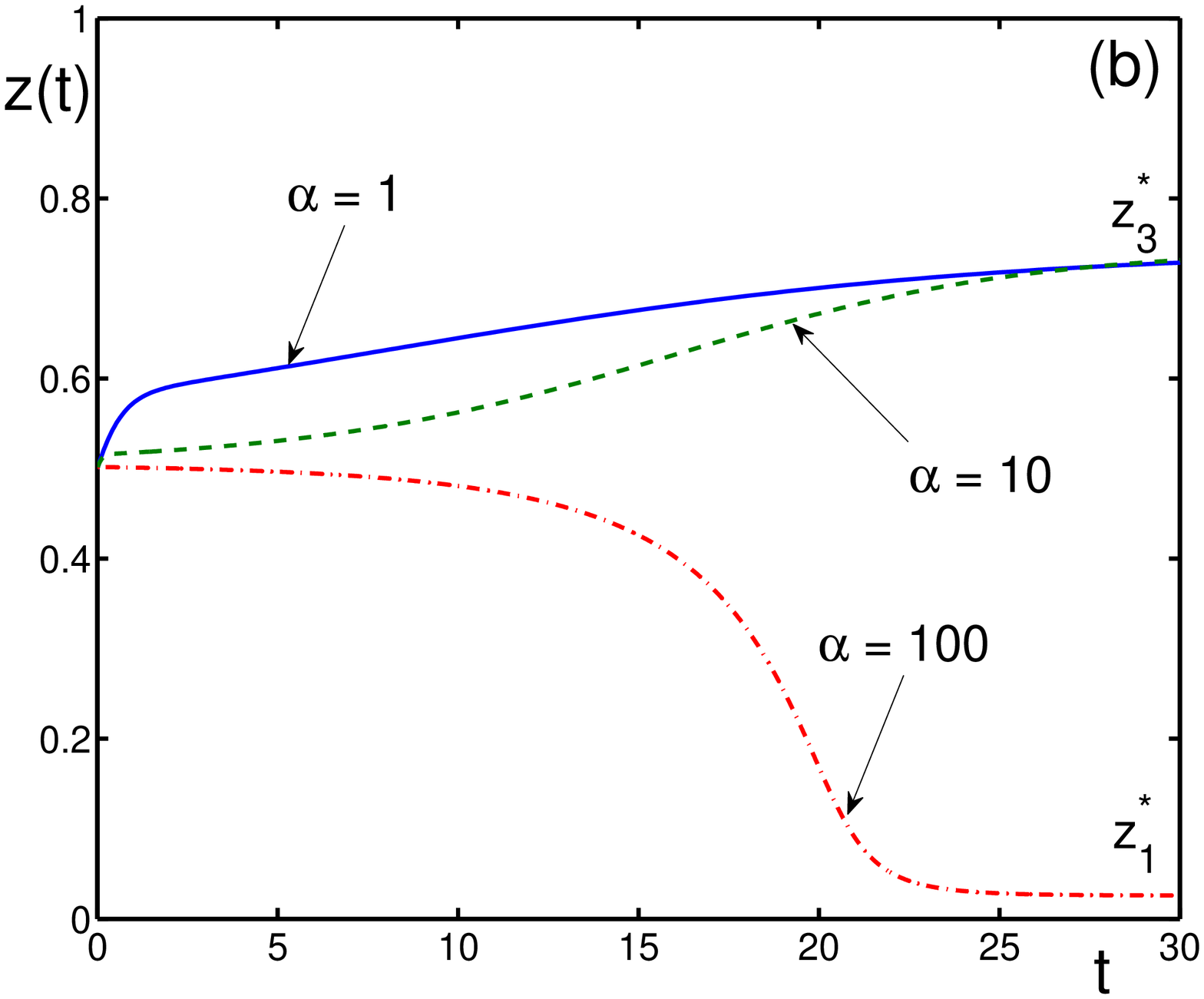} }  }
\vskip 9pt
\centerline{
\hbox{ \includegraphics[width=5.5cm]{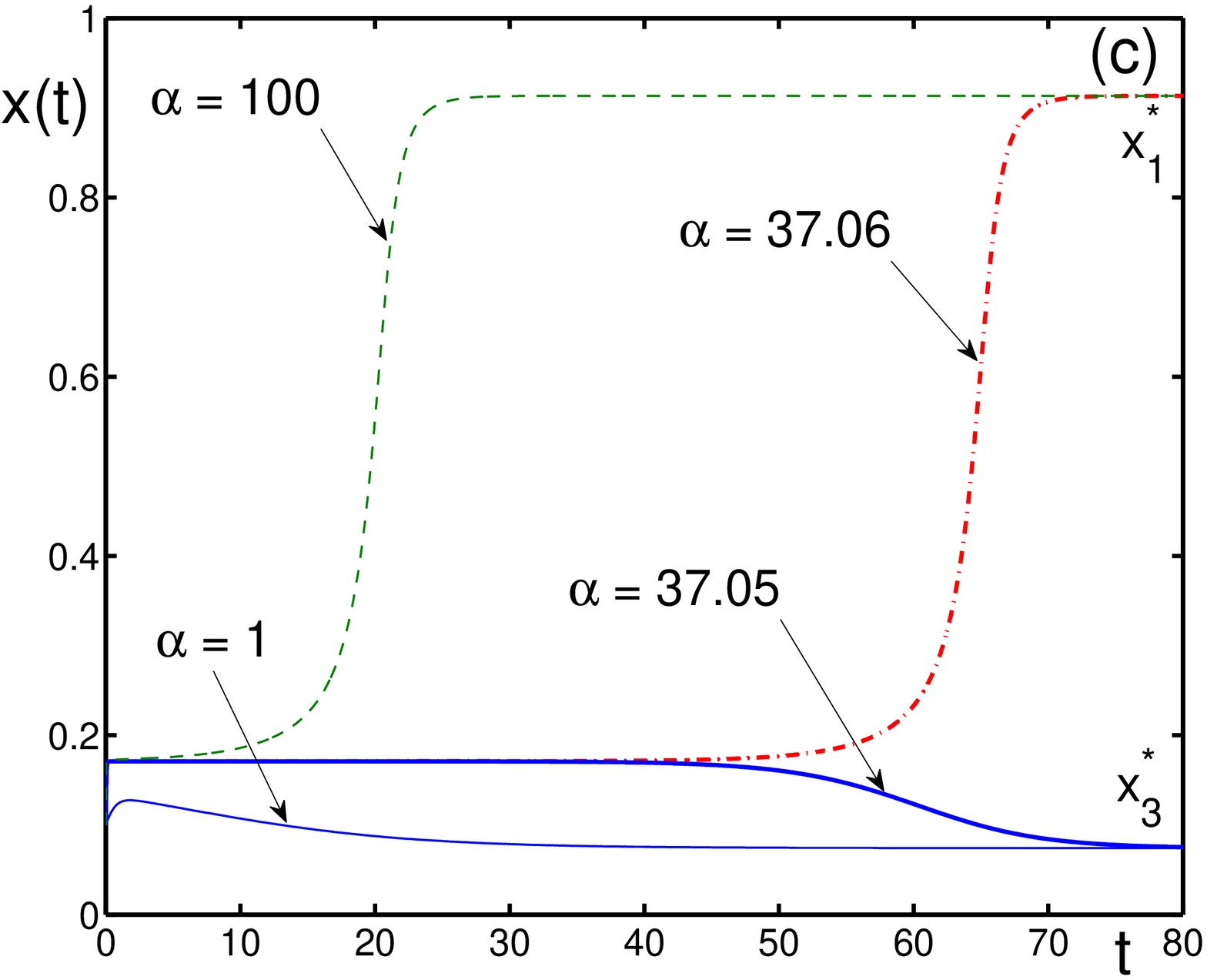} \hspace{2.5cm}
\includegraphics[width=5.5cm]{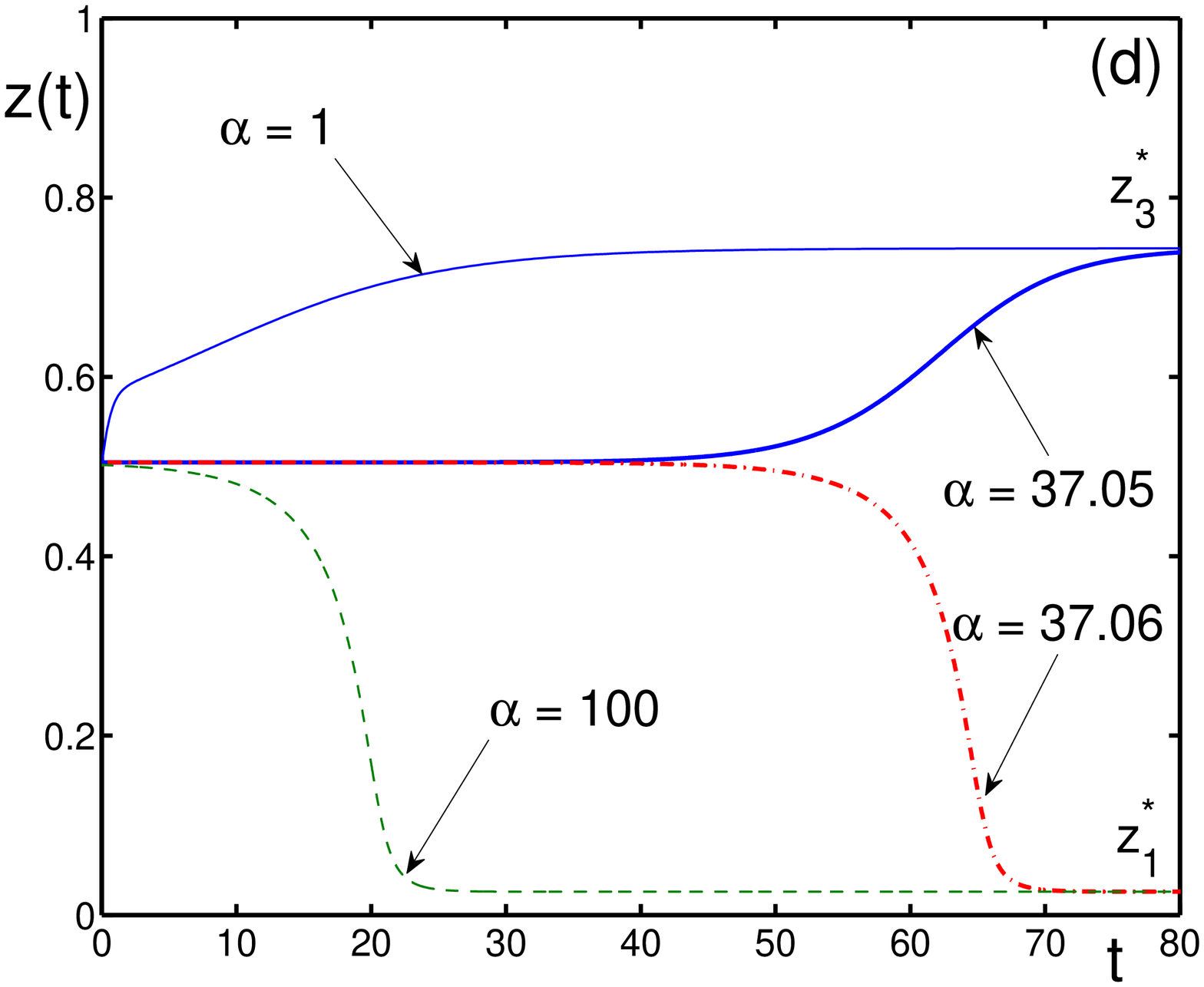} } }
\caption{Dynamics of populations $x(t)$ and $z(t)$, in the parametric
region $D_2$, with $b = - 3.5$ and $g = - 4$, the initial conditions
$x_0 = 0.1$ and $z_0 = 0.5$, for different $\al$. The stable nodes are
$\{x_1^* = 0.91332, z_1^* = 0.02591\}$ and $\{x_3^* = 0.074141, z_3^* = 0.74337\}$ 
and the saddle is $\{x_2^* = 0.17099, z_2^* = 0.50461\}$, as in figs. 11 and 
12. (a) Population $x(t)$ for $\al = 1$ (solid line), $\al = 10$ (dashed line), 
and $\al = 100$ (dashed-dotted line);
(b) population $z(t)$ for $\al = 1$ (solid line), $\al = 10$ (dashed line),
and $\al = 100$ (dashed-dotted line); (c) population $x(t)$ for $\al = 1$
(thin solid line), $\al = 37.05$ (solid line), $\al = 37.06$
(dashed-dotted line), and $\al = 100$ (thin dashed line); (d) population $z(t)$
for $\al = 1$ (thin solid line), $\al = 37.05$ (solid line), $\al = 37.06$
(dashed-dotted line), and $\al = 100$ (thin dashed line).
There exists the critical growth rate $\al_{crit}$ in the interval
$37.05 < \al_{crit} < 37.06$, such that, when $1 \leq \al < \al_{crit}$,
the populations tend to the node $\{x_3^* = 0.074141, z_3^* = 0.74337\}$,
while if $\al > \al_{crit}$, the populations converge to the node
$\{x_1^* = 0.91332, z_1^* = 0.02591\}$.
}
\label{fig:Fig.13}
\end{figure}

\begin{figure}[ht]
\centerline{
\hbox{ \includegraphics[width=5cm]{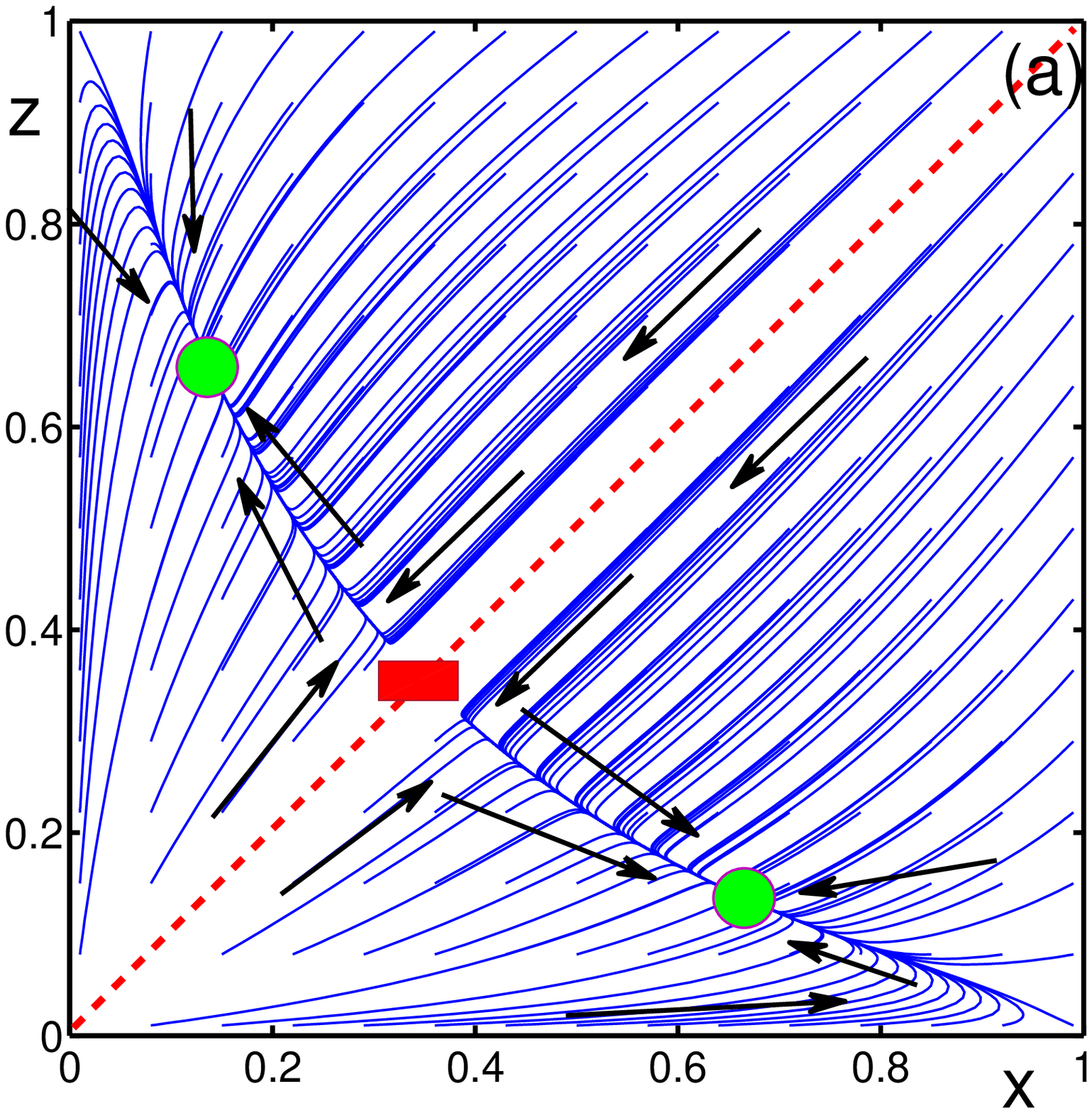} \hspace{1cm}
\includegraphics[width=5cm]{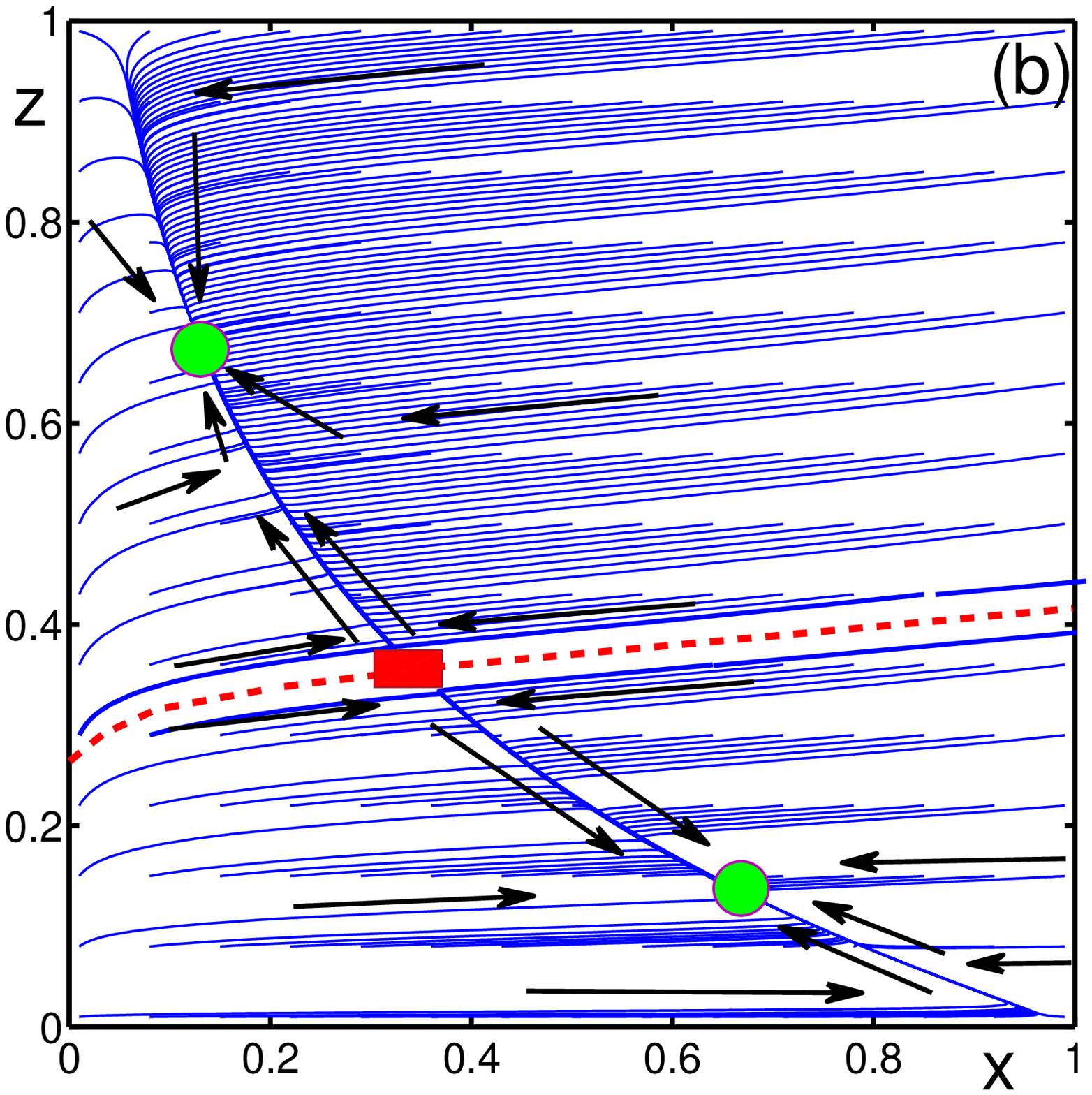} \hspace{1cm}
\includegraphics[width=5cm]{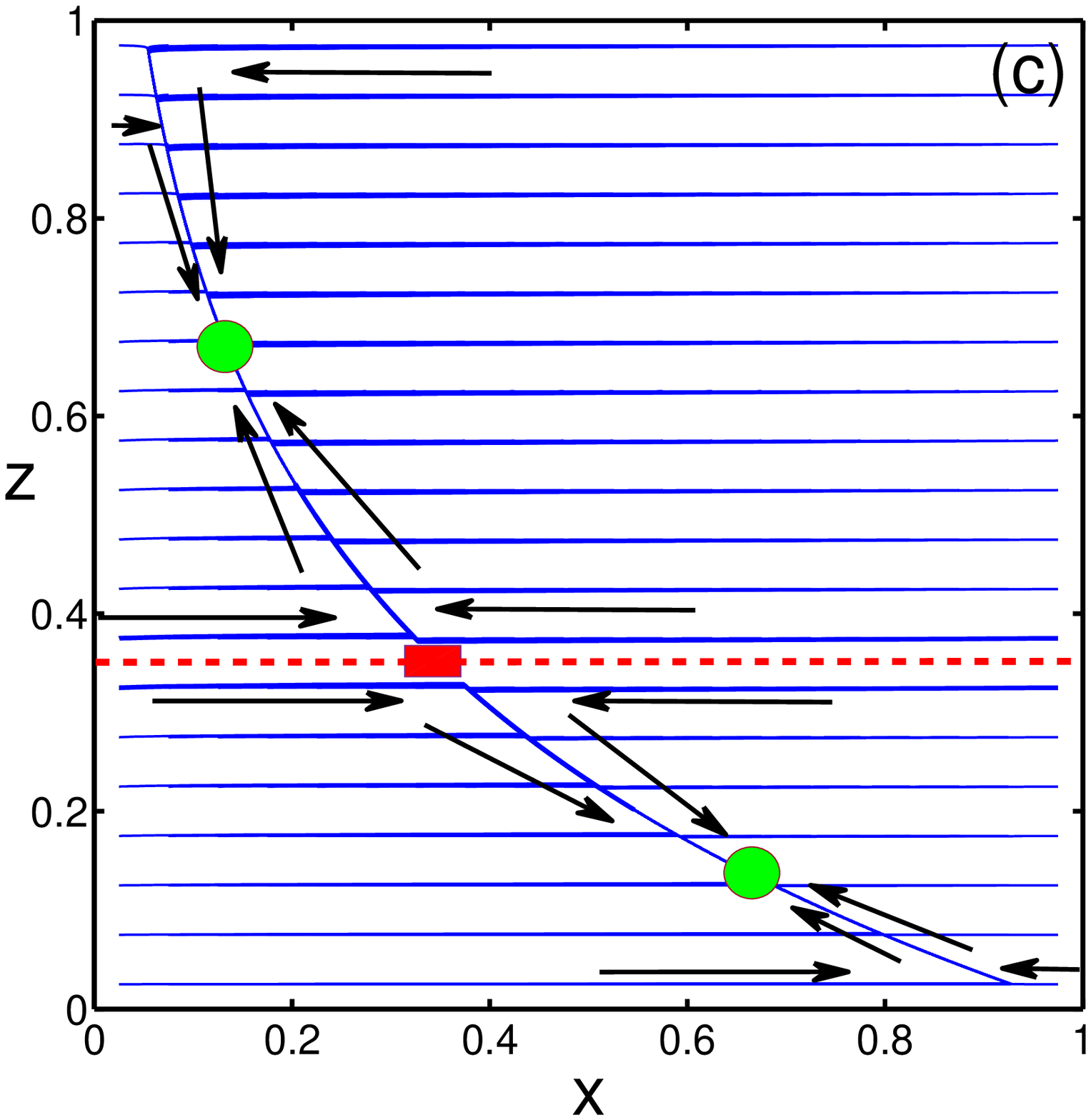} }  }
\caption{Phase portraits on the plane $x-z$ for the parametric region $D_2$,
with $b = g = -3$, for different $\al$. The dashed line shows the boundary
between the attraction basins of the stable nodes
$\{x_1^* = 0.66474, z_1^*=0.13612\}$ and $\{x_3^* = 0.13612, z_3^*=0.66474\}$,
which are represented by the filled green discs. The saddle point 
$\{x_2^*= z_2^*= 0.34997\}$ is shown by the filled red rectangle. 
(a) Phase portrait for $\al = 1$, which exhibits
a symmetric boundary line given by the equation $x = z$ (dashed line);
(b) phase portrait for $\al = 10$; (c) phase portrait for $\al = 200$.
For large growth rates $\al \ra \infty$, the boundary between the
attraction basins tends to the line $z=z_2^*$.
}
\label{fig:Fig.14}
\end{figure}

\begin{figure}[ht]
\centerline{
\hbox{ \includegraphics[width=5cm]{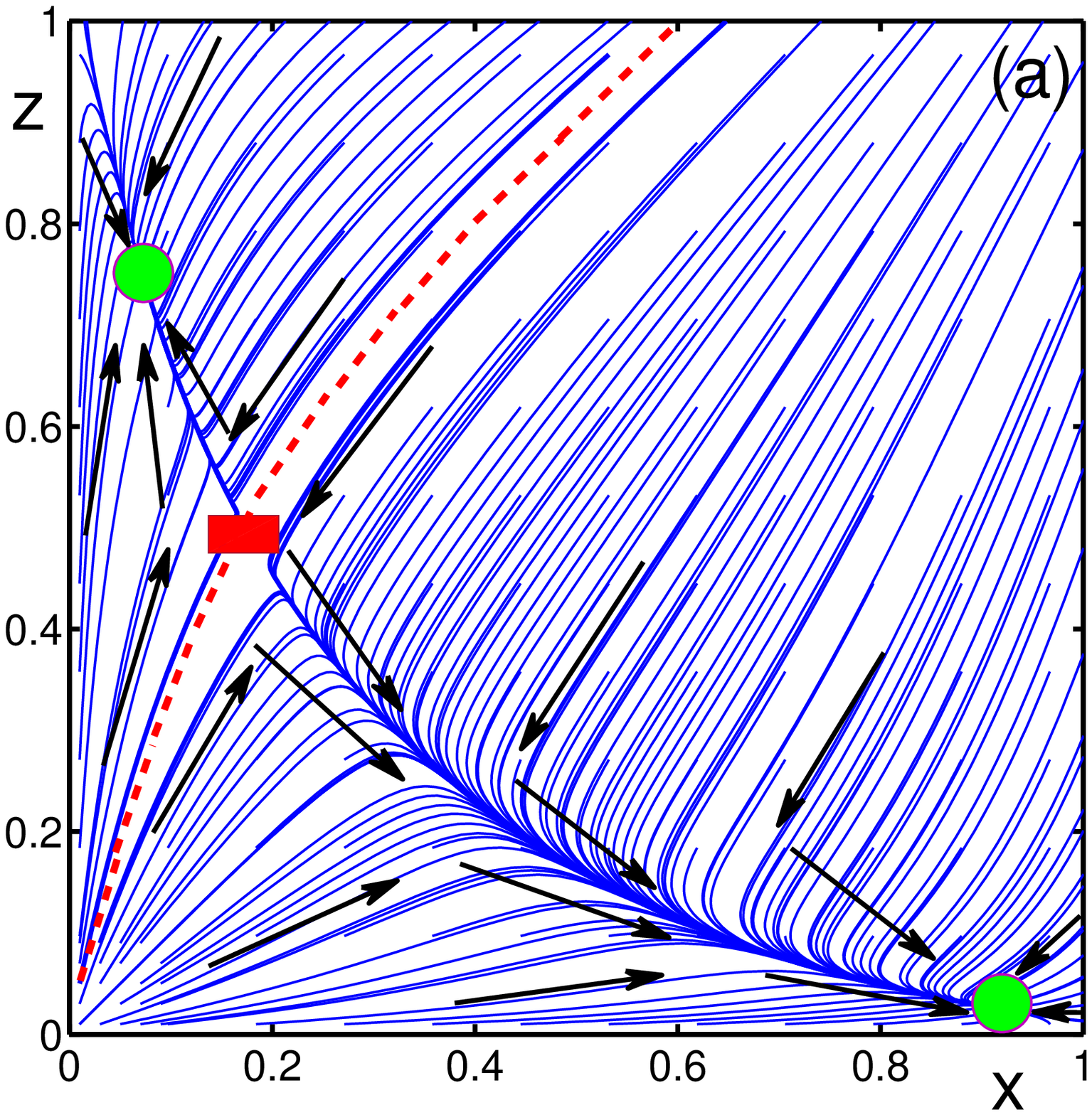} \hspace{1cm}
\includegraphics[width=5cm]{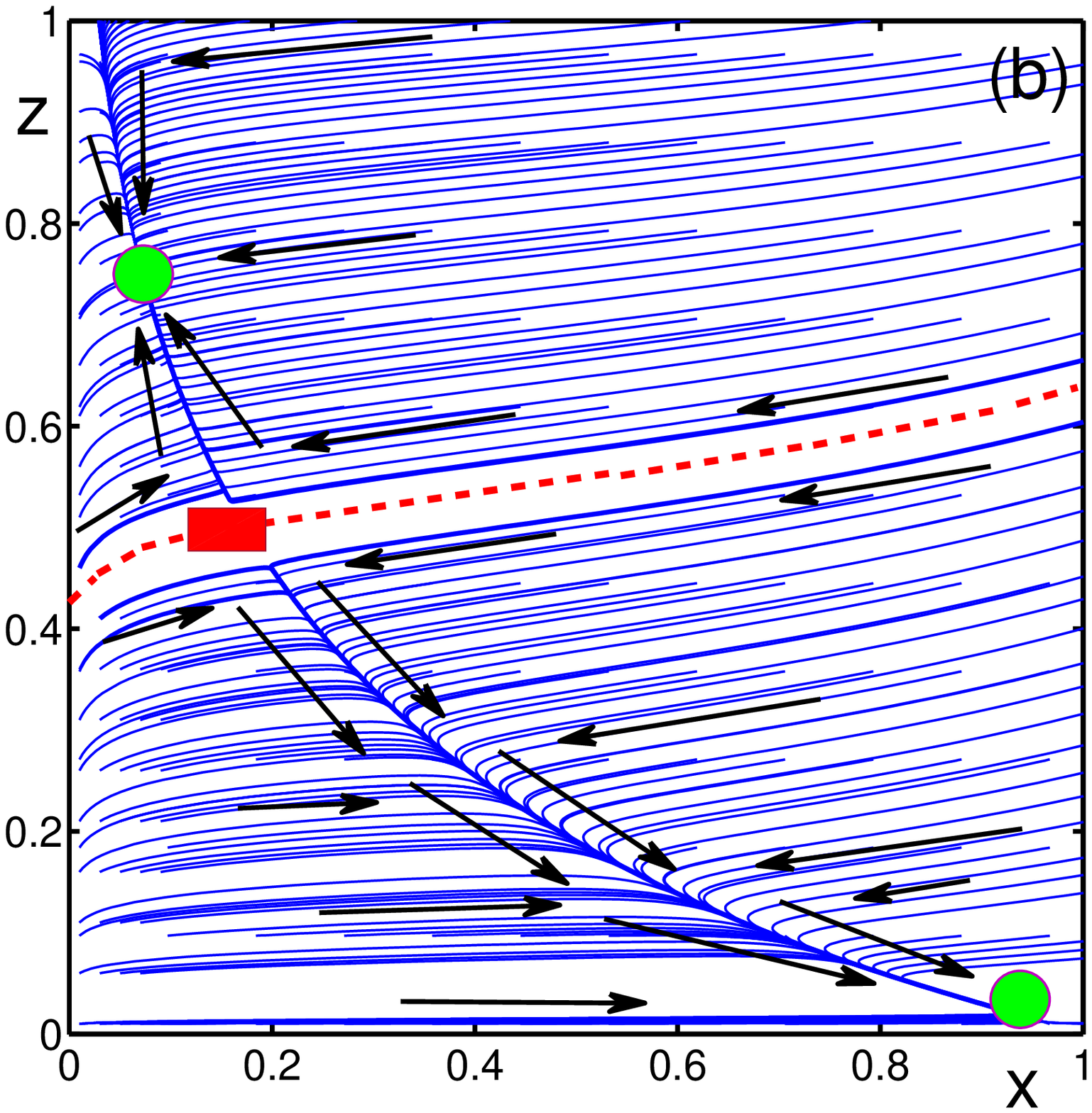} \hspace{1cm}
\includegraphics[width=5cm]{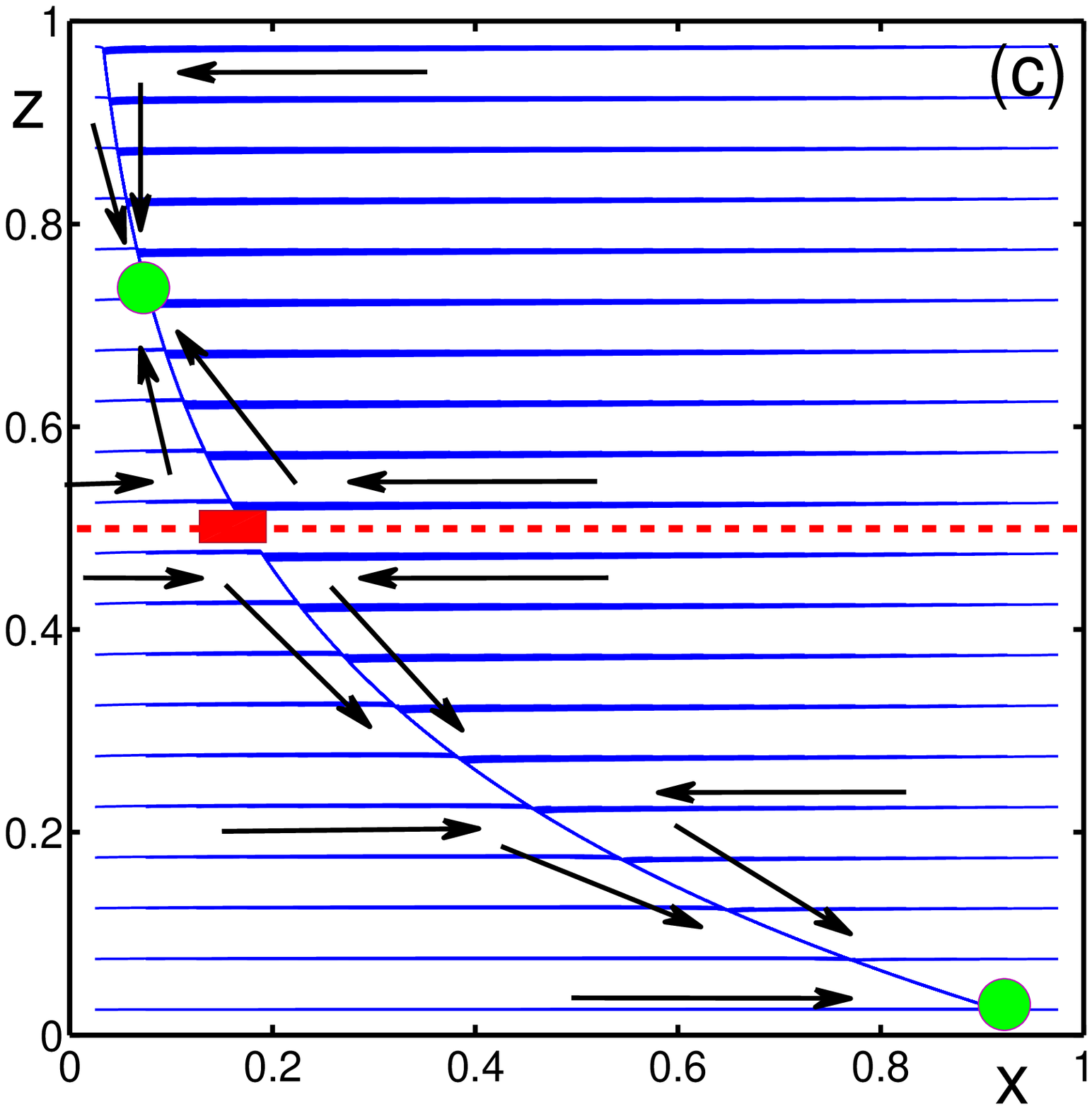} }  }
\caption{Phase portraits on the plane $x-z$ for the parametric region $D_2$,
with $b = - 3.5$ and $g = - 4$, for different $\al$. The dashed line shows
the boundary between the attraction basins of the stable nodes
$\{x_1^* = 0.91332, z_1^* = 0.02591\}$ and $\{x_3^* = 0.074141, z_3^* = 0.74337\}$,
which are represented by the filled green discs. The saddle point
$\{x_2^* = 0.17099, z_2^* = 0.50461\}$ is indicated by the filled red rectangle.
(a) Phase portrait for $\al = 1$; (b) phase portrait for $\al = 10$;
(c) phase portrait for $\al = 200$. In the limit of large $\al$'s, the boundary
between the attraction basins tends to the line $z=z_2^*$.
}
\label{fig:Fig.15}
\end{figure}

\section{Approximate Solutions of Symbiotic Equations in the Presence 
of Coexisting Fast and Slow Populations}

For large growth rates $\al \gg 1$, equations (\ref{11}) imply that the variable 
$x$ is fast while the variable $z$ is slow. In this case, the analysis of the 
evolution equations can be done by resorting to the Bogolubov-Krylov 
averaging techniques \cite{Bogolubov_20}. As is described 
in the scale-separation approach \cite{Yukalov_21}, we solve the equation for 
the fast variable, keeping the slow variable as a quasi-integral of motion, 
which gives
\be
\label{21}
 x = \frac{x_0}{x_0(1-e^{-\al t})e^{-bz} +e^{-\al t}} \; .
\ee
This expression is substituted into the equation for the slow variable, which 
is averaged over time, resulting in the equation 
\be
\label{22}
\frac{dz}{dt} = z - z^2 \exp \left ( -g e^{bz} \right ) \;  .
\ee
Equations (\ref{21}) and (\ref{21}) define the so-called guiding centers 
of the solutions to equations (\ref{11}). 

In Figures 16 and 17, we demonstrate that the guiding-center solutions, 
prescribed by equations (\ref{21}) and (\ref{22}), provide rather good 
approximations for the exact solutions following from the initial
equations (\ref{11}). Surprisingly, the approximate solutions are already 
rather close to the true solutions even for $\al = 1$. The stationary states 
are identical for the approximate solutions and for exact ones.  

\begin{figure}[ht]
\centerline{
\hbox{ \includegraphics[width=5.5cm]{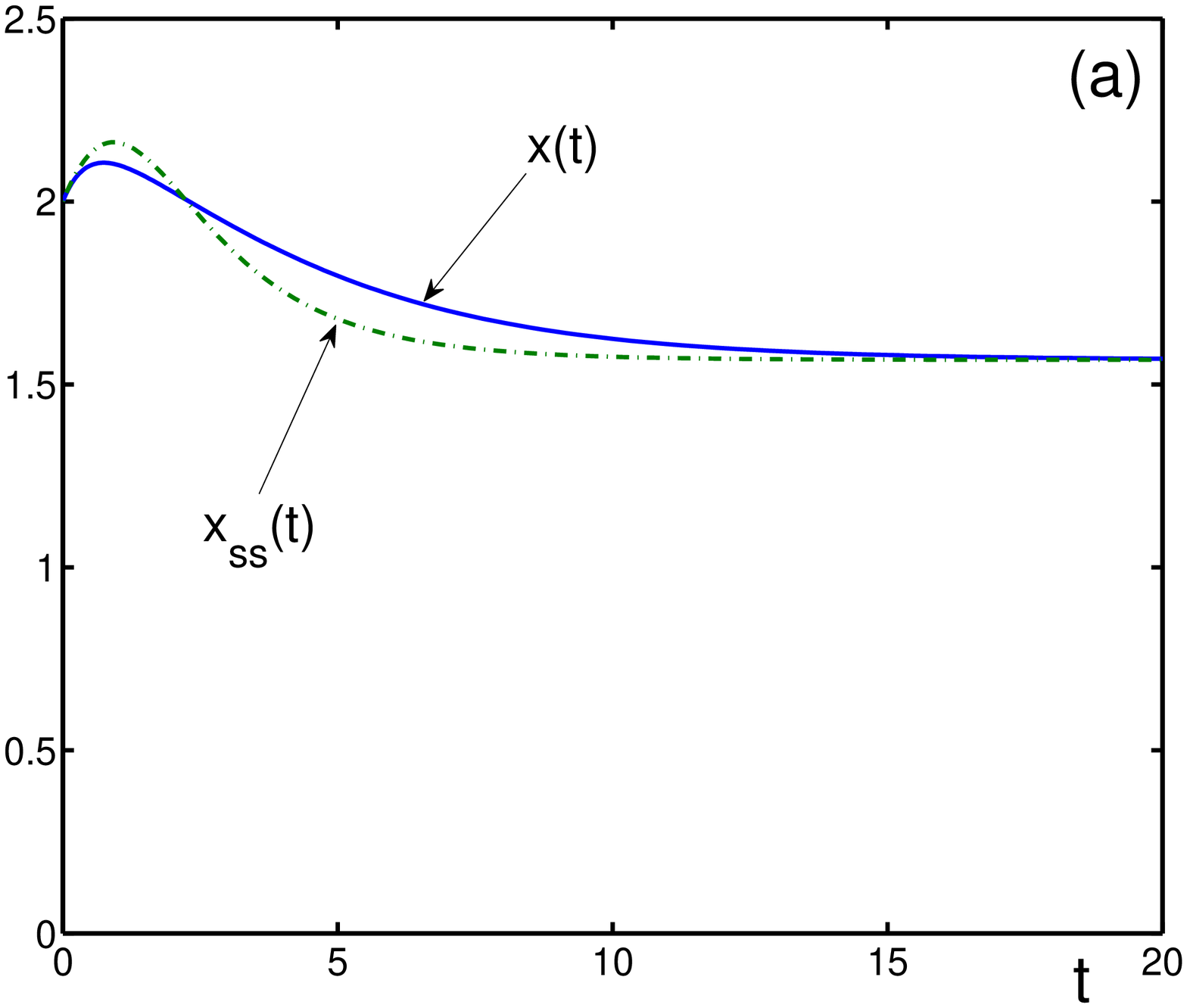} \hspace{2.5cm}
\includegraphics[width=5.5cm]{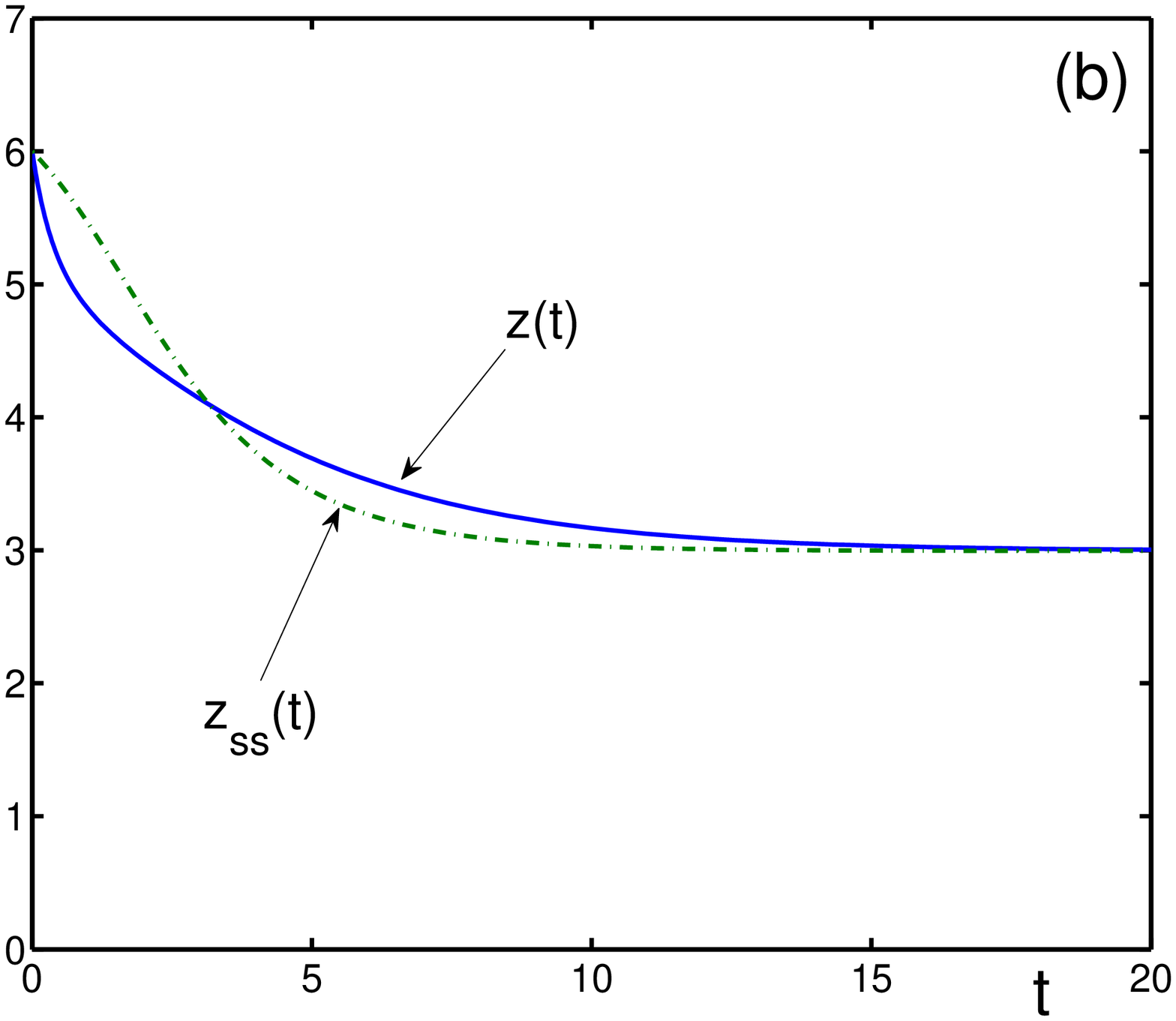} }  }
\vskip 9pt
\centerline{
\hbox{ \includegraphics[width=5.5cm]{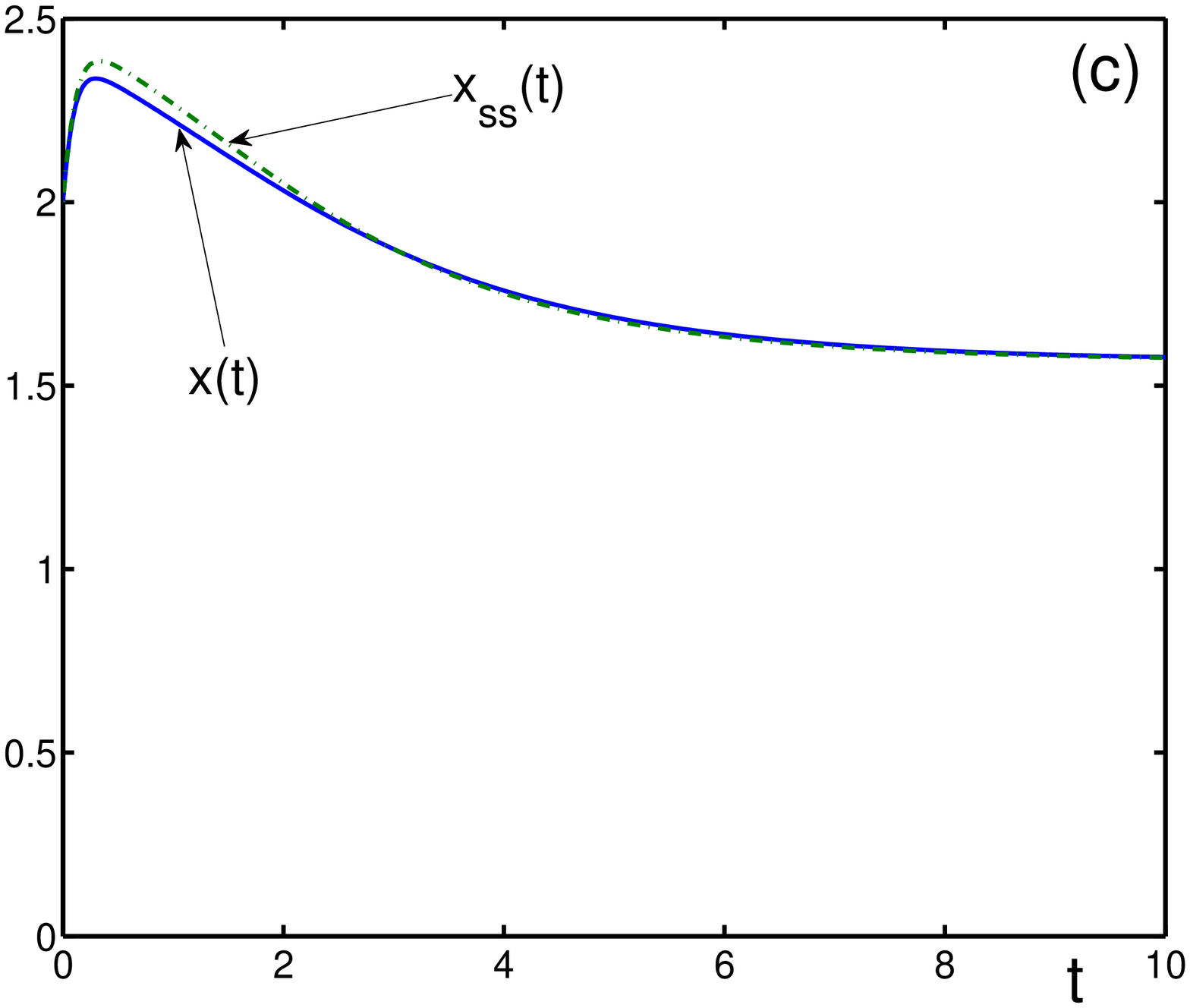} \hspace{2.5cm}
\includegraphics[width=5.5cm]{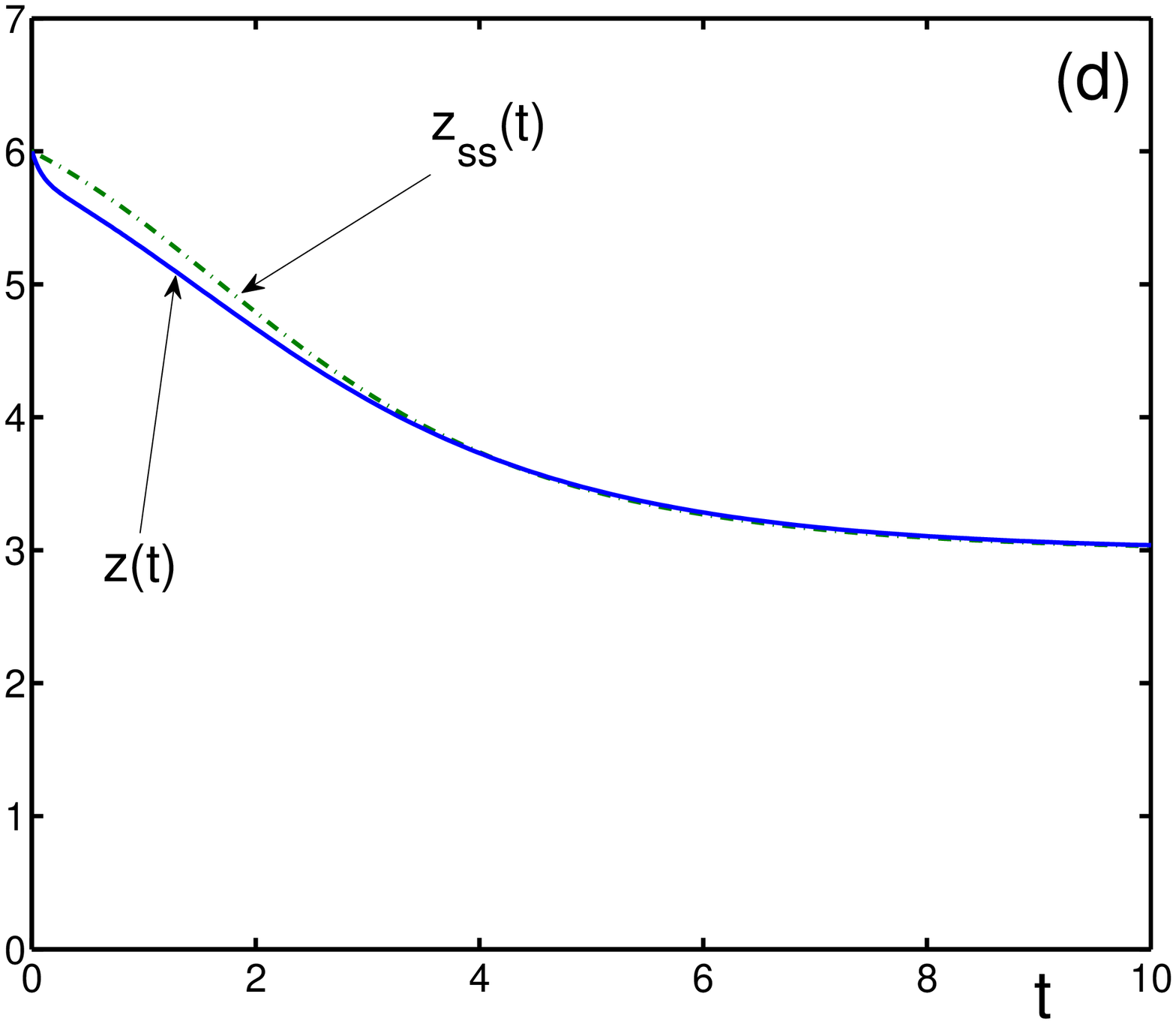} } }
\caption{Temporal behavior of approximate solutions $x_{ss}(t)$ and
$z_{ss}(t)$, obtained by the scale separation approach, as compared to the
exact solutions $x(t)$ and $z(t)$, for the case of mutualism, with the
symbiotic parameters $b = 0.15$ and $g = 0.7$, with the initial conditions
$\{x_0 = 2, z_0 = 6\}$, for different growth rates $\al$. The stable
stationary state is $\{x^* = 1.56721, z^* = 2.9953\}$.
(a) Population $x(t)$ (solid line) and its approximation $x_{ss}(t)$
(dashed-dotted line) for $\al = 1$; (b) population $z(t)$ (solid line) and
its approximation $z_{ss}(t)$ (dashed-dotted line) for $\al = 1$;
(c) population $x(t)$ (solid line) and its approximation $x_{ss}(t)$
(dashed-dotted line) for $\al = 10$; (d) population $z(t)$ (solid line) and
its approximation $z_{ss}(t)$ (dashed-dotted line) for $\al = 10$.
}
\label{fig:Fig.16}
\end{figure}

\begin{figure}[ht]
\centerline{
\hbox{ \includegraphics[width=5.5cm]{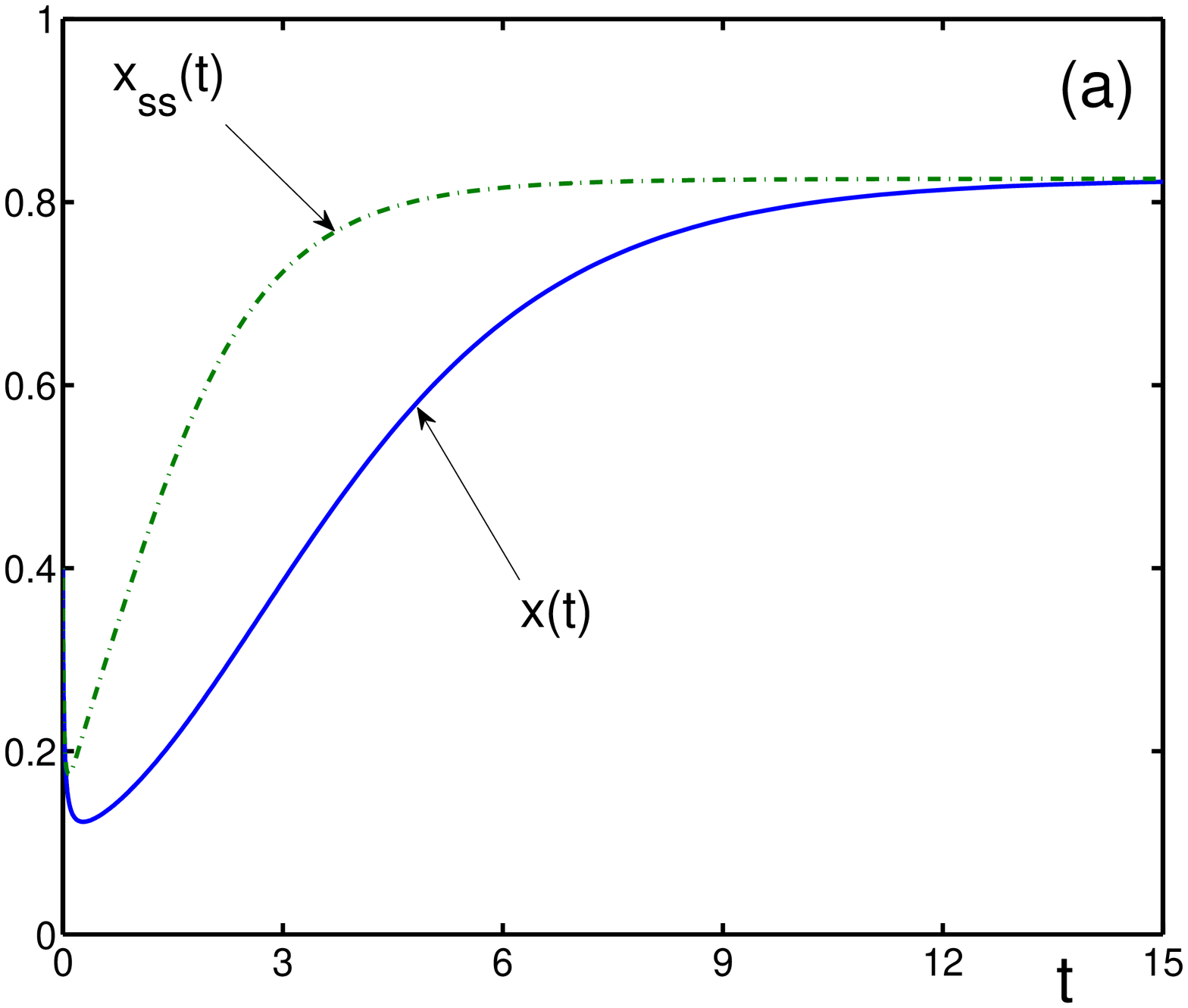} \hspace{2.5cm}
\includegraphics[width=5.5cm]{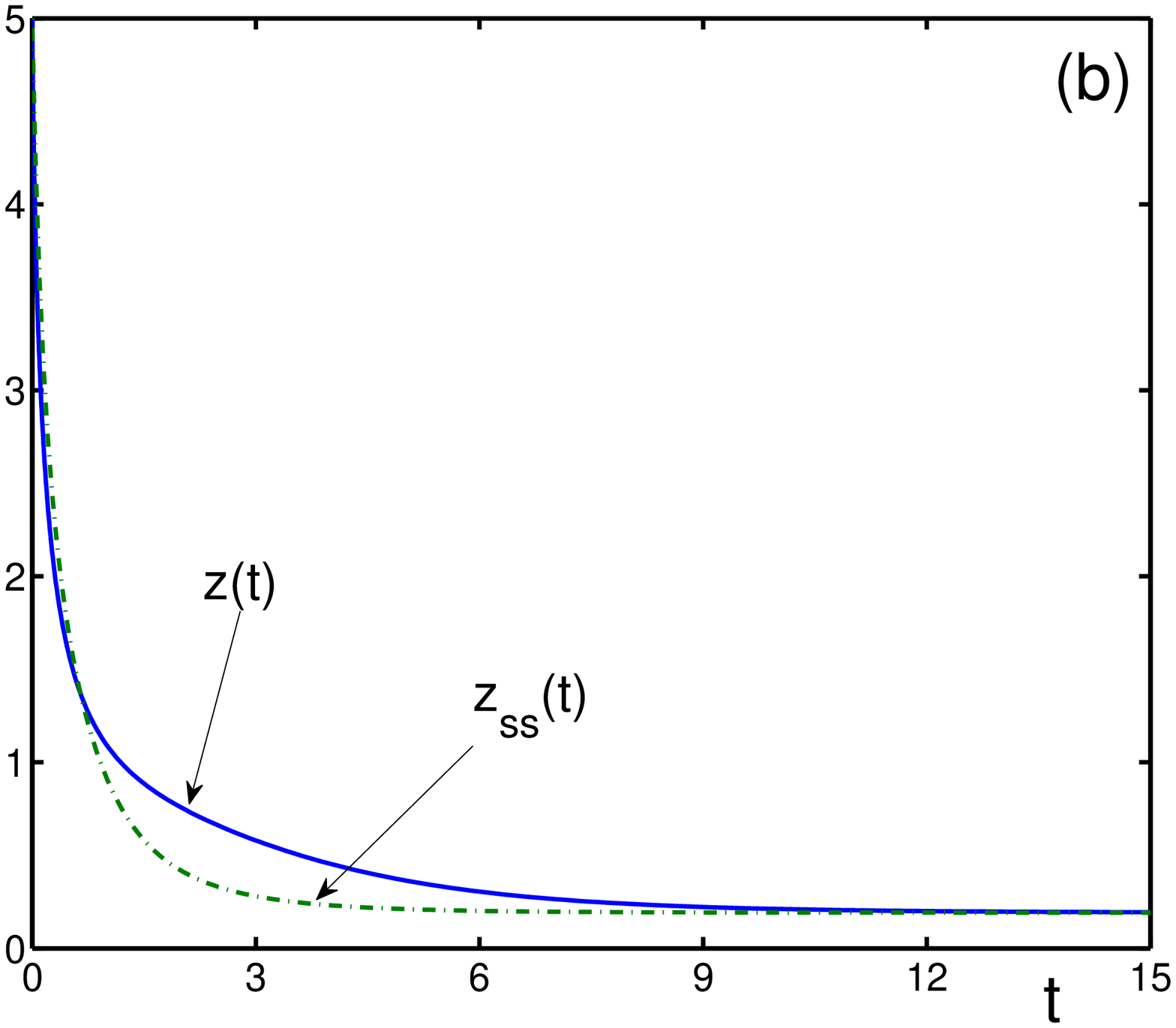} }  }
\vskip 9pt
\centerline{
\hbox{ \includegraphics[width=5.5cm]{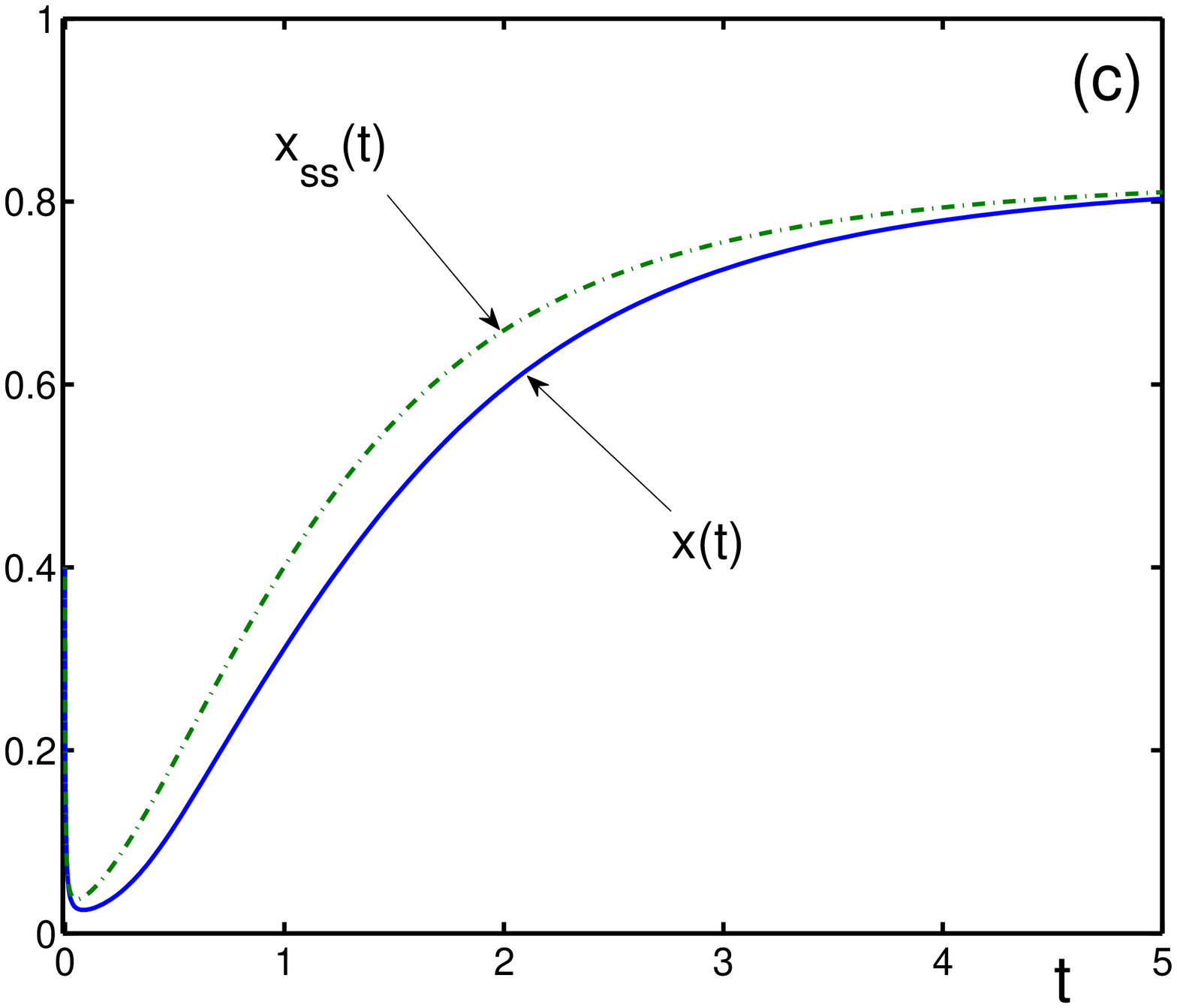} \hspace{2.5cm}
\includegraphics[width=5.5cm]{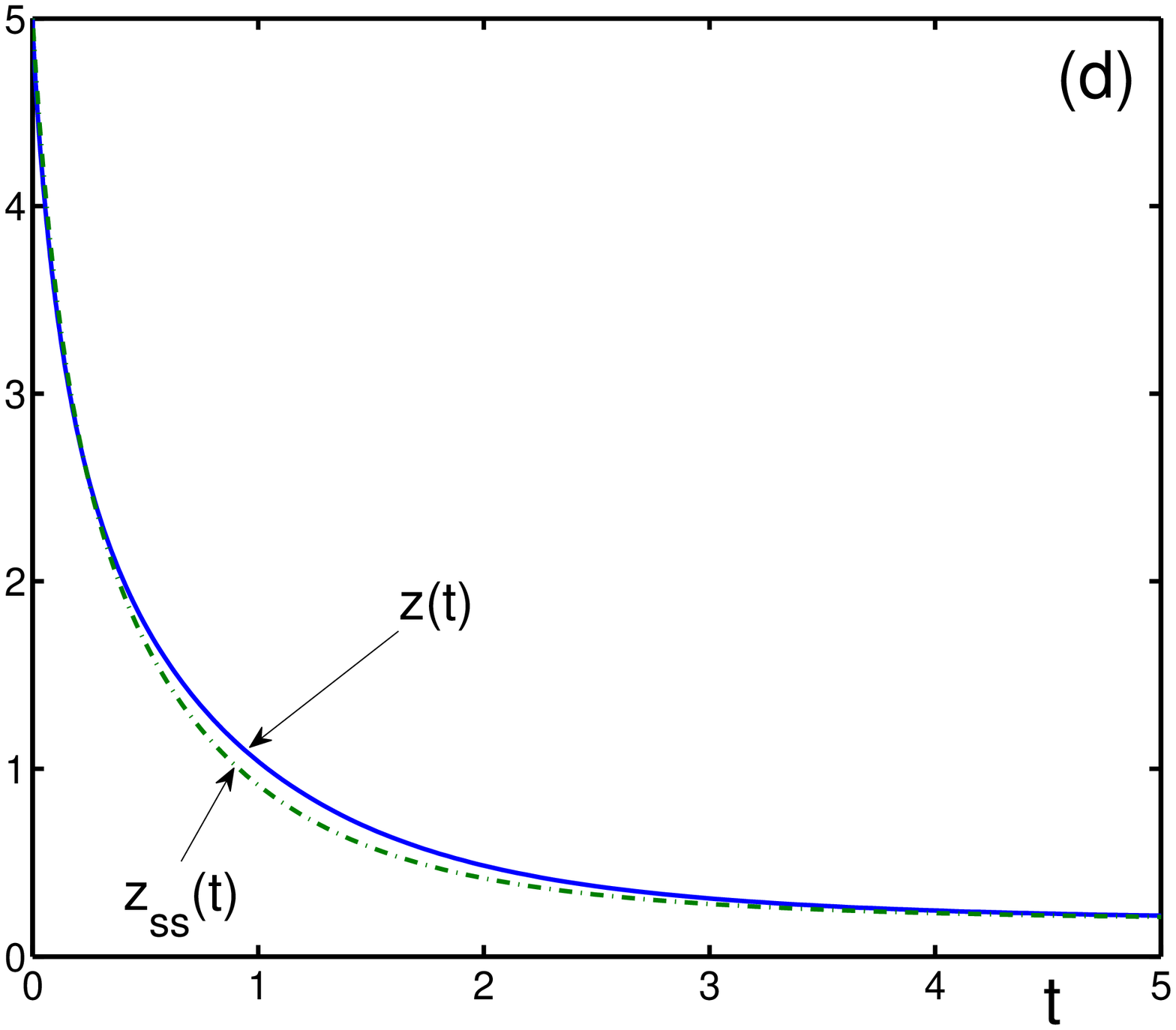} } }
\caption{Temporal behavior of approximate solutions $x_{ss}(t)$ and
$z_{ss}(t)$, obtained by the scale separation approach, as compared to the
exact solutions $x(t)$ and $z(t)$, for the case of parasitism, with the
symbiotic parameters $b = - 1$ and $g = - 2$, with the initial conditions
$x_0 = 0.4$ and $z_0 = 5$, for different growth rates $\al$.
The stable stationary state is $\{x^* = 0.82539, z^* = 0.1919\}$.
(a) Population $x(t)$ (solid line) and its approximation $x_{ss}(t)$
(dashed-dotted line) for $\al = 1$; (b) population $z(t)$ (solid line) and
its approximation $z_{ss}(t)$ (dashed-dotted line) for $\al = 1$;
(c) population $x(t)$ (solid line) and its approximation $x_{ss}(t)$
(dashed-dotted line) for $\al = 10$; (d) population $z(t)$ (solid line)
and its approximation $z_{ss}(t)$ (dashed-dotted line) for $\al = 10$.
}
\label{fig:Fig.17}
\end{figure}

\section{Conclusion}

In the case of a standard dynamic transition, a qualitative change of 
dynamical behavior occurs when a system parameter reaches a bifurcation 
point, where the nature of fixed points changes. We have demonstrated 
the existence of a non-standard dynamic transition, in which a qualitative 
change of dynamical behavior occurs as a result of the variation of the 
growth rate that does not influence the fixed points. The sharp change in 
dynamical behavior happens because the varying growth rate shifts the 
boundary of the basins of attraction of the fixed points, while the fixed 
points themselves do not change. Typically, the initial point of a trajectory, 
which was inside the attraction basin of the stable point for a first
value of the growth rate, can happen to be found outside of it
due to the change of the shape of the attraction basin for a different
value of the growth rate, or vice versa.

We have illustrated this dynamic transition, caused by the distortion of the 
shape of the basin of attraction, using a dynamical system describing the 
evolution of symbiotic species with different growth rates. The effect can 
happen under mutualism as well as under parasitism of the co-evolving species. 
As has been explained earlier \cite{YYS_13,YYS_14,YYS_30,YYS_16},
the considered symbiotic equations can characterize various biological and 
social systems. Biological systems have also much in common with economical 
systems \cite{Trenchard_28} as well as with structured human societies
\cite{Perc_27}. Therefore the described effect can occur in different nonlinear
dynamical systems.  

As an example, where the described effect does happen in nature, it is possible 
to mention the ubiquitous symbiosis between fungi and plants. The proliferation 
of the Arbuscular Mycorrhizal fungi network at a late stage in plant life is well 
established to be beneficial for plant growth and reproduction 
\cite{Kapulnik_22,Smith_23}. However, a too fast 
proliferation of this fungi at the early stage of the plant life cycle can lead to 
the suppression of plant seedling due to the carbon cost associated with 
sustaining the fungi \cite{Varga_8,Koide_24,Johnson_25,Ronsheim_26}. Here, the 
early or late plant life stage correspond to different initial conditions, when 
the plant is either still small or already mature. Depending on these initial 
conditions, the same fungi growth rate can be either beneficial for the plant 
or suppressing it, similarly to the cases considered in our article.

\nonumsection{Acknowledgments} \noindent 
We acknowledge financial support from the ETH Z\"{u}rich Risk Center.

\end{document}